\documentclass[twocolumn,trackchanges]{aastex62}
\usepackage[section]{placeins}
\usepackage{longtable}
\usepackage{hyperref}
\usepackage{url}
\usepackage{amsmath}
\graphicspath{{./}{figures/}}

\usepackage{enumitem}
\usepackage{comment}
\usepackage{fp}
\usepackage{fontawesome}
\usepackage{listings}
\usepackage{xspace}
\newcommand{\mosfit}{{\tt MOSFiT}\xspace}
\newcommand{\Ni}{$^{56}$Ni }

\definecolor{codegreen}{rgb}{0,0.6,0}
\definecolor{codegray}{rgb}{0.5,0.5,0.5}
\definecolor{codepurple}{rgb}{0.58,0,0.82}
\definecolor{backcolour}{rgb}{0.95,0.95,0.92}

\lstdefinestyle{mystyle}{
    backgroundcolor=\color{backcolour},   
    commentstyle=\color{codegreen},
    keywordstyle=\color{magenta},
    numberstyle=\tiny\color{codegray},
    stringstyle=\color{codepurple},
    basicstyle=\ttfamily\footnotesize,
    breakatwhitespace=false,         
    breaklines=true,                 
    captionpos=b,                    
    keepspaces=true,                 
    numbers=left,                    
    numbersep=5pt,                  
    showspaces=false,                
    showstringspaces=false,
    showtabs=false,                  
    tabsize=2
}

\newcommand{\total}{262}
\newcommand{\gold}{168}
\newcommand{\silver}{70}
\FPeval{\full}{clip(\gold + \silver)}
\newcommand{\bronze}{24}
\newcommand{\svrlc}{26}
\newcommand{\svrunc}{12}
\newcommand{\svrz}{2}
\newcommand{\svrext}{2}
\newcommand{\svrlsn}{21}
\newcommand{\svrphot}{7}
\newcommand{\brzic}{4}
\newcommand{\brzspec}{12}
\newcommand{\brzlc}{4}
\newcommand{\brzunc}{2}
\newcommand{\brzz}{2}

\usepackage[hang,flushmargin]{footmisc} 

\newcommand{\LCO}{\affiliation{Las Cumbres Observatory, 6740 Cortona Drive, Suite 102, Goleta, CA 93117-5575, USA}}
\newcommand{\UCSB}{\affiliation{Department of Physics, University of California, Santa Barbara, CA 93106-9530, USA}}

\newcommand{\OAPD}{\affiliation{INAF-Osservatorio Astronomico di Padova, Vicolo dell'Osservatorio 5, I-35122 Padova, Italy}}

\newcommand{\STScI}{\affiliation{Space Telescope Science Institute, 3700 San Martin Drive, Baltimore, MD 21218, USA}}

\newcommand{\QUB}{\affiliation{Astrophysics Research Centre, School of Mathematics and Physics, Queen's University Belfast, Belfast BT7 1NN, UK}}

\newcommand{\CfA}{\affiliation{Center for Astrophysics \textbar{} Harvard \& Smithsonian, 60 Garden Street, Cambridge, MA 02138-1516, USA}}

\newcommand{\CIERA}{\affiliation{Center for Interdisciplinary Exploration and Research in Astrophysics (CIERA) and Department of Physics and Astronomy, Northwestern University, Evanston, IL 60208, USA}}

\newcommand{\Steward}{\affiliation{Steward Observatory, University of Arizona, 933 North Cherry Avenue, Tucson, AZ 85721, USA}}

\newcommand{\IAIFI}{\affiliation{The NSF AI Institute for Artificial Intelligence and Fundamental Interactions, USA}}
\newcommand{\JHU}{\affiliation{Department of Physics and Astronomy, Johns Hopkins University, 3400 North Charles Street, Baltimore, MD 21218, USA}}
\newcommand{\UVA}{\affiliation{Department of Astronomy, University of Virginia, Charlottesville, VA 22904, USA}}
\newcommand{\Konkoly}{\affiliation{Konkoly Observatory, Research Center for Astronomy and Earth Sciences, H-1121 Budapest Konkoly Th. M. út 15-17., Hungary; MTA Centre of Excellence}}
\newcommand{\Konkolyb}{\affiliation{CSFK, MTA Centre of Excellence, Budapest, Konkoly Thege út 15-17., H-1121, Hungary}}
\newcommand{\Szeged}{\affiliation{Department of Experimental Physics, Institute of Physics, University of Szeged, D\'om t\'er 9, Szeged, 6720 Hungary}}
\newcommand{\Gothard}{\affiliation{ELTE Eötvös Loránd University, Gothard Astrophysical Observatory, Szombathely, Hungary}}
\newcommand{\Goethe}{\affiliation{Institut f\"ur Theoretische Physik, Goethe Universit\"at, Max-von-Laue-Str. 1, 60438 Frankfurt am Main, Germany}}

\shorttitle{The SLSN Catalog}
\shortauthors{Gomez et al.}

\begin{document}

\title{The Type I Superluminous Supernova Catalog I: Light Curve Properties, Models, and Catalog Description}

\correspondingauthor{Sebastian Gomez}
\email{sgomez@stsci.edu}

\author[0000-0001-6395-6702]{Sebastian Gomez}
\STScI

\author[0000-0002-2555-3192]{Matt Nicholl}
\QUB

\author[0000-0002-9392-9681]{Edo Berger}
\CfA\IAIFI

\author[0000-0003-0526-2248]{Peter K. Blanchard}
\CfA\CIERA

\author[0000-0002-5814-4061]{V. Ashley Villar}
\CfA\IAIFI

\author[0000-0002-3825-0553]{Sofia Rest}
\JHU

\author[0000-0002-0832-2974]{Griffin Hosseinzadeh}
\Steward

\author[0000-0002-9085-8187]{Aysha Aamer}
\QUB

\author[0009-0007-8764-9062]{Yukta Ajay}
\JHU

\author[0000-0002-2866-6416]{Wasundara Athukoralalage}
\CfA

\author[0000-0003-4263-2228]{David C. Coulter}
\STScI

\author[0000-0003-0307-9984]{Tarraneh Eftekhari}
\altaffiliation{NHFP Einstein Fellow}
\CIERA

\author[0000-0002-0403-3331]{Achille Fiore}
\Goethe\OAPD

\author[0000-0003-4537-3575]{Noah Franz}
\Steward

\author[0000-0003-2238-1572]{Ori Fox}
\STScI

\author[0000-0003-4906-8447]{Alexander Gagliano}
\CfA\IAIFI

\author[0000-0002-1125-9187]{Daichi Hiramatsu}
\CfA\IAIFI

\author[0000-0003-4253-656X]{D. Andrew Howell}
\LCO\UCSB

\author[0000-0002-9454-1742]{Brian Hsu}
\Steward

\author[0000-0003-2495-8670]{Mitchell Karmen}
\JHU

\author[0000-0003-2445-3891]{Matthew R. Siebert}
\STScI

\author[0000-0002-8770-6764]{R\'eka K\"onyves-T\'oth}
\Konkoly\Konkolyb\Szeged\Gothard

\author[0000-0003-0871-4641]{Harsh Kumar}
\CfA\IAIFI

\author[0000-0001-5807-7893]{Curtis McCully}
\LCO

\author[0000-0002-7472-1279]{Craig Pellegrino}
\UVA

\author[0000-0002-2361-7201]{Justin Pierel}
\altaffiliation{NHFP Einstein Fellow}
\STScI

\author[0000-0002-4410-5387]{Armin Rest}
\JHU\STScI

\author[0000-0001-5233-6989]{Qinan Wang}
\JHU

\begin{abstract}

We present the most comprehensive catalog to date of Type I Superluminous Supernovae (SLSNe), a class of stripped envelope supernovae (SNe) characterized by exceptionally high luminosities. We have compiled a sample of \total\ SLSNe reported through 2022 December 31. We verified the spectroscopic classification of each SLSN and collated an exhaustive data set of UV, optical and IR photometry from both publicly available data and our own FLEET observational follow-up program, totaling over 30,000 photometric detections. Using these data we derive observational parameters such as the peak absolute magnitudes, rise and decline timescales, as well as bolometric luminosities, temperature and photospheric radius evolution for all SLSNe. Additionally, we model all light curves using a hybrid model that includes contributions from both a magnetar central engine and the radioactive decay of $^{56}$Ni. We explore correlations among various physical and observational parameters, and recover the previously found relation between ejecta mass and magnetar spin, as well as the overall progenitor pre-explosion mass distribution with a peak at $\approx 6.5$ M$_\odot$. We find no significant redshift dependence for any parameter, and no evidence for distinct sub-types of SLSNe. We find that $< 3$\% of SLSNe are best fit with a significant contribution from radioactive decay $\gtrsim 50$\%, representing a set of relatively dim and slowly declining SNe. We provide several analytical tools designed to simulate typical SLSN light curves across a broad range of wavelengths and phases, enabling accurate K-corrections, bolometric scaling calculations, and inclusion of SLSNe in survey simulations or future comparison works. The complete catalog, including all of the photometry, models, and derived parameters, is made available as an open-source resource on GitHub \href{https://github.com/gmzsebastian/SLSNe}{\faGithub}.

\end{abstract}

\keywords{supernovae: general -- methods: statistical -- surveys -- stars: massive}

\section{Introduction}\label{sec:intro}

Core-collapse supernovae (CCSNe) result from the deaths of massive stars with zero-age main sequence (ZAMS) masses $\gtrsim 8$M$_\odot$ \citep{Woosley86}. Some of these massive stars lose their hydrogen, and in some cases helium, envelopes before undergoing core-collapse, which can lead to either a Type Ib SN if the star is deprived of hydrogen, or a Type Ic SN if it has lost both its hydrogen and helium \citep{Woosley95, Filippenko97}. Given their light curve evolution and spectral properties, it has been established that stripped-envelope SNe (SESNe) are powered by the radioactive decay of \Ni synthesized during the explosion \citep{Arnett82}.

Over a decade ago, a new class of SESNe was discovered and dubbed Type I superluminous supernovae (SLSNe), given a lack of hydrogen in their spectra and luminosities up to 100 times larger than normal SESNe \citep{Chomiuk11, Quimby11, Gal-Yam12, Howell17, Gal-Yam19}. Unlike normal SNe Ib/c, SLSNe cannot be powered by radioactive decay alone, but instead require an additional or alternative power source \citep{Dessart12, Inserra13, Nicholl13, Sukhbold16, Blanchard18, Margalit18}. Possible alternative models include the pair-instability or pulsational pair-instability (PISN or PPISN) mechanism \citep{Heger02, Woosley07, Gal-Yam09, Kasen11, Kozyreva15, Woosley17}, interaction with circumstellar material (CSM) around the progenitor star \citep{Chevalier11, Chatzopoulos13, Vreeswijk17, Yan17}, a contribution from jet launching \citep{Soker17, Soker22}, energy injected from a central engine such as a millisecond magnetar \citep{Kasen10, Woosley10, Inserra13, Mazzali16, Jerkstrand17_long, Liu17, Nicholl17_mosfit, Yu17, Cia18, Dessart19, Blanchard21_16inl, Lin20, Hsu21}, or fallback accretion from a black hole \citep{Dexter13, Kasen16, Moriya18}. It has also been suggested that some or all SLSNe could be powered by a combination of these models \citep{Inserra17, Gomez22_LSN, Chen22a, Chen22b, Chen17}.

\begin{figure*}
	\begin{center}
		\includegraphics[width=\textwidth]{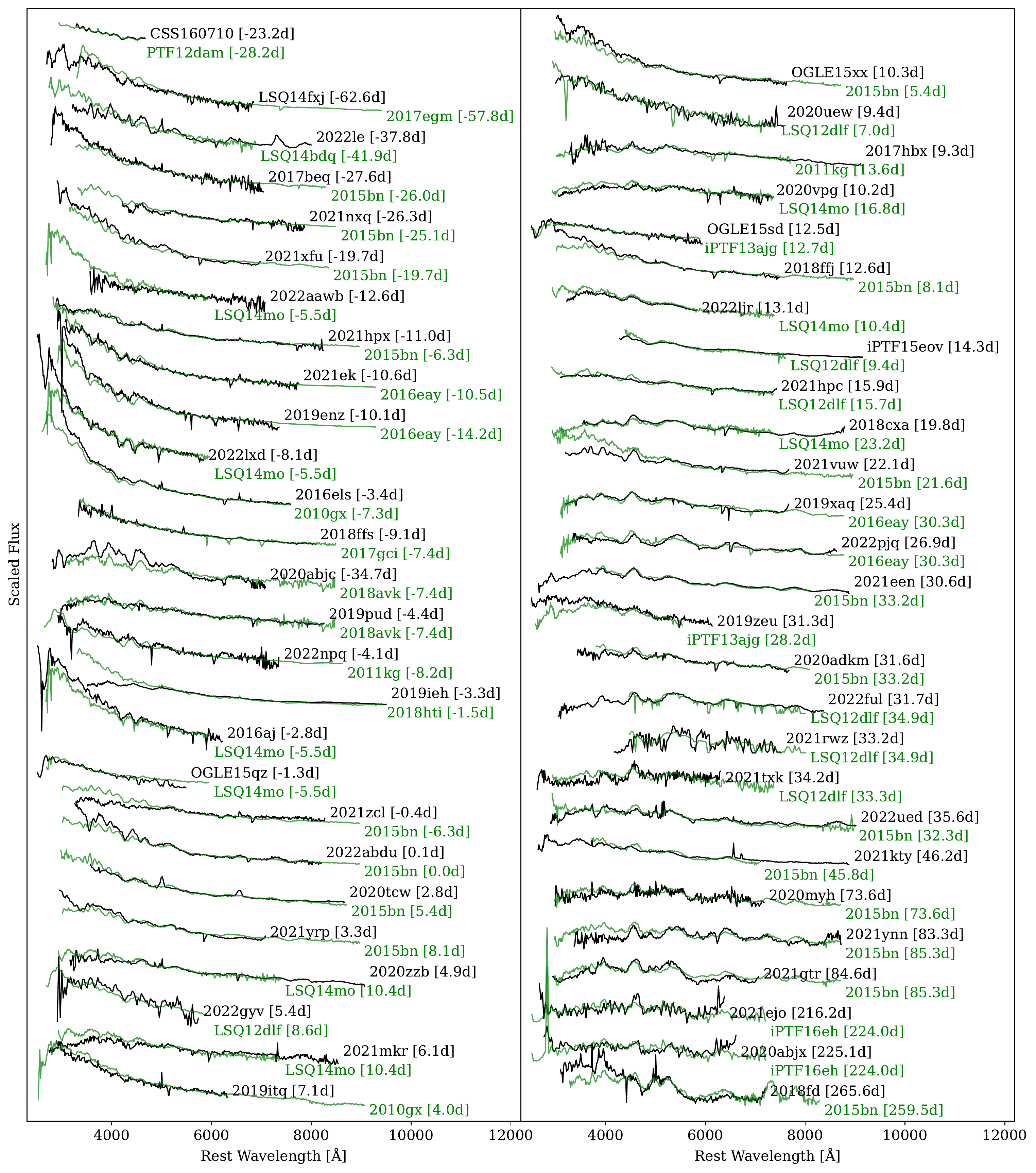}
		\caption{A representative spectrum for each SLSN in our sample that has not yet been presented in a refereed publication is shown in black, along with the corresponding best-matching spectrum from known SLSNe in green, used to verify their classification. The phase in brackets represents rest-frame days from peak. Individual sources and references for each SLSN are listed in the Appendix. \label{fig:spectra}}
	\end{center}
\end{figure*}

As SLSNe fade, their spectra tend to resemble those of SNe Ic and broad-lined SNe Ic (SNe Ic-BL), suggesting these populations are closely related \citep{Pastorello10, Inserra13, Quimby18, Blanchard19, Nicholl19_nebular}. In \cite{Gomez22_LSN} we showed how luminous supernovae (LSNe), or SESNe with peak magnitudes between those of SNe Ic/Ic-BL and SLSNe covering the range of $M_r = -19$ to $-20$ mag, are likely powered by a combination of a magnetar central engine and radioactive decay.

\begin{figure}
	\begin{center}
		\includegraphics[width=\columnwidth]{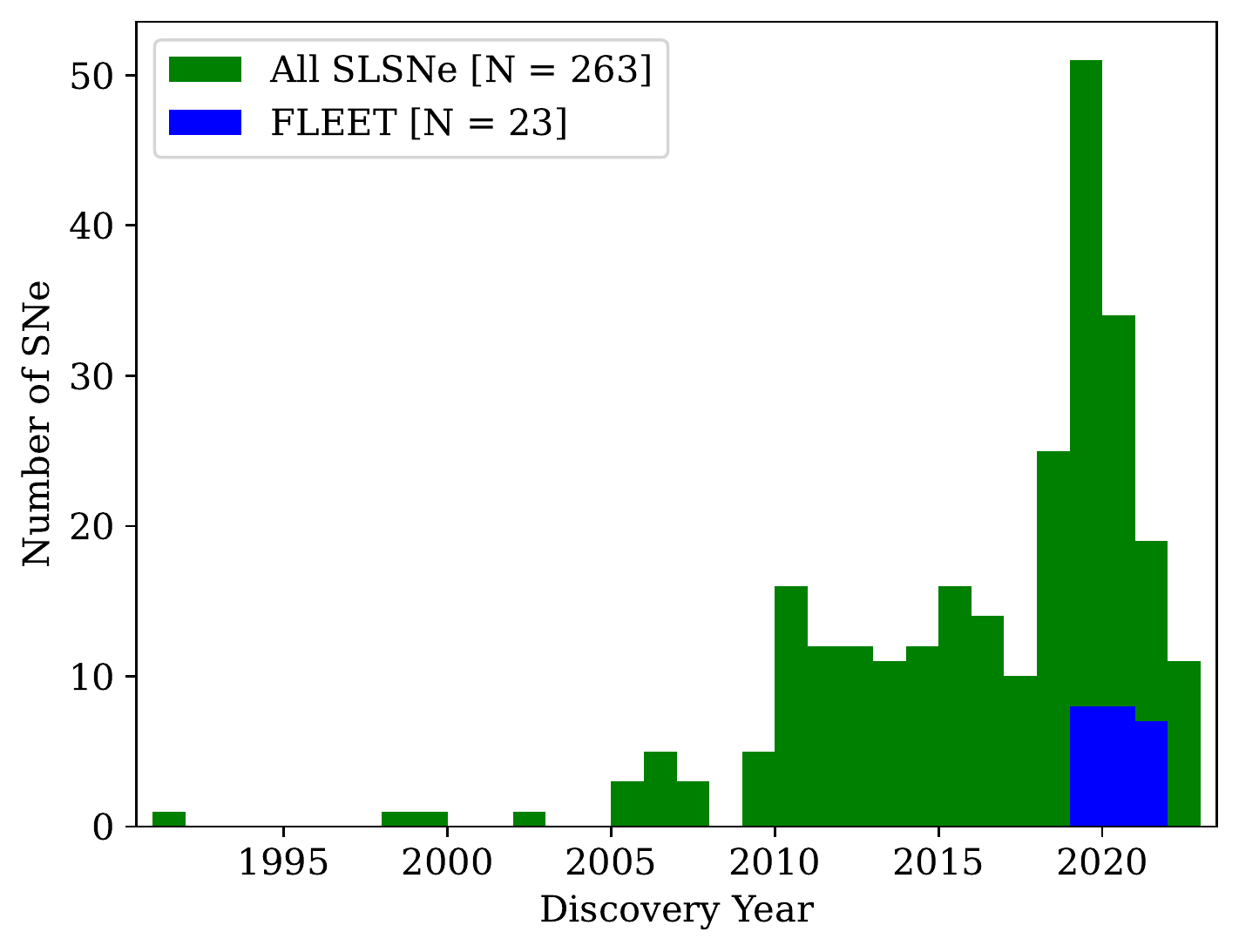}
		\caption{Histogram of the discovery year of all SLSNe in this work, including the bronze sample. In blue we show the SLSNe discovered as part of our FLEET observational program, which contributed to the peak of SLSN discoveries in 2020 to 2022. \label{fig:discovery}}
	\end{center}
\end{figure}

Besides their power source, there are several areas of active investigation regarding the nature of SLSNe, including their spectroscopic and photometric diversity \citep{Nicholl15_diversity, Inserra17, Inserra18, Gal-Yam19_spec, Konyves20, Konyves23}, progenitors \citep{Lunnan15, Aguilera18, Blanchard20}, connection to other transients \citep{Justham14, Japelj16, Metzger17, Margutti18, Margalit18b, Eftekhari21, Gomez21_2019stc, Gomez22_LSN, Liu22}, studies of their environments and host galaxies \citep{Lunnan14, Leloudas15, Angus16, Perley16, Chen17_host, Hatsukade18, Schulze18, Orum20, Hsu23}, whether they can be used as cosmological probes \citep{Inserra14, Scovacricchi16, Hsu21, Inserra21}, how to find them in large surveys \citep{Tanaka12, Villar18, Hsu22, Hosseinzadeh20, Villar20, Gagliano23, Gomez23_Roman, Sheng23}, and analyses of early and late time ``bumps'' in their light curves \citep{Hosseinzadeh22, Moriya22, Dong23, Zhu24}. Some of these topics have been addressed in studies of large samples of SLSNe \citep{Nicholl15_diversity, Perley16, Cia18, Nicholl17_mosfit, Liu17, Quimby18, Lunnan18, Angus19, Hinkle23}, with the largest study totaling 78 events \citep{Chen22a,Chen22b}. Here, we provide a compilation that is not limited to a single survey, but encompasses the full parameter space of known SLSNe with a sample size more than triple that of the previous largest sample.

In this work we include all known SLSNe discovered any time before 31 Dec 2022, totalling \total\ events. We verify their spectroscopic classification as SLSNe, and include all their publicly available photometry, in addition to photometry from our own Finding Luminous and Exotic Extragalactic Transients (FLEET) follow-up program \citep{Gomez20, Gomez23}. We model their light curves with a magnetar plus radioactive decay model, and provide both physical and observational parameters for the entire sample. This represents the first data release (DR1) of what will be a series of data releases on SLSNe, including a study on the photospheric spectra of SLSNe (Aamer, A. et al., in prep.) and their late-time nebular spectra (Blanchard, P. et al., in prep.). All data and products are publicly available on GitHub\footnote{\url{https://github.com/gmzsebastian/SLSNe}}, the supplementary materials of this journal, and Zenodo \citep{slsne}. The GitHub repository also contains Python scripts that can be used to either reproduce the plots in this paper, or to include the parameters of all known SLSNe in other works.

The structure of this paper is as follows. In \S\ref{sec:sample} we describe the sample of SLSNe and how their data were compiled. In \S\ref{sec:methods} we present our light curve modeling and population properties, and describe the results of this analysis in \S\ref{sec:population} and \S\ref{sec:results}. We outline the main conclusions in \S\ref{sec:conclusions}. In \S\ref{sec:catalog} we include a detailed description of the open-source catalog and Python examples on how to use the tools provided. In the Appendix we include details about each SN used in this sample. Throughout the paper we assume a flat $\Lambda$CDM cosmology based on the Planck 2018 results with \mbox{$H_{0} = 67.8$ km s$^{-1}$ Mpc$^{-1}$} and $\Omega_{m} = 0.308$ \citep{Planck18}.

\begin{figure}
	\begin{center}
		\includegraphics[width=1.025\columnwidth]{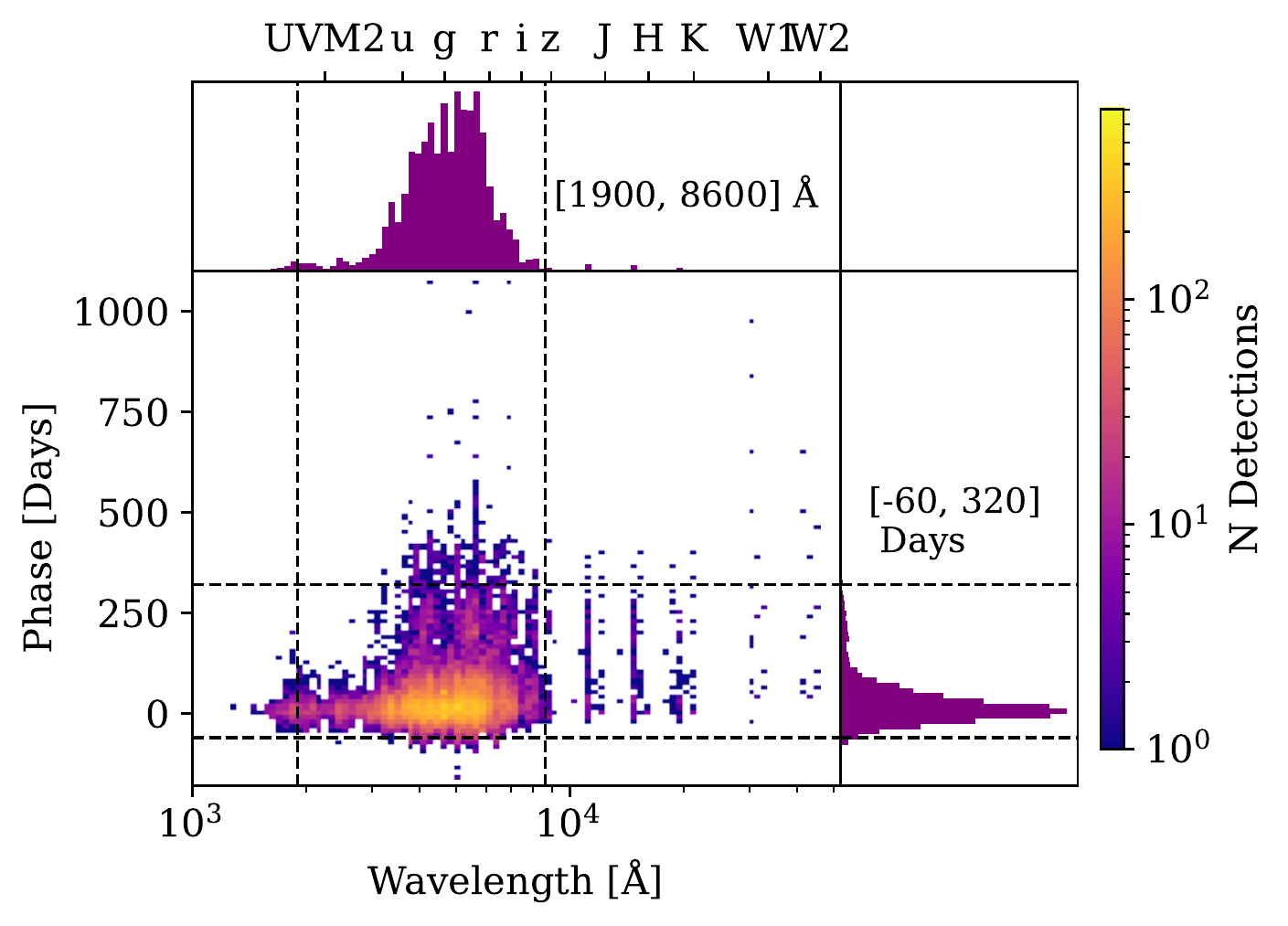}
            \caption{2D histogram of rest-frame phase in days versus rest-frame wavelength for all photometric measurements of the full sample of SLSNe. The dashed lines represent the 1st and 99th percentile of each histogram, emphasizing the abundance of optical data within a few months of peak brightness. \label{fig:counts}}
	\end{center}
\end{figure}

\section{SLSN Sample and Data}\label{sec:sample}

\subsection{Sample Definition}

We compile a sample of all known SLSNe discovered before 31 Dec, 2022, and aim to include all their available photometry. The SNe in this sample are compiled from the Open Supernova Catalog\footnote{\label{ref:osc}\url{https://github.com/astrocatalogs/supernovae}} (OSC; \citealt{Guillochon17}), the Transient Name Server (TNS)\footnote{\label{ref:tns}\url{https://www.wis-tns.org/}}, the Weizmann Interactive Supernova Data Repository (WISeREP; \citealt{Yaron12})\footnote{\url{https://www.wiserep.org/}}, a literature search, and our FLEET follow-up program \citep{Gomez20,Gomez23}.

\begin{figure*}
	\begin{center}
		\includegraphics[width=\textwidth]{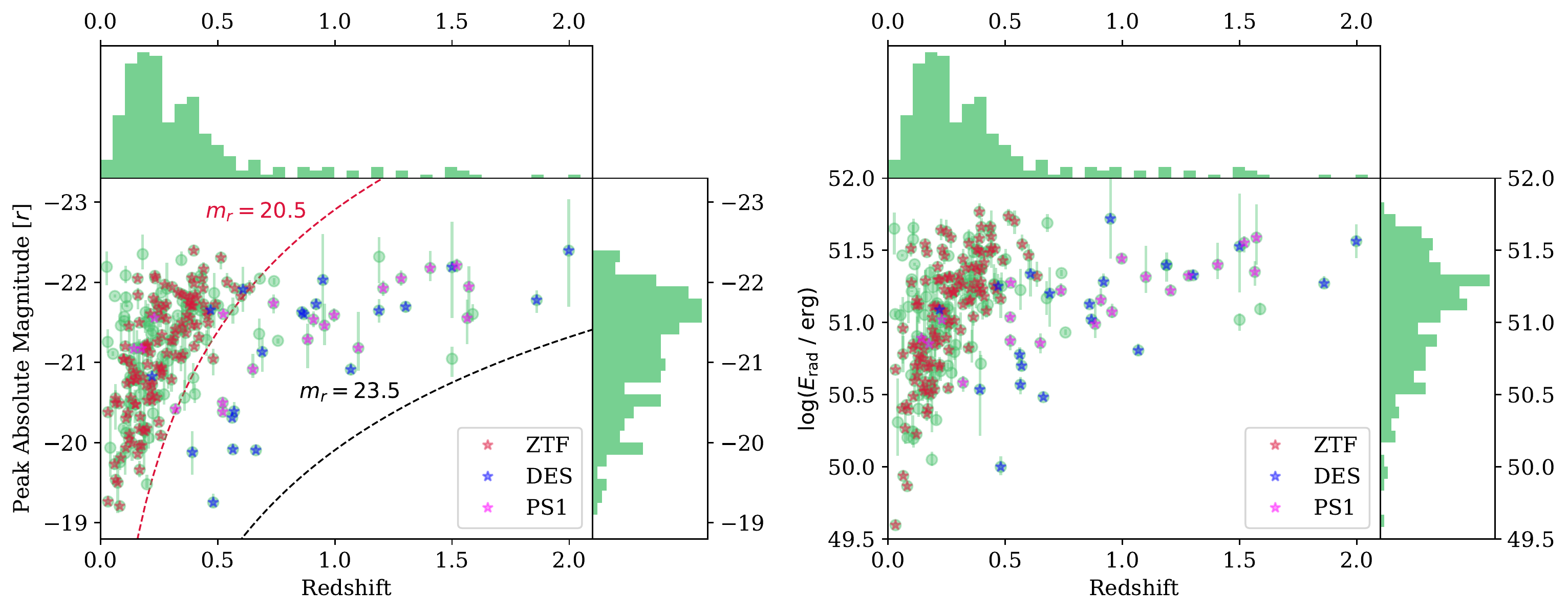}
            \caption{\textit{Left}: Peak absolute magnitude in rest-frame $r$-band as a function of redshift for the full SLSN sample. In red, blue, and pink we show SLSNe detected in several large surveys: ZTF, DES, and PS1, respectively. The red dashed line indicates the nominal magnitude limit of ZTF, and the black line the limit for PS1 and DES (we note that SLSNe from PS1 and DES are further limited by a rough spectroscopic follow-up limit of $\sim 22.5$ mag). \textit{Right}: Total radiated energy $E_{\rm rad}$ during the first 200 days of the light curve for the full SLSN sample. \label{fig:absmag_redshift}}
	\end{center}
\end{figure*}

To confirm their nature as SLSNe we require at least one public spectrum for each SN. We obtain these spectra from either the OSC, TNS, WISeREP, published works, or our FLEET program. To verify that the SNe in this sample are consistent with being SLSNe we visually match their spectra to reference spectra from known SLSNe. Special care is taken for SNe that have not yet been presented in a refereed publication. We show the spectra of all 54 unpublished SLSNe, as well as their best matching templates in Figure~\ref{fig:spectra}. The individual references for the spectra used are listed in the Appendix.

We divide the sample of \total\ SNe into \gold\ ``gold'', \silver\ ``silver'', and \bronze\ ``bronze'' SLSNe. Gold SLSNe have spectra consistent with a SLSN and photometry available before and after peak in more than one band. We assign a silver label if any of the following conditions are met: there is no photometry before or after peak (\svrlc\ events); the transient was also classified as a LSN (\svrlsn); the spectra are too noisy to provide a confident classification, but are nevertheless consistent with a SLSN (\svrunc); photometry is only available in one band (\svrphot); the redshift measurement is uncertain but the closest redshift still places the peak absolute magnitude brighter than $M_r = -20$ mag (\svrz); or if there are signs of high extinction with an uncertain measurement (\svrext). Alternatively, we assign a bronze label if any of the following conditions are met: no public spectra of the source are available (\brzspec); the transient has been classified in the literature as a SLSN, but we find it to be most consistent with a SN Ic based on its spectra and peak magnitude (\brzic); the redshift measurement is uncertain enough such that the lowest redshift places the peak absolute magnitude dimmer than $M_r = -20$ mag (\brzz); there are less than three data points of photometry available (\brzlc); the spectroscopic classification cannot be accurately determined due to low signal-to-noise spectra (\brzunc).

In addition to the sample of confirmed SLSNe, we include a list of nine objects that have been previously suggested to be SLSNe, but which we argue are likely not SLSNe. These objects, along with the list of all SLSNe are included in the Appendix with individual notes on each object, including their peculiarities, as well as the sources of their discovery and data. In Figure~\ref{fig:discovery} we show a histogram of all SLSNe included in this work (gold, silver, and bronze) as a function of their discovery year. The number of SLSNe discovered after 2022 declined since both the FLEET and ZTF follow-up programs decreased their focus on classifying SLSNe after this point. For all subsequent analyses and plots we include only the \full\ gold and silver SLSNe and refer to this as the ``full sample'' or ``all SLSNe'', unless otherwise stated.

\begin{figure*}
    \begin{center}
        \includegraphics[width=\textwidth]{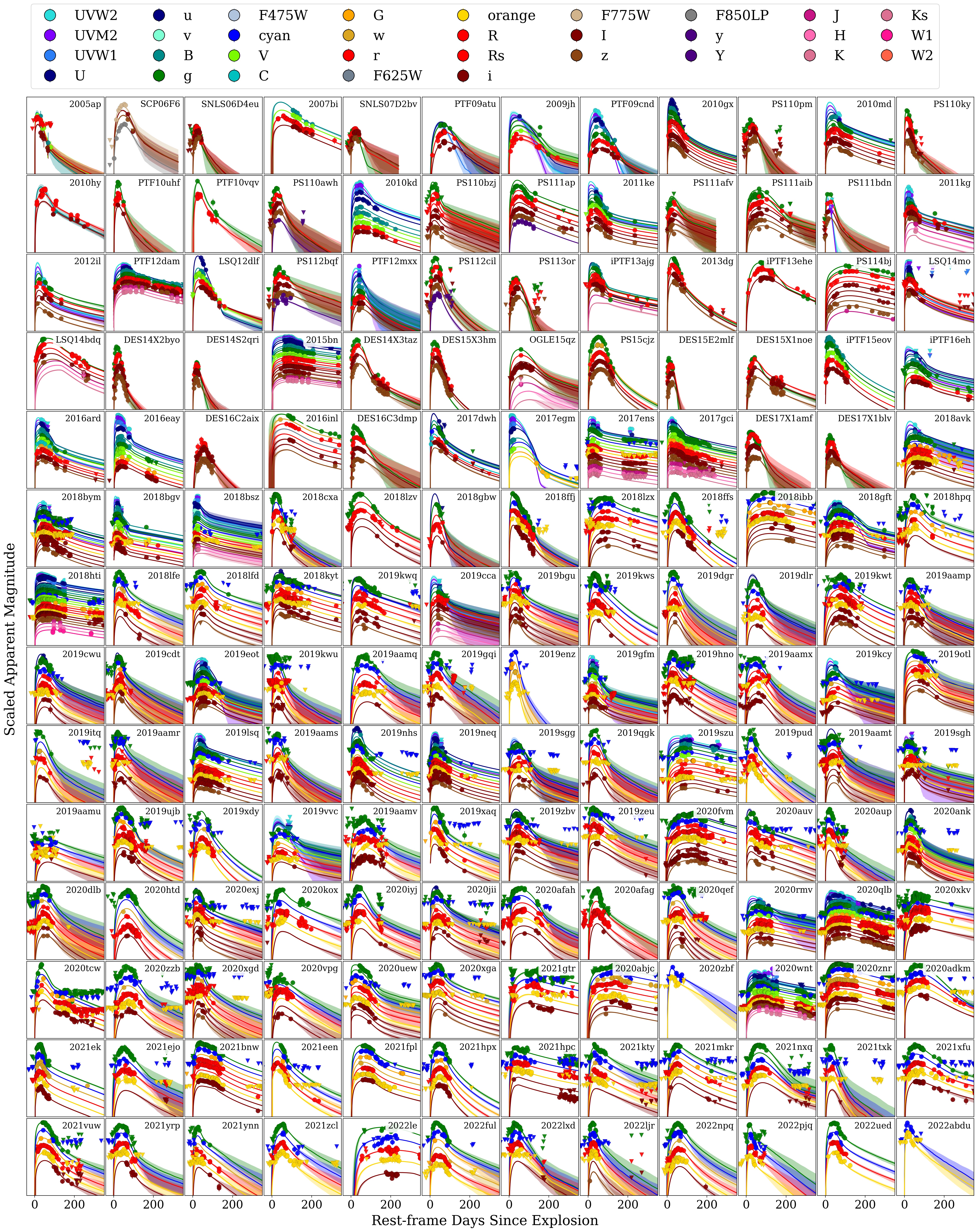}
        \caption{Light curves and best-fit \mosfit models for all \gold\ gold SLSNe, shown in order of discovery date. Upper limits are shown as inverted triangles. \label{fig:all_lightcurves}}
	\end{center}
\end{figure*}

\begin{figure}
    \begin{center}
        \includegraphics[width=\columnwidth]{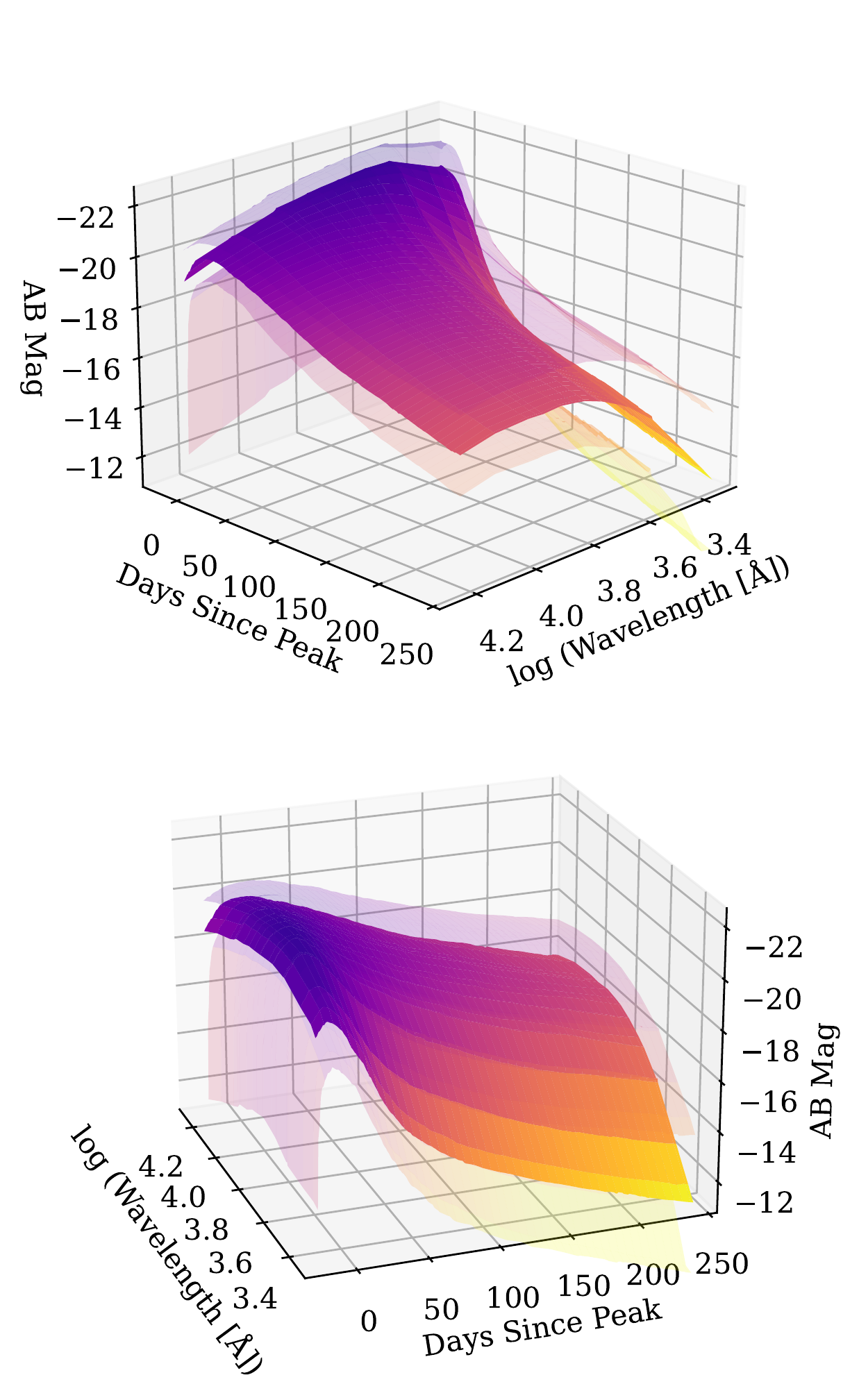}
        \caption{3D visualization seen from two vantage points of the mean absolute AB magnitude as a function of phase and rest-frame wavelength for the full sample of SLSNe. The solid contour shows the mean of the distribution, while the shaded contours represent the $\pm 1 \sigma$ range in the population. The color represents the AB magnitude. \label{fig:heatmap_mosfit}}
	\end{center}
\end{figure}

\subsection{Photometry}\label{sec:photometry}

We compiled UV, optical, and IR photometry of all SLSNe from a variety of sources. Photometry from the Zwicky Transient Facility (ZTF; \citealt{Bellm19}) is taken either from the public archive using the Automatic Learning for the Rapid Classification of Events (ALeRCE) broker \citep{Forster20}, or from our own photometry done on raw ZTF images obtained from the NASA/IPAC Infrared Science Archive\footnote{\url{https://irsa.ipac.caltech.edu/Missions/ztf.html}}. We include photometry from the Asteroid Terrestrial-impact Last Alert System (ATLAS; \citealt{Tonry18, Smith20}), the All Sky Automated Survey for SuperNovae (ASAS-SN; \citealt{Shappee14, Kochanek17, Hart23}), the \textit{Gaia} Science Alerts (GSA; \citealt{Wyrzykowski16}), the Optical Gravitational Lensing Experiment (OGLE; \citealt{Wyrzykowski14}), the Catalina Real Time Transient Survey (CRTS; \citealt{Drake09}), the Pan-STARRS Survey for Transients (PSST; \citealt{Huber15}), the Dark Energy Survey (DES; \citealt{Angus19}, and the Swift Optical/Ultraviolet Supernova Archive (SOUSA; \citealt{Brown14}). In addition to data available from public repositories, we collect photometry from individual publications that have these data available, either from the supplementary materials of their journals, the TNS, WISeREP, the OSC, or private communication with the authors. The individual data sources, as well as any details pertaining to the photometry of each individual SN are listed in the Appendix.

\begin{deluxetable}{lllp{3cm}}
	\tablecaption{\mosfit Parameter Definitions \label{tab:parameters}}
	\tablehead{ & \colhead{Prior} & \colhead{Units} & \colhead{Definition}}
	\startdata
	$M_{\text{ej}}$        & $[0.1, 100]$              &  M$_\odot$     & Ejecta mass  \\
	$f_{\text{Ni}}$        & $\log((0, 0.5])$          &                & Nickel mass as a fraction of the ejecta mass  \\
	$V_{\text{ej}}$        & $\log([10^3, 10^5])$      &  km s$^{-1}$   & Ejecta velocity  \\
	$M_{\text{NS}}$        & $1.7 \pm 0.2$             &  M$_\odot$     & Neutron star mass   \\
	$P_{\text{spin}}$      & $[0.7, 30]$               &  ms            & Magnetar spin   \\
	$B_{\perp}$            & $\log((0, 15])$           &  $10^{14}$ G   & Magnetar magnetic field strength \\
	$\theta_{\text{BP}}$   & $[0, \pi/2]$              &  rad           & Angle of the dipole moment \\
	$t_{\text{exp}}$       & $[0, 200]$                &  days          & Explosion time relative to first data point  \\
	$T_{\text{min}}$       & $[3000, 10000]$           &  K             & Photosphere temperature floor  \\
	$\lambda_0$            & $[2000, 6000]$            &  \AA           & Flux below this wavelength is suppressed by $\alpha$ \\
	$\alpha$               & $[0, 5]$                  &                & Slope of the wavelength suppression \\
	$n_{H,\text{host}}$    & $\log([10^{16},10^{23}])$ &  cm$^{-2}$     & Column density in the host galaxy \\
	$\kappa$               & $[0.01, 0.34]$            &  cm$^2$g$^{-1}$& Optical opacity \\
	$\kappa_{\gamma}$      & $\log([0.01, 0.5])$       &  cm$^2$g$^{-1}$& Gamma-ray opacity  \\
	$\sigma$               & $[10^{-3}, 10^2]$         &                & Uncertainty required for $\chi^2_r=1$ \\
	\enddata
    \tablecomments{Parameters used in the \mosfit model, their priors, units, and definitions. Priors noted in $\log$ have a log-flat prior, priors without it are flat in linear space, and priors with a center and error bars have a Gaussian distribution.}
\end{deluxetable}

Additionally, we include photometry from our FLEET follow-up program \citep{Gomez20,Gomez23}. Images from this program were taken with either the KeplerCam imager on the 1.2-m telescope at the Fred Lawrence Whipple Observatory (FLWO), the Low Dispersion Survey Spectrograph (LDSS3C; \citealt{stevenson16}) or Inamori-Magellan Areal Camera and Spectrograph (IMACS; \citealt{dressler11}), both on the Magellan Clay 6.5-m telescopes at Las Campanas Observatory, or Binospec \citep{Fabricant19} on the MMT 6.5-m telescope. We also include $gri$ images taken by the Global Supernova Project (GSP) with the Las Cumbres Observatory's global telescope network (Las Cumbres; \citealt{Brown13}).

\begin{figure*}
	\begin{center}
		\includegraphics[width=0.9\textwidth]{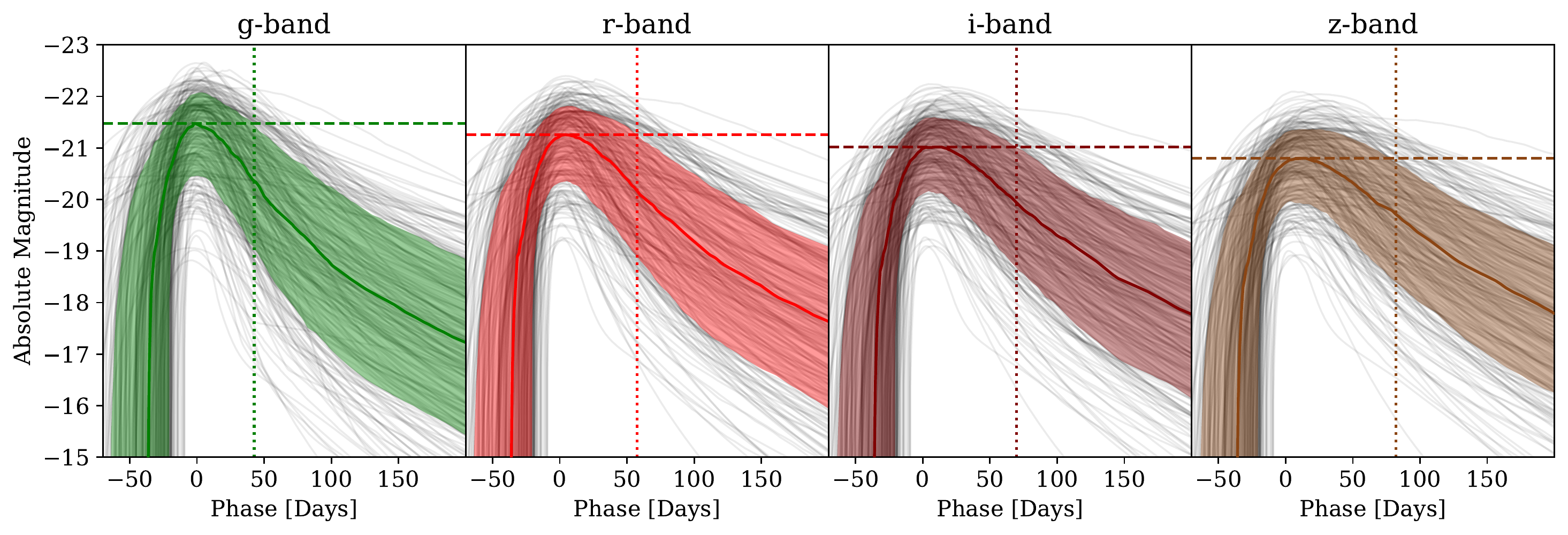}
		\caption{Rest-frame absolute magnitude light curves in the $griz$ bands for the full sample of SLSNe as a function of phase from their respective peaks. The gray lines show the light curves of each individual SLSN, while the solid color curves and shaded regions represent the mean evolution of the full sample and their 1$\sigma$ ranges, respectively. The horizontal dashed and vertical dotted lines demarcate the mean peak magnitude in each band and the mean $\tau_e$ decline timescale in each band. \label{fig:SLSNe_LCs}}
	\end{center}
\end{figure*}

We perform photometry on all images in a uniform way. Instrumental magnitudes were measured by modeling the point-spread function (PSF) of each image using field stars and fitting the model PSF to the target. The magnitudes are then calibrated to AB magnitudes from the PS1/$3\pi$ catalog \citep{Chambers16}. For the majority of sources, we separate the flux of the SN from its host galaxy by doing difference imaging using a pre-explosion PS1/$3\pi$ template for comparison. We subtract the PS1/$3\pi$ template from the science images using {\tt HOTPANTS} \citep{Becker15}. To process the ATLAS photometry we use {\tt ATClean}, a new pipeline designed to use the fluxes and uncertainties from the ATLAS forced photometry service \citep{Shingles21} to produce binned and statistically cleaned photometry \citep{Rest24}. For some sources we determine the host galaxy contribution to be negligible and report PSF photometry taken directly from the science images without subtracting a template. Notes on the data reduction procedure for individual SN are listed in the Appendix.

For the photometry that is not corrected for foreground Milky Way extinction, we correct it using the dust maps from \cite{Schlafly11} and the {\tt Astropy} \citep{astropy} implementation of the \cite{Gordon23} extinction law. This recent extinction law measurement represents the only one that provides accurate measurements extending from 912\AA\ to 32$\mu$m \citep{Gordon09, Fitzpatrick19, Decleir22}, with the most significant deviations from the more commonly used \cite{Cardelli89} extinction law occurring in the IR, above $\sim 1 \mu$m. In a few cases, the existing SLSN photometry is already provided corrected for extinction, in which case we do not apply any additional corrections.

In Figure~\ref{fig:counts} we show a 2D-histogram of all detections of the full sample as a function of rest-frame wavelength and phase. Throughout this work we define phase as rest-frame days from $r$-band peak, unless otherwise stated. The histogram contains more than 33,265 detections for the \full\ SLSNe in the full sample. We include lines to demarcate the 1st and 99th percentile coverage in terms of wavelength and phase. While there are ample optical observations of SLSNe during peak, there are very few observations of SLSNe bluewards of 1900~\AA, redwards of 8600~\AA, or later than 320 days after peak. The sample has a mean value of 130 detections per SN, and a median of 78.

The peak of the redshift distribution of the full sample of SLSNe is $z \approx 0.26$, which depends on the depth of the surveys finding these SLSNe. Most SLSNe were discovered by relatively shallow surveys like ZTF and ATLAS, which have a magnitude limit of $m_r \sim 20.5$, and only a few SLSNe were found by deeper surveys like DES or the PS1 Medium Deep Survey (MDS; \citealt{Lunnan18}), which have a limit of $m_r \sim 23.5$. In Figure~\ref{fig:absmag_redshift} we show the peak absolute magnitude in $r$-band of the full sample as a function of redshift, with representative surveys marked with different colors. The sample of spectroscopically confirmed SLSNe does not extend down to the photometric limit of $\sim 23.5$ given that we are usually constrained by the shallower limit of spectroscopic observations. On the right panel of Figure~\ref{fig:absmag_redshift} we also show the total radiated energy during the first 200 days of the light curve as a function of redshift for the full sample. Of the full sample, 99\% of SLSNe have a total radiated energy $E_{\rm rad} < 4.7 \times 10^{51}$ erg, above which there is a sharp drop-off, seen in the histogram on the right panel of Figure~\ref{fig:absmag_redshift}.

\section{Methods}\label{sec:methods}

\subsection{Light Curve Modeling}\label{sec:modeling}

\begin{figure*}
	\begin{center}
		\includegraphics[width=\textwidth]{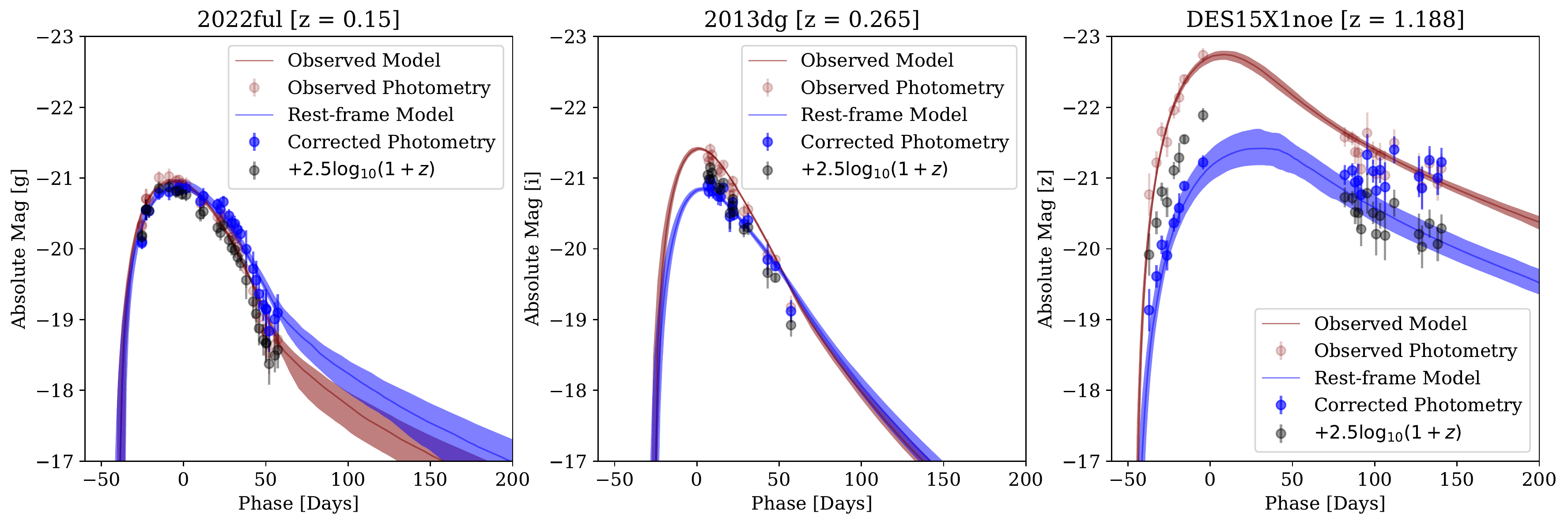}
		\caption{Example of applying our optimal K-correction model to a few representative SLSNe, spanning a broad range of redshifts. The brown points and shaded regions represent the observed photometry and best-fit model, respectively. The blue points and shaded regions are the same data but after applying our K-correction model. The black points represent the same photometry but modified using only a simple K-correction of $+2.5 \log_{10} (1 + z)$. Our method outperforms the simple K-correction for most cases. \label{fig:K_correction_map}}
	\end{center}
\end{figure*}

We model the light curves of all SLSNe using the Modular Open-Source Fitter for Transients (\mosfit) package, a flexible Python code that uses the {\tt emcee} \citep{Foreman13} implementation of Markov chain Monte Carlo (MCMC) to fit the light curves of transients using a variety of different power sources \citep{guillochon18}. Here, we assume a magnetar central engine to be the dominant power source of SLSNe. Additionally, we account for contributions from radioactive decay, recognized as the primary power source in Type Ic SNe, the less luminous analogs to SLSNe. The choice to model the light curves with a magnetar central engine comes from the fact that this model has been able to accurately reproduce the light curves (e.g., \citealt{Nicholl17_mosfit}), late-time evolution (e.g., \citealt{Kasen10}), and spectra (e.g., \citealt{Mazzali16}) of most SLSNe. A significant advantage of this model is its ability to replicate the light curves of SLSNe with a minimal set of parameters, particularly when compared to more complex models such as those powered by CSM interaction (e.g., \citealt{Chevalier11, Chatzopoulos13, Chen22b}). Modeling the entire dataset with distinct models is beyond the scope of this work. Regardless of the model choice, some physical parameters such as the ejecta mass and velocity, should still correlate with the diffusion time-scale, which is related to the duration of the light curve. Additionally, we use these models to derive model-independent properties from the light curves of SLSNe.

The parameters being fit in the models, their prior ranges, units, and definitions are listed in Table~\ref{tab:parameters}. The original magnetar model used in \mosfit is defined in \cite{Nicholl17_mosfit}. The models, priors, and parameter constraints used here are identical to those used to model LSNe in \cite{Gomez22_LSN}. The main difference from the original \cite{Nicholl17_mosfit} models and the ones used to model LSNe is the inclusion of an additional radioactive decay component. We also allow for the suppression $\alpha$ of the flux blue-wards of some wavelength $\lambda_0$ to vary between 0 and 5, as opposed to being fixed to 1. Given that the value of the neutron star mass $M_{\rm NS}$ has little to no effect on the output light curve, we impose a Gaussian prior of $M_{\rm NS} = 1.7 \pm 0.2 $ M$_\odot$ motivated by the typical masses of neutron stars \citep{Ozel16}. We run each model with 150 walkers and test for convergence by ensuring that the models reach a potential scale reduction factor of $<1.3$ \citep{gelman92}, which corresponds to a few\,$\times 10^4$ steps, depending on the SN. We show the light curves and best-model fits of all Gold SLSNe in Figure~\ref{fig:all_lightcurves}.

We use the posterior distribution of all the best-fit parameters from \mosfit to calculate the values of additional physical parameters. We use the individual values of all walkers to obtain the most accurate estimate for the mean value and correlations between these derived parameters. We include measurements for the total nickel mass $M_{\rm Ni}$, as well as for the initial magnetar spin-down luminosity $L_0$ and spin-down time $t_{\rm SD}$. The latter two parameters are included to allow for a direct comparison to the model of \cite{Omand23}, where these parameters are defined. We also include a measurement of the kinetic energy KE $= (3/10) M_{\rm ej} V_{\rm ej}^2$, the radiative efficiency between luminosity $L$ and kinetic energy $\epsilon = L / {\rm KE}$, and a measurement of $f_{\rm mag}$, or the fraction of the total luminosity during the first 200 days that comes from the magnetar contribution, as opposed to radioactive decay.

Lastly, we use the \mosfit light curve models to calculate a series of observational parameters. We provide rest-frame light curves for all SLSNe by generating \mosfit models based on the best-fit parameters of each SLSN, but at a fixed distance of 10 pc and with no host extinction. These models represent rest-frame SLSN light curves in $ugrizy$, UVBRIJHKs, and all \textit{Swift} bands. We do not include models redder than $\lambda = 1.58\ \mu$m, since observations in these bands have been shown to be heavily affected by emission from dust \citep{Sun22}, and the models would therefore not be representative of real SLSN observations. We use these rest-frame models to measure values for $\Delta m_{15}$, the magnitudes by which a SN fades 15 days after maximum in B-band; $\tau_e$, the number of days it takes for a SN to fade from bolometric peak by a factor of $e$; $\tau_{\rm 1}$, the number of days it takes for a SN to fade by 1 mag in $r$-band; $\tau_{\rm rise}$, the number of days it takes a SN to go from explosion to $r$-band peak; $M_{\rm max}$, the peak absolute magnitude in rest-frame $r$-band; $L_{\rm max}$, the bolometric luminosity at peak; and $E_{\rm rad}$, the total radiated energy during the first 200 days after explosion.

\startlongtable
\begin{deluxetable*}{cc|cc|cc|cc}
    \tablecaption{Mean and Standard Deviation of Various Observational and Physical Parameters \label{tab:mean}}
    \tablehead{\colhead{Parameter}   & \colhead{Value}      &   \colhead{Parameter} & \colhead{Value} &   \colhead{Parameter} & \colhead{Value}  &  \colhead{Parameter} & \colhead{Value}  }
    \startdata
    $z$ & $0.26^{+0.3}_{-0.13}$ & $\log(L_{\rm max} /{\rm erg\ s}^{-1})$ & $44.3^{+0.3}_{-0.5}$ & $M_{\rm Ni} /{\rm M}_\odot$ & $0.3^{+1.9}_{-0.2}$ & $\log(n_{H,\text{host}} / {\rm cm}^{-2})$ & $18.2^{+0.5}_{-0.4}$  \\ 
    $t_{\rm rise} / $days & $27^{+25}_{-13}$ & $P_{\rm spin} /{\rm ms}$ & $2.4^{+3.0}_{-1.2}$ & $\log({\rm KE} /{\rm erg\ s}^{-1})$ & $51.4 \pm 0.4$ & $\lambda /$\AA & $3400^{+1000}_{-700}$  \\ 
    $\tau_{\rm 1} / $days & $53^{+35}_{-20}$ & $\log(B_{\perp} / G)$ & $14.2 \pm 0.4$ & $\log(L_0 /{\rm erg\ s}^{-1})$ & $45.8^{+1.9}_{-1.3}$ & $\alpha$ & $1.7^{+1.5}_{-1.3}$  \\ 
    $\tau_{\rm e} / $days & $44^{+38}_{-18}$ & $T_{\text{min}} /$K & $6500^{+1700}_{-1400}$ & $\log(\rm t_{\rm SD} / s)$ & $5.7 \pm 1.0$ & ($M_{\text{NS}} /{\rm M}_\odot$)\tablenotemark{\dag} & $1.72 \pm 0.04$  \\ 
    $\Delta m_{15} / $mag & $0.2^{+0.2}_{-0.1}$ & $V_{\rm ej} / 1000 {\rm\ km\ s}^{-1}$ & $6.8^{+3.4}_{-2.0}$ & $f_{\rm mag}$ & $1.0^{+0.0}_{-0.1}$ & ($\theta_{\text{BP}} /{\rm rad}$)\tablenotemark{\dag} & $1.0^{+0.1}_{-0.2}$  \\ 
    $M_{\rm max}/$mag & $-21.3^{+0.9}_{-0.6}$ & $M_{\rm ej} /{\rm M}_\odot$ & $9.3^{+12.9}_{-4.8}$ & $\epsilon$ & $0.5^{+0.6}_{-0.3}$ & ($\kappa /$cm$^2$g$^{-1}$)\tablenotemark{\ddag} & $0.08^{+0.11}_{-0.05}$  \\ 
    $m_r /$mag & $19.2^{+1.4}_{-1.2}$ & f$_{\rm Ni}$ & $0.02^{+0.14}_{-0.02}$ & $A_V$/mag & $0.0009^{+0.0022}_{-0.0005}$ & ($\kappa_\gamma /$cm$^2$g$^{-1}$)\tablenotemark{\ddag} & $0.06^{+0.07}_{-0.05}$  \\ 
    $E_{\rm rad}$/erg & $51.1^{+0.3}_{-0.5}$ & & & & & & \\ 
    \enddata
    \tablecomments{List of all parameters for SLSNe, their mean value, and 1$\sigma$ ranges. We include observational and physical parameters. For definitions of the physical parameters, see Table~\ref{tab:parameters}. $z$ is the redshift, $\tau_{\rm rise}$ is the time from explosion to peak, $\tau_{\rm 1}$ is the time it takes the SN to decline by 1-magnitude, $\tau_{\rm e}$ is the time it takes the SN to decline by a factor of $e$, $\Delta m_{15}$ is the magnitudes by which a SN fades 15 days after maximum in B-band, $M_{\rm max}$ is the peak rest-frame $r$-band magnitude, $m_r$ is the peak observed $r$-band magnitude, $E_{\rm rad}$ is the total radiated energy of the SN during the first 200 days, $L_{\rm max}$ is the luminosity at peak, $M_{\rm Ni}$ is the nickel mass, KE is the kinetic energy, $L_0$ and $t_{\rm SD}$ are the initial magnetar spin-down luminosity and spin-down time from \cite{Omand23}, $f_{\rm mag}$ is the fraction of the total luminosity due to the magnetar contribution, $\epsilon$ is the radiative efficiency, and $A_V$ is the intrinsic host extinction in V-band.}
    \tablenotetext{\dag}{These parameters are mostly unconstrained by the data and the posterior is highly dependent on the choice of prior.}
    \tablenotetext{\ddag}{These parameters are only constrained for a fraction events.}
\end{deluxetable*}

\subsection{Extrabol}\label{sec:extrabol}

To measure the bolometric properties of each SLSN, including their luminosities, temperatures, and radii, we model each light curve using {\tt extrabol} \citep{Thornton24}, a Gaussian Process (GP) implementation of the {\tt George} \citep{Foreman-Mackey15} package, based on the original {\tt SuperBol} \citep{Nicholl18_superbol} code. Unlike \mosfit, which produces smoothly evolving light curves, the {\tt extrabol} model allows us to account for short-term variability, such as bumps or undulations, that the \mosfit model cannot reproduce. We limit the {\tt extrabol} models only to SLSNe that have at least three detections in three distinct bands to obtain a more robust fit to the data in terms of phase and wavelength. We find that {\tt extrabol} tends to over-predict UV-magnitudes when no UV observations are available, particularly at very early or late phases. To mitigate this issue we include additional model photometry derived from the \mosfit model, corresponding to the model magnitude of the SN in every observed band at the time of the first and last detection. We find this addition helps {\tt extrabol} produce much more reliable results in the interpolation.

Additionally, we measure the rate of change of the photospheric radius as a function of phase by fitting a straight line to the {\tt extrabol} models from explosion to peak bolometric luminosity. We use this value to approximate the photospheric velocity of the SLSNe, and refer to it as the ``blackbody velocity'' to distinguish it from a direct measurement of the photospheric velocity.

\begin{figure}
	\begin{center}
		\includegraphics[width=1.05\columnwidth]{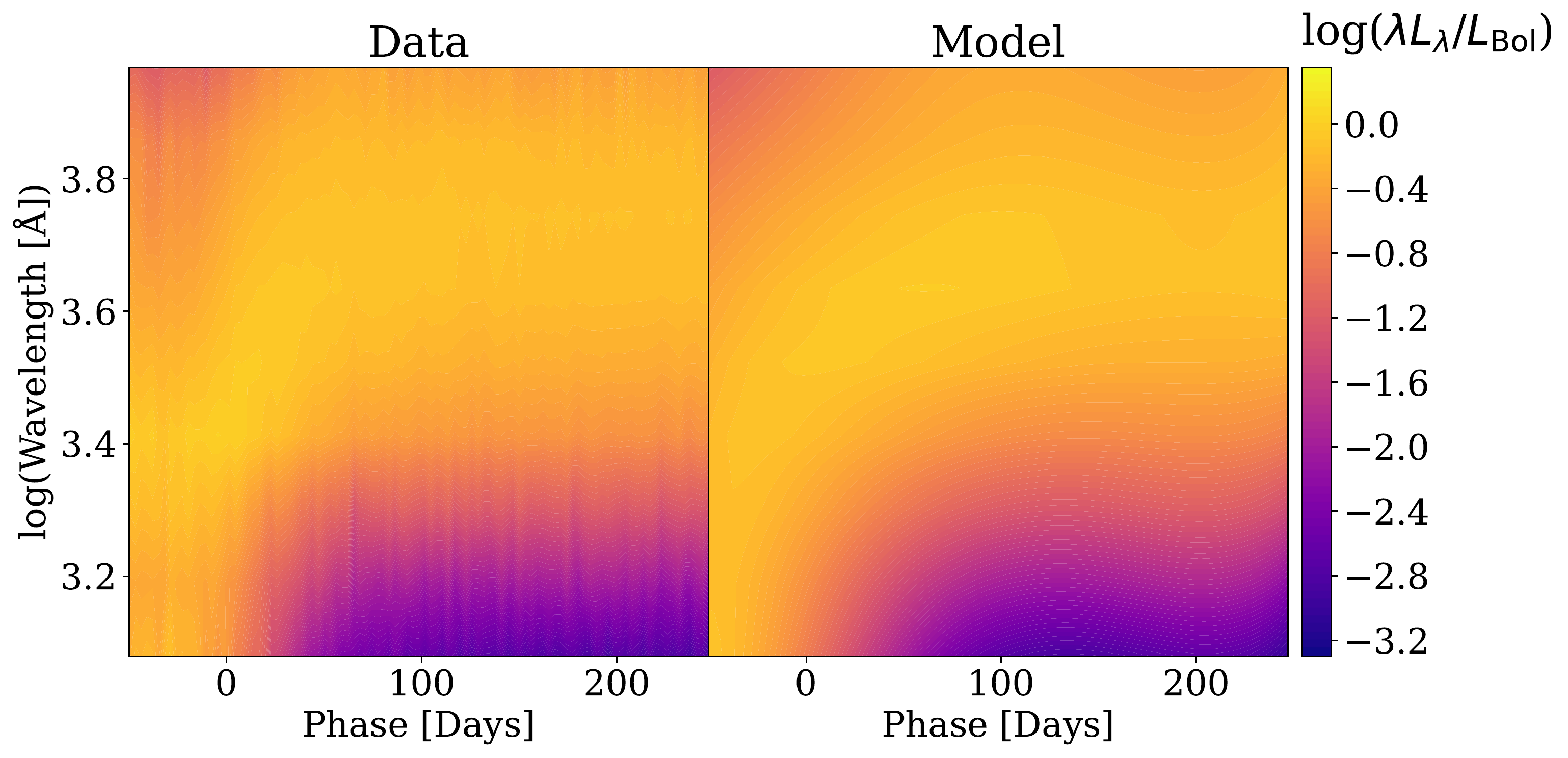}
		\caption{Bolometric correction ratio $\lambda L_\lambda / L_{\rm Bol}$ as a function of phase and rest-frame wavelength. The left panel shows the data representing the mean value of $\lambda L_\lambda / L_{\rm Bol}$ for the full sample of SLSNe, while the right panels shows a 4th order 2D polynomial fit to these data. \label{fig:corr_bol_small}}
	\end{center}
\end{figure}

\section{Population Properties}\label{sec:population}

\begin{figure}
	\begin{center}
		\includegraphics[width=\columnwidth]{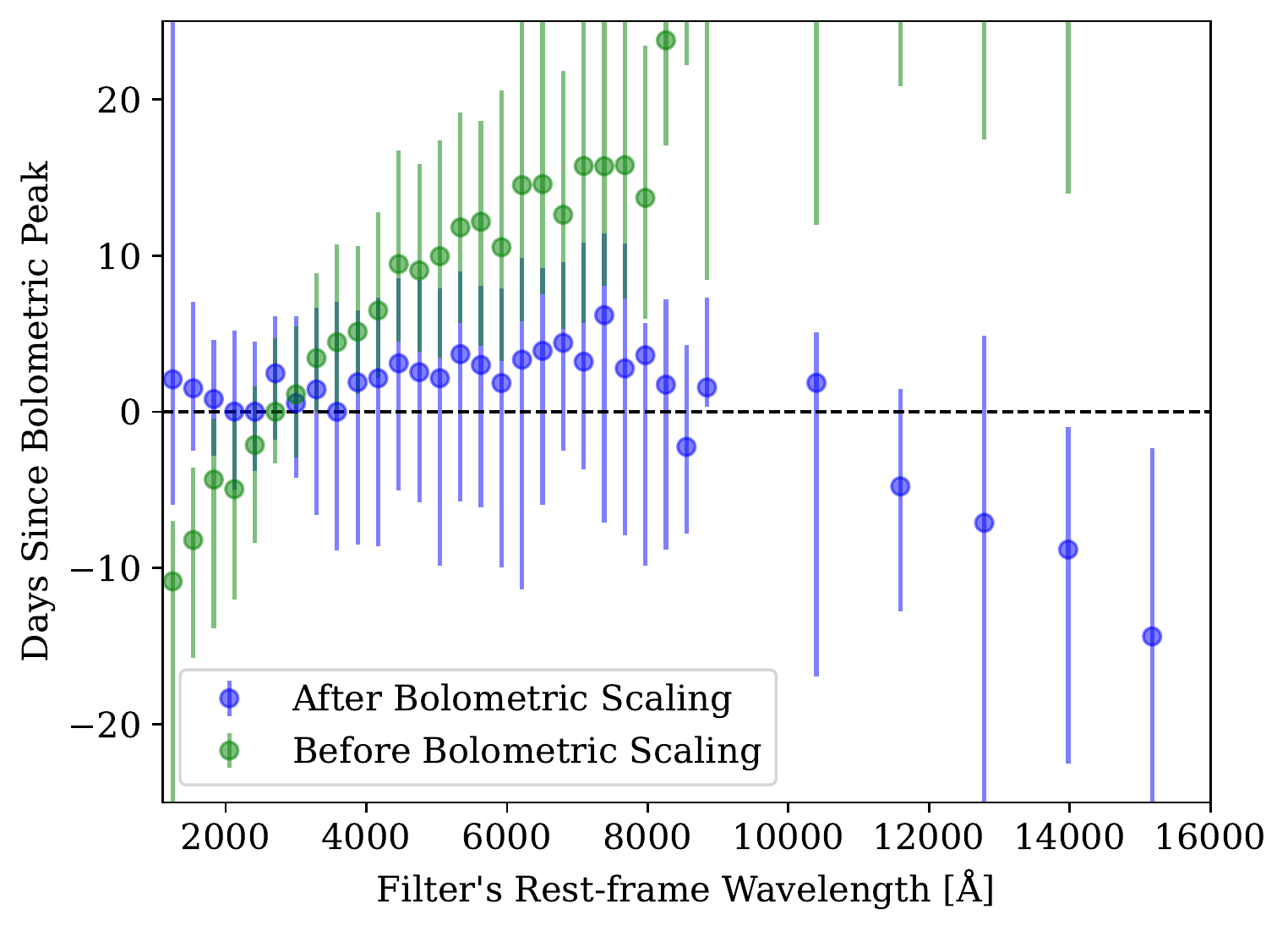}
		\caption{The mean time of light curve peak relative to the peak of the bolometric light curve as a function of the rest-frame wavelength for the full sample, shown before (green) and after (blue) applying a bolometric scaling. Before bolometric scaling, it is evident the peak time evolves as a function of wavelength. \label{fig:check_bol_wave}}
	\end{center}
\end{figure}

In this section we describe basic observational properties of the SLSN population based on the parameters described in \S\ref{sec:methods}. Since some of the properties described here, such as the peak luminosity or rise time, are derived from the light curve models, we consider them to be largely model independent.

\begin{figure}
	\begin{center}
		\includegraphics[width=\columnwidth]{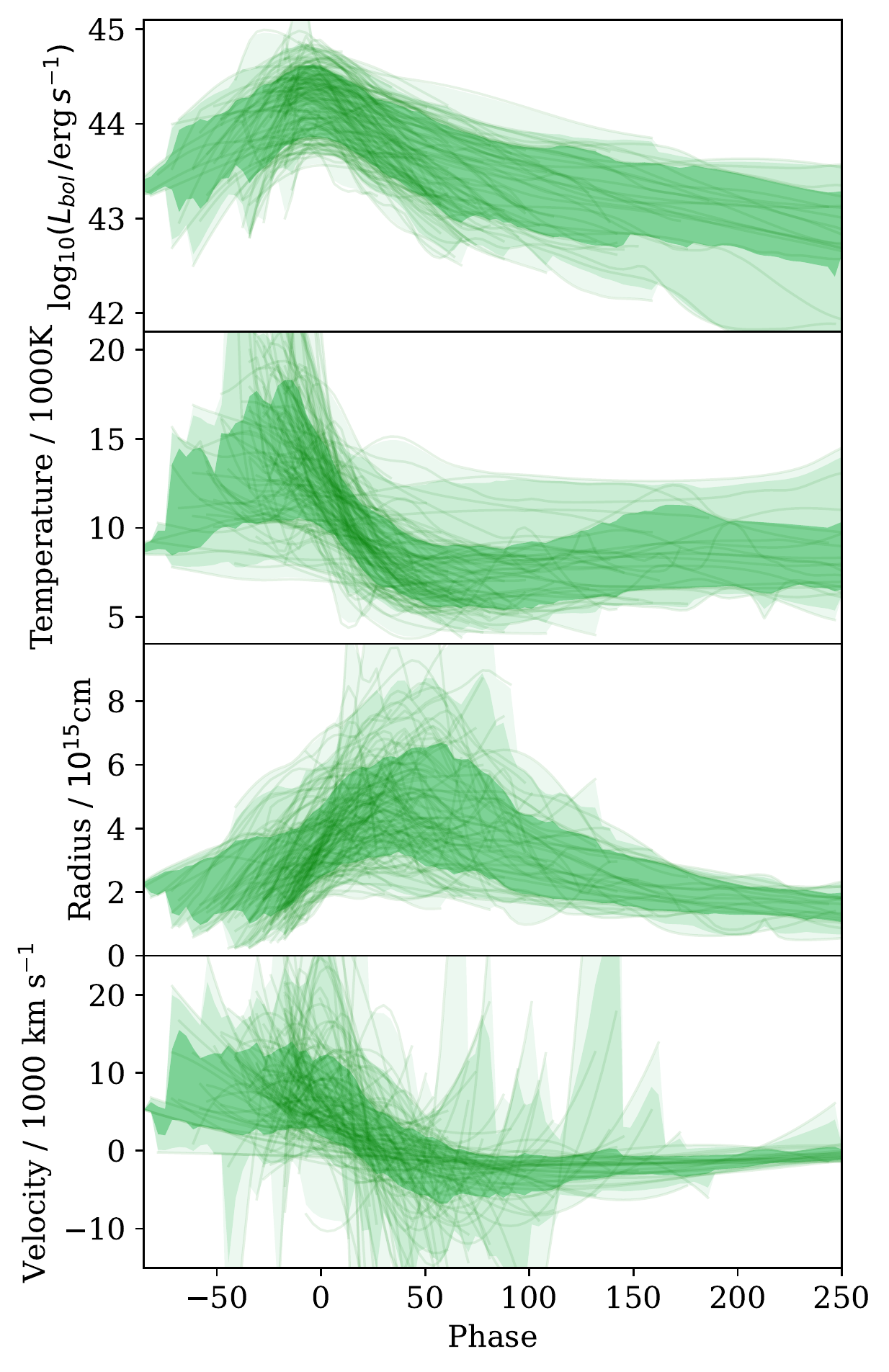}
		\caption{Bolometric luminosity, and photospheric temperature, radius, and velocity as a function of phase for the full sample. These parameters are derived from the {\tt extrabol} model fits to the light curves. The shaded regions represent the $\pm$ 1, 2, and 3$\sigma$ ranges. \label{fig:SEDs}}
	\end{center}
\end{figure}

\subsection{The Mean SLSN}\label{sec:mean}

We use the rest-frame \mosfit light curve models of all SLSNe to create a map of their mean evolution as a function of phase and wavelength. The mean is calculated by averaging the magnitudes of all models of all SLSNe at each phase and wavelength. The resulting mean values and corresponding $\pm 1 \sigma$ range in magnitudes are shown in Figure~\ref{fig:heatmap_mosfit}. This map provides an estimate of the mean AB magnitude and scatter of SLSNe at any wavelength between 2,100 and 16,000 \AA\ and any phase between $-23$ and 243 days. For ease of use, we provide a version of this map as a function of days from peak, and another as a function of days after explosion in the data repository. 

While the light curves used to generate this map were generated using the common filters listed in \S\ref{sec:methods}, we provide an option to estimate the mean magnitude of SLSNe at any arbitrary wavelength and phase by interpolating this map. For example, in Figure~\ref{fig:SLSNe_LCs} we show the mean evolution of SLSNe in $griz$ bands, as well as the individual light curves of each SLSN in gray.

\subsection{K-corrections}\label{sec:kcorr}

If one observes a SN at a redshift $z$ through a filter with an effective wavelength $\lambda_{\rm obs}$ and measures a magnitude $m_{\rm obs}$, these observations are actually probing the SED of the SN at $\lambda_{\rm rest} = \lambda_{\rm obs} / (1 + z)$ with a magnitude $m_{\rm rest}$. If the goal is to measure the absolute magnitude of the SN in the rest frame, this can be calculated as $M_{\rm rest} = m_{\rm obs} - \mu - K$, where $\mu$ is the distance modulus and $K$ is the K-correction \citep{Hogg02}. If the SN were to have a flat SED with constant luminosity $\lambda L_\lambda$, the K-correction would be simply $K = - 2.5 \log_{10} (1 + z)$. Real SNe however do not have a flat SED, which requires us to know, or assume, the color of the SED to account for the difference in magnitude between $m_{\rm rest}$ and $m_{\rm obs}$ and derive an accurate K-correction. A more accurate K-correction is $K = (m_{\rm rest} - m_{\rm obs}) - 2.5 \log_{10} (1 + z) $. While the most accurate K-correction value will depend on the specific SED of each SN, we can estimate the mean K-correction value for the SLSN population at large and use this to obtain a rapid approximation of their rest-frame magnitude. We use the 3D map shown in Figure~\ref{fig:heatmap_mosfit} to derive these mean K-correction values for SLSNe as a function of wavelength and phase.

\begin{figure*}
	\begin{center}
        \includegraphics[width=\textwidth]{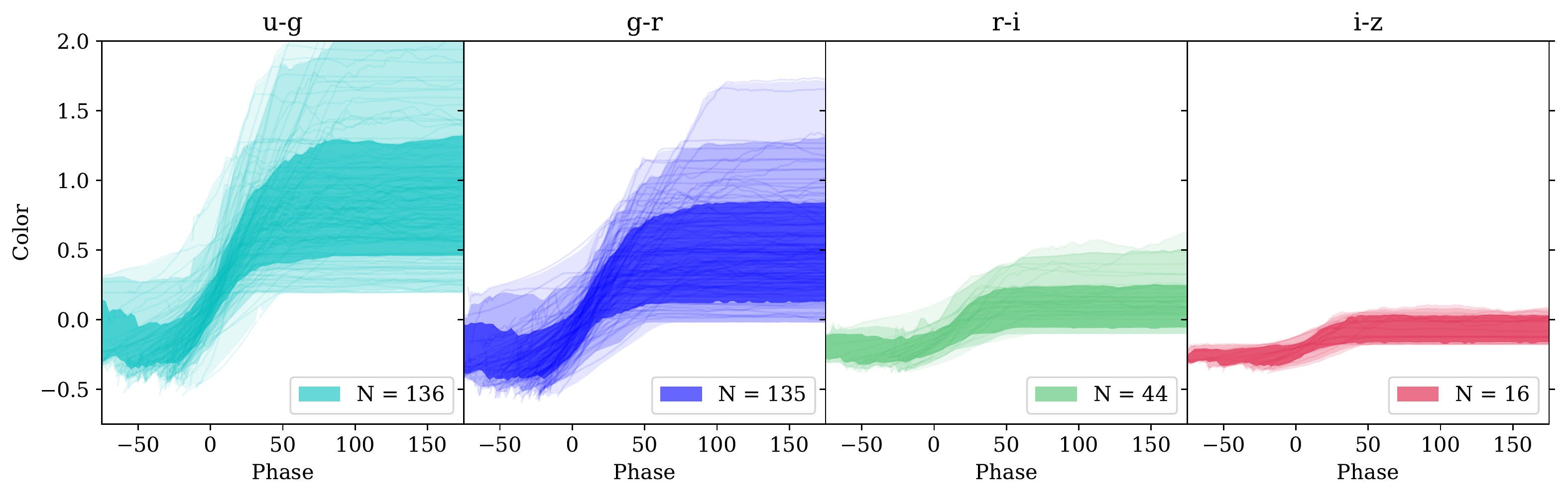}
		\caption{Rest-frame colors as a function of phase, derived from an interpolation of the SLSNe light curves using \mosfit. We only include N objects for which there are observations available in the interpolated filters. The shaded regions represent the $\pm$ 1, 2, and 3$\sigma$ ranges. The redder filter pairs have a smaller span in colors in part because the shape of the SED flattens out at redder wavelengths, but these are also less well measured given that there are fewer observations available in these filters. \label{fig:color}}
	\end{center}
\end{figure*}

We provide a Python script for calculating K-corrections that takes in values for $\lambda_{\rm obs}$, $z$, and phase, as well as an optional output wavelength to which the data will be corrected $\lambda_{\rm output}$. If no value for $\lambda_{\rm output}$ is provided, this is assumed to be $\lambda_{\rm output} = \lambda_{\rm rest} = \lambda_{\rm obs} / (1 + z)$. The code will calculate the mean color difference between $m_{\rm rest}$ and $m_{\rm obs}$ to derive the K-correction. A demonstration of how to run the code is provided in \S\ref{sec:catalog}. We show an example of these corrections applied to three representative SLSNe in Figure~\ref{fig:K_correction_map}. We show how our method generally provides more accurate rest-frame light curves than the simplistic $+ 2.5 \log_{10} (1 + z)$ correction, with the exception of SNe with a redshift $\gtrsim 1$. For these high-redshift SNe we recommend using a filter with a $\lambda_{\rm obs}$ closer to the desired $\lambda_{\rm rest}$, if available. For example, for a SN at $z = 0.9$, $z$-band observations with $\lambda_{\rm obs} \sim 8920 $ \AA\ are actually probing $\lambda_{\rm rest} \sim 4690 $ \AA. This value is much closer to the rest-frame wavelength of $g$-band at $\sim 4670$ \AA. In this case, using $z$-band observations would provide a much better estimate of the rest-frame $g$-band magnitude than observed $g$-band would.

\subsection{Bolometric Scaling}\label{sec:bolometric}

We use the \mosfit light curve models to derive a relation between the observed photometry and the bolometric luminosity of SLSNe. For each SN, we calculate the ratio between its monochromatic luminosity measured in an observed filter $\lambda L_\lambda [{\rm erg}\ {\rm s}^{-1}]$ and its bolometric luminosity $L_{\rm Bol} [{\rm erg}\ {\rm s}^{-1}]$ as a function of phase. This way we create a map of the mean ratio between $L_\lambda$ and $L_{\rm Bol}$ as a function of phase and wavelength for all SLSNe. We show this map in Figure~\ref{fig:corr_bol_small}, which we fit with a two-dimensional 4th degree polynomial shown in Eq.~\ref{eq:poly}, where $\phi$ is the phase and $\lambda$ is the wavelength. Using this equation we derive a functional form of the mean value of $L_\lambda / L_{\rm Bol}$ as a function of phase and wavelength. This scaling can then be applied to any observed photometry between 1200 and 16000 \AA\ and any phase between $-50$ and 250 days to obtain an estimate of the bolometric luminosity of a SLSN.

\begin{equation}\label{eq:poly}
    P(\phi, \lambda) = \sum_{i=0}^{4} \sum_{j=0}^{4-i} a_{ij} \phi^i \lambda^j
\end{equation}

Applying this bolometric scaling to the light curves of SLSNe also helps shifts the peak of the monochromatic light curve closer to the time of bolometric peak. In Figure~\ref{fig:check_bol_wave} we show how the mean time of light curve peak varies as a function of rest-frame wavelength. Light curves observed in filters with rest-frame wavelengths bluer than $\sim 2700$ \AA\ peak before bolometric peak, and redder filters peak after bolometric peak. This also means that filters with rest-frame wavelengths close to $\sim 2700$ \AA\ provide the most accurate estimates of bolometric peak for SLSNe. After applying our bolometric scaling correction to the light curves of all SLSNe, we see this relation becomes much flatter, particularly for wavelengths $\lesssim 10000$ \AA. In \S\ref{sec:catalog} we show code examples on how users can apply these corrections to any SLSN.

\begin{figure}
	\begin{center}
		\includegraphics[width=\columnwidth]{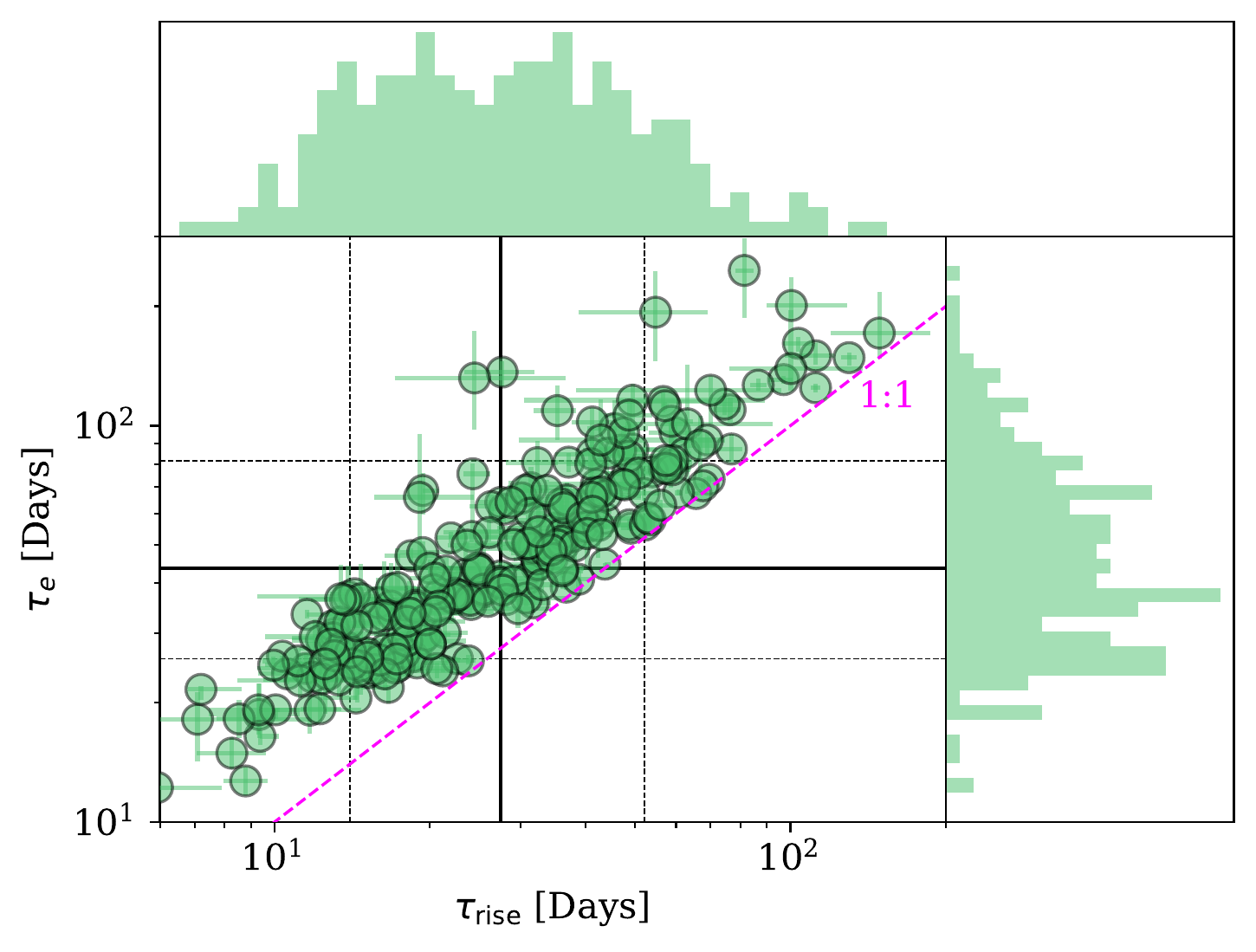}
		\caption{The e-fold decline time $\tau_{\rm e}$ as a function of rise time $\tau_{\rm rise}$ for the full sample of SLSNe. The solid and dashed lines show the mean and $1\sigma$ values of these parameters, respectively. The pink line shows the one-to-one correspondence. \label{fig:Rise_Time_E_fold}}
	\end{center}
\end{figure}

\begin{figure*}
	\begin{center}
		\includegraphics[width=\columnwidth]{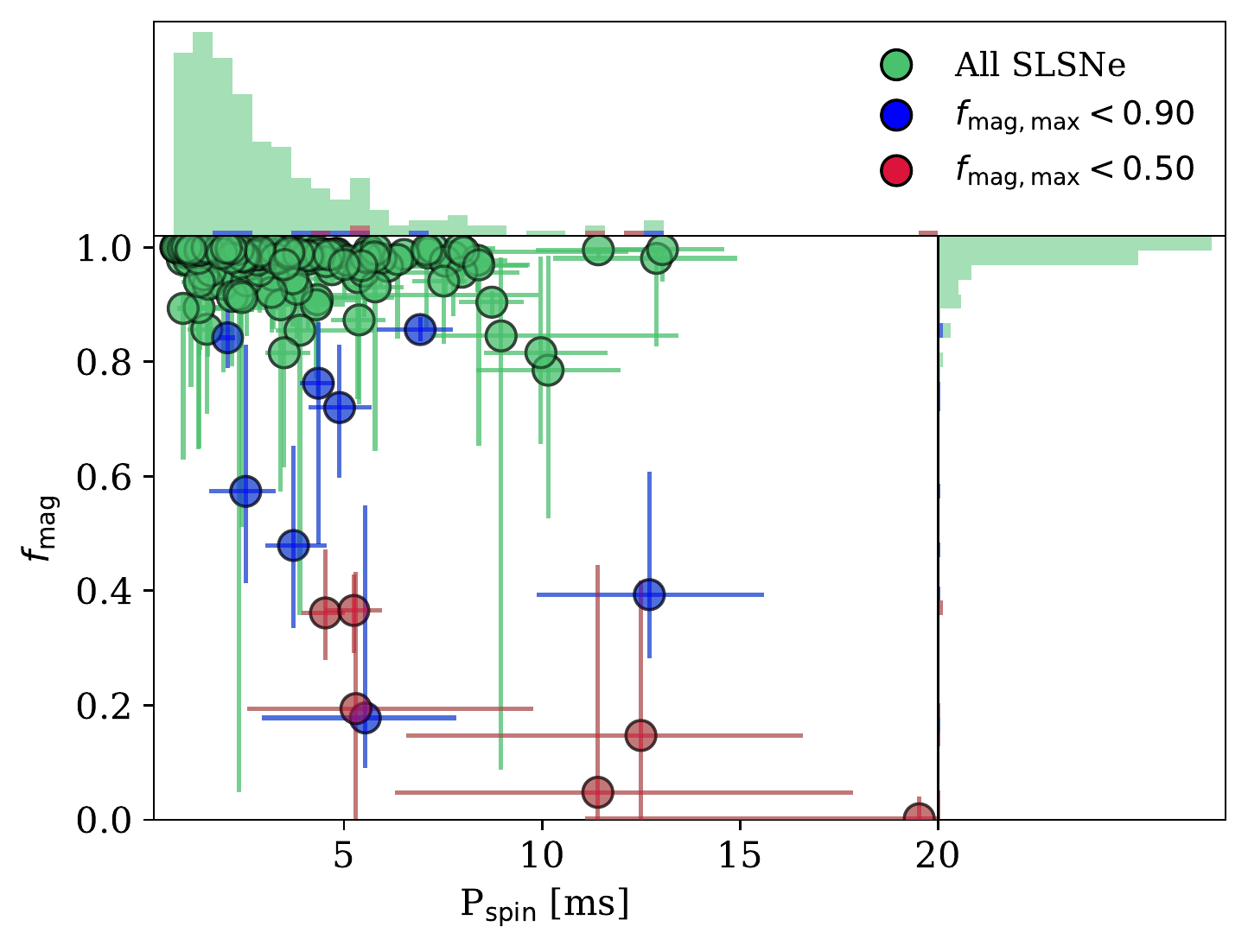}
		\includegraphics[width=\columnwidth]{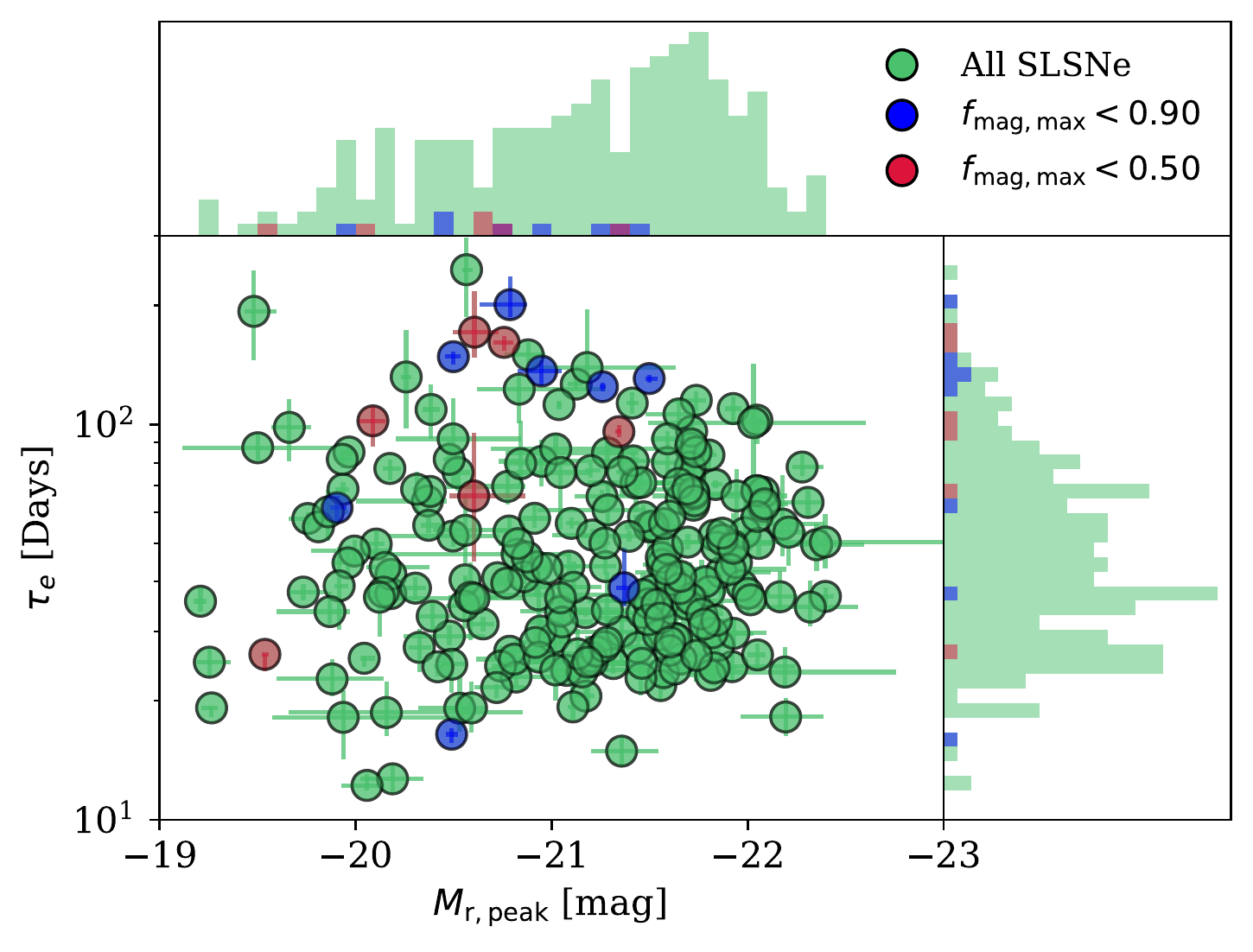}
		\caption{\textit{Left}: The fraction of total radiated energy produced by a magnetar ($f_{\rm mag}$), as opposed to radioactive decay, as a function of $P_{\rm spin}$. The blue and red points mark SLSNe in which more than 10\% and 50\%, respectively, of the radiated energy is from radioactive decay. \textit{Right}: The e-fold decline time $\tau_{\rm e}$ as a function of peak r-band magnitude $M_r$ for the full sample of SLSNe. The SLSNe with the highest fraction of radioactive decay input are slowly-declining and relatively dim. \label{fig:bol_lum_frac}}
	\end{center}
\end{figure*}

\subsection{Evolution}\label{sec:evolution}

In Figure~\ref{fig:SEDs} we show the evolution of the bolometric luminosity, photospheric radius and temperature for all SLSNe, as well as a blackbody velocity derived from the {\tt extrabol} models in \S\ref{sec:extrabol}. We see that SLSNe have initial temperatures of $T \sim 10000 - 15000$ K, and then settle into a mean temperature of $T \sim 7000$ K. The typical radius of SLSNe expands from an initial $R \sim 2\times10^{15}$ cm to $R \sim 5\times10^{15}$ cm before receding back to a radius of $R \sim 2\times10^{15}$ cm, in line with expectations from models of homologous expansion of SNe \citep{Liu18_expansion}. While these measurements are consistent with previous results (e.g. \citealt{Lunnan18, Chen22a, Hinkle23}), the larger sample size used here allows us to constrain the decrease in photospheric radius $\sim 50$ days after peak, not clearly evident in previous studies, which in turn allows us to measure the velocity at which the blackbody radius evolves.

We use the rest-frame light curve models from \mosfit to calculate the mean color evolution for SLSNe for a series of filter pairs. In Figure~\ref{fig:color} we show this evolution, where we see SLSNe tend to be blue before peak with a mean color of $g-r \sim -0.2$ before reddening with time to typical values of $g {\rm -} r\sim 0.2 - 0.8$.

We find the mean duration for a SLSN to reach peak from explosion is $\tau_{\rm rise} = 27^{+25}_{-13}$ days. This parameter is tightly correlated with the decline time $\tau_e$, which has a mean value of $\tau_e = 44^{+38}_{-18}$ days, as shown in Figure~\ref{fig:Rise_Time_E_fold}. These parameters are similarly correlated with both $\tau_1$ and $\Delta m_{15}$. We fit the relation between various rise and decline timescales and find the following correlations:
\begin{align*}
\log(\Delta m_{15}) &= (-0.51 \pm 0.01) \times \log(\tau_{\rm 1}) + (1.32 \pm 0.01) \\
\log(\tau_{e}) &= (0.82 \pm 0.02) \times \log(\tau_{\rm 1}) + (0.35 \pm 0.04) \\
\log(\Delta m_{15}) &= (-0.59 \pm 0.02) \times \log(\tau_{\rm rise}) + (1.00 \pm 0.02) \\
\log(\tau_{e}) &= (1.00 \pm 0.03) \times \log(\tau_{\rm rise}) - (0.20 \pm 0.05)
\end{align*}

\cite{Cia18} studied a sample of 26 SLSNe and was able to measure the decline rates of 13 of these. The authors found all measured decline rates to be consistent with the rate of decay of $^{56}$Co to $^{56}$Fe of 0.0098 mag day$^{-1}$, assuming full trapping of the decay process. We perform a similar experiment here, and manage to measure the decline rates of 105 SLSNe at 100 days post explosion but find only $\sim 17$\% of these to have a decline rate consistent with the radioactive decay of $^{56}$Co, while all other SLSNe decline faster than this. Therefore, we conclude these slow decline rates are most likely the low end of a distribution of SLSNe decline rates and do not necessarily imply these are radioactively dominated tails.

\subsection{Extreme SLSNe}\label{sec:extreme}

In this section we outline the properties of the SLSNe with the most extreme observational parameters. The SLSNe with the brightest rest-frame $r$-band magnitudes are DES16C2nm with $m_r = -22.67 \pm 0.53$ and SCP06F6 with $m_r = -22.55 \pm 0.21$, each $\sim 2\sigma$ brighter than the mean of the distribution. However, these two SLSNe are found at relatively large redshifts of $z = 1.998$ and $z = 1.189$ and their peak $r$-band estimates have a large uncertainty since we have to extrapolate their observed magnitudes into rest-frame $r$-band. SN\,2020dlb has a similarly bright peak magnitude of $m_r = -22.44 \pm 0.05$, or $\sim 1.7 \sigma$ brighter than the mean SLSN, but with a much better constrained measurement and at a redshift of only $z = 0.398$, making it the brightest relatively nearby SLSN to date.

The SLSN with the highest peak bolometric luminosity is SN\,2017ens, which reached a luminosity of $\log(L) = 45.3 \pm 0.1$ erg s$^{-1}$, or $\sim 2.7 \sigma$ brighter than the mean of the SLSN distribution. The SLSNe with the highest total integrated radiated energy during the first 200 days after explosion are PTF12dam and SN\,2020afag, both with a total radiated energy of $\log(E_{\rm rad}) \approx 51.8$ erg, or $\sim 1.9 \sigma$ above the mean of the distribution. For a discussion on the dimmest SLSNe see \cite{Gomez22_LSN}, where we describe the continuum formed between SLSNe and Type Ic/Ic-BL SNe. The furthest SLSN found to date is DES16C2nm at a redshift of $z = 1.998$, while the closest one is SN\,2018bsz at a redshift of $z = 0.0267$.

The SLSN with the slowest rise time in our sample is PS1-14bj, which took $\tau_{\rm rise} = 89^{+15}_{-5}$ days to reach its peak, or $\sim 2.4 \sigma$ slower than the mean of the SLSN distribution. The SLSN with the slowest ejecta velocity is SN\,2022le, with a best-fit of $V_{\rm ej} = 1030 \pm 30$ km s$^{-1}$ and a corresponding blackbody velocity of $V_{\rm BB} \approx 1190$ km s$^{-1}$, each around $\sim 1.9 \sigma$ slower than the mean of the population. SN\,2022le took a notably long $\tau_{\rm 1} = 210 \pm 9$ days to fade from peak by 1 magnitude, a significant $\sim 4.5 \sigma$ longer than the mean SLSN population.

\section{Results}
\label{sec:results}

In this section we discuss the results derived from the \mosfit model fits. Additionally, we include comparisons to the observational parameters derived in \S\ref{sec:population} and previous studies.

\begin{figure}
	\begin{center}
		\includegraphics[width=\columnwidth]{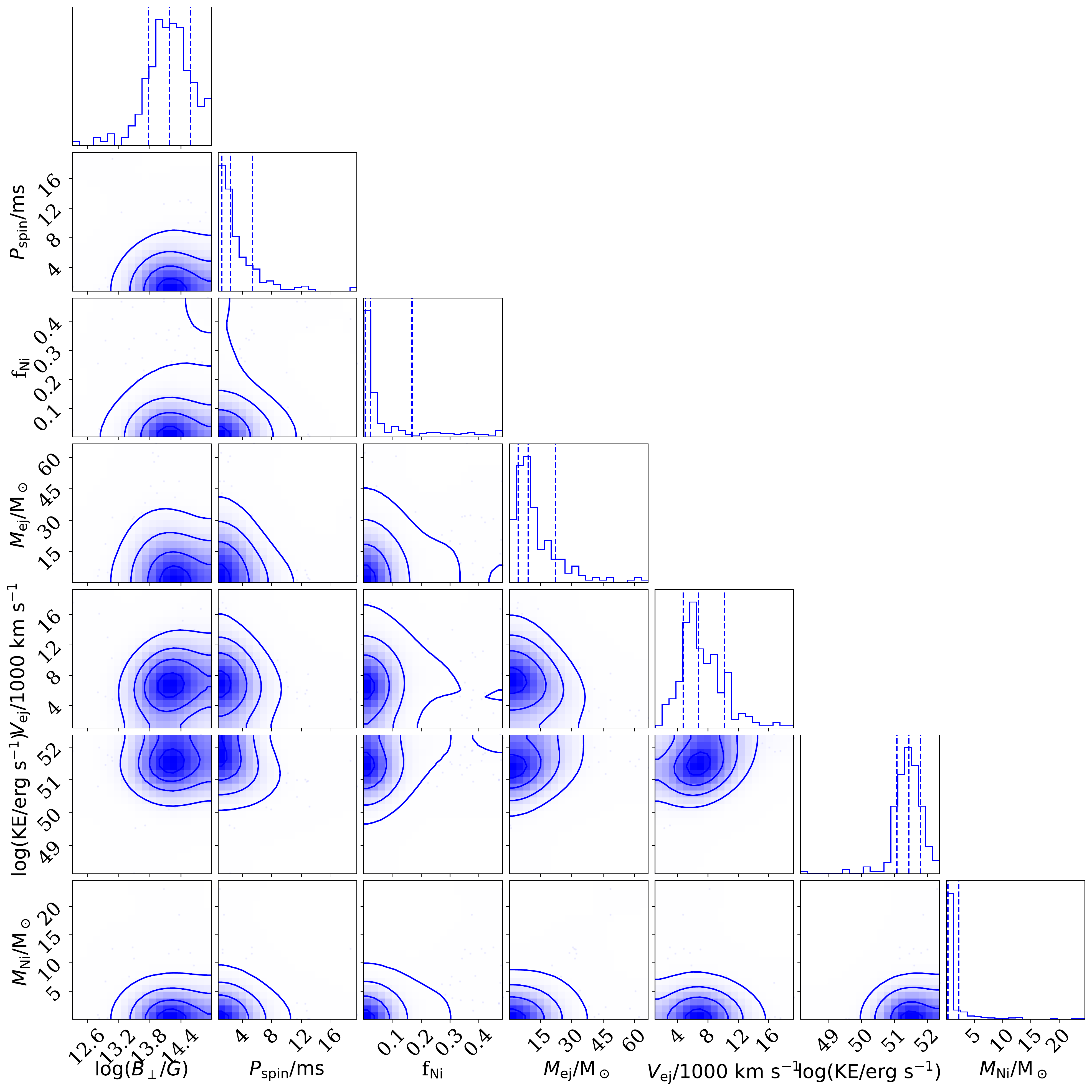}
		\caption{Corner plot showing the correlation between the most critical physical parameters for the full sample, defined in Table~\ref{tab:parameters}. The histograms show the marginalized distribution of each parameter, as well as the mean and $\pm 1 \sigma$ range. \label{fig:corner}}
	\end{center}
\end{figure}

\subsection{Parameter Distributions}\label{sec:parameters}

In Table~\ref{tab:mean} we list the mean values and 1$\sigma$ scatter for all physical and observational parameters presented in this work. The distributions of the best-fit values as well as their correlations for the full SLSN population are shown in Figure~\ref{fig:corner}. We use the full sample of SLSNe, including an exploration of all their physical and observational parameters, to attempt to separate SLSNe into distinct groups with the aim of determining whether there are distinct classes of SLSNe. We use the {\tt Scikit-learn} \citep{Scikit-learn} implementation of the K-means clustering algorithm, a popular method for partitioning a dataset into distinct, non-overlapping clusters which assigns each data point to its nearest cluster center, with the goal of minimizing the variance within each cluster. The most significant clustering separation we find has two clusters with a silhouette coefficient of 0.13, which roughly separates the most luminous SLSNe from the least luminous ones. A silhouette coefficient this low suggests that the clusters are weakly distinguished from each other \citep{Rousseeuw87}. This means we are unable to find any significant separation in the existing SLSN population, and SLSNe appear to be drawn from a mostly continuous distribution, at least based on photometry alone. Future studies that include spectroscopic parameters of SLSNe might find a different conclusion.

We caution that the value of $M_{\text{NS}}$ has very little effect on the light curve, and its posterior distribution is therefore largely unconstrained and dominated almost entirely by the choice of prior. Similarly, any value of $n_\mathrm{H,host}$ below $\sim 18$ cm$^{-2}$ reflects effectively no extinction and is therefore unconstrained.

\subsection{Magnetar Parameters}\label{sec:magnetar}

\begin{figure}
	\begin{center}
		\includegraphics[width=1.025\columnwidth]{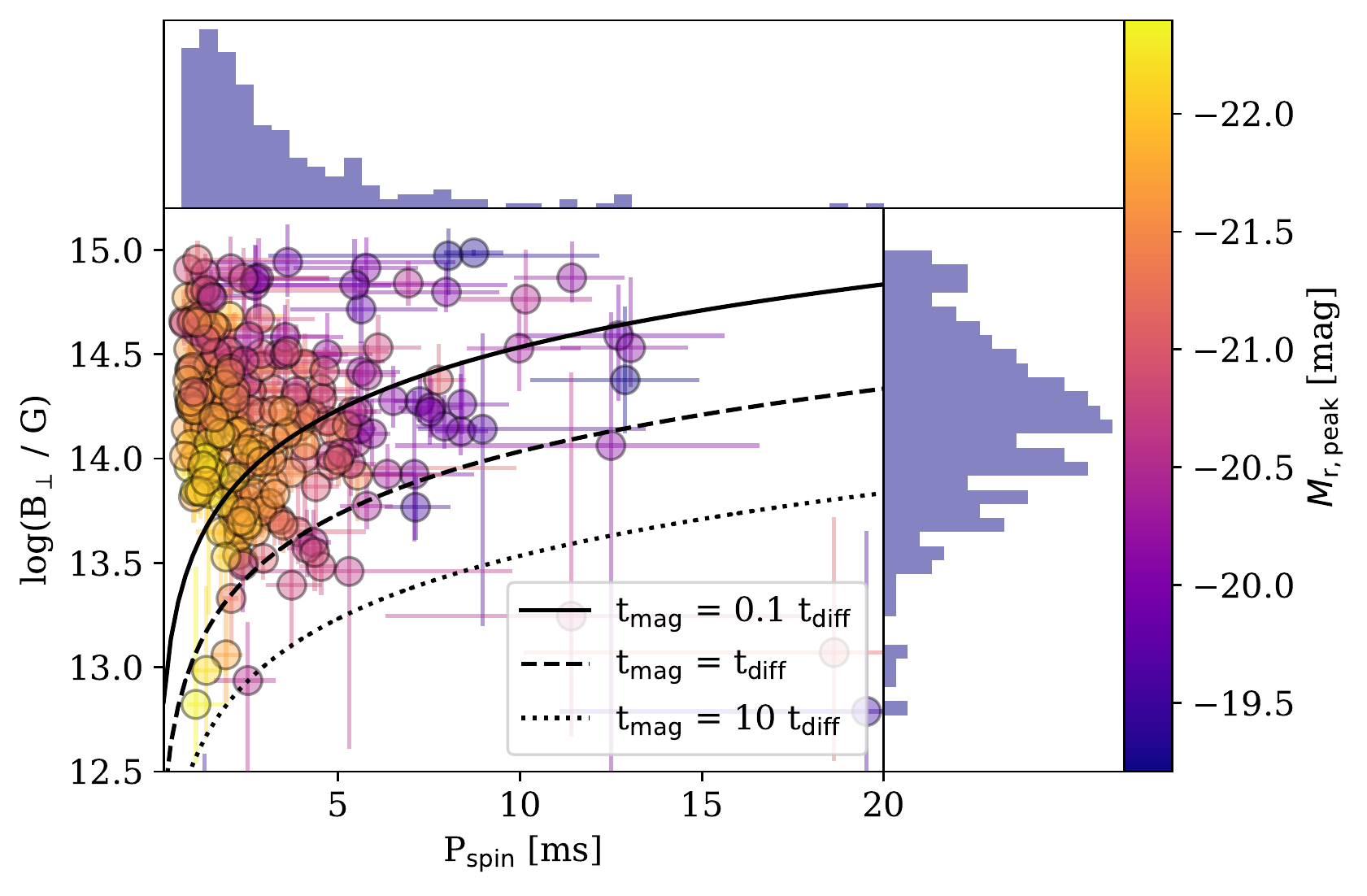}
		\caption{Magnetic field versus spin period for the full sample, color coded by peak absolute magnitude in rest-frame $r$-band. The black lines show trends for different ratios of the magnetar to diffusion timescales, assuming the mean ejecta mass and kinetic energy for the sample, as listed in Table~\ref{tab:mean}. For very fast spin periods, the magnetar time scale can be much shorter than the diffusion timescale. \label{fig:Pspin_log_Bfield}}
	\end{center}
\end{figure}

In Figure~\ref{fig:Pspin_log_Bfield} we show the key magnetar parameters $P_{\text{spin}}$ and $B_{\perp}$ for the full population of SLSNe. We include the trend lines from \cite{Nicholl17_mosfit} that show different ratios of magnetar timescale $t_{\rm mag}$ to diffusion timescale $t_{\rm diff}$ for a fixed ejecta mass of $M_{\rm ej} = 9.3$~M$_\odot$, equal to the mean of the SLSN population. Most of the SLSNe lie above the $t_{\rm mag} / t_{\rm diff} = 0.1$ line, suggesting that for fast spin periods, the magnetar loses energy quickly, but that enough rotational power is left to produce a long diffusion timescale. While almost all of the brightest SLSNe with $M_r \sim -22$ mag have a strong preference for short spin periods $\lesssim 4$ ms, not all SLSNe with short spin periods are brighter than this. This implies that while high rotational energy is key to producing luminous SLSNe, other parameters such as having a large ejecta mass can contribute as well. \cite{Hinkle23} studied a sample of 27 SLSNe and found a tentative correlation between ejecta mass and magnetic field strength. We determine that this correlation was most likely an artifact of the small sample size, given that we see no evidence for a correlation between these two parameters.

\begin{figure}
	\begin{center}
		\includegraphics[width=\columnwidth]{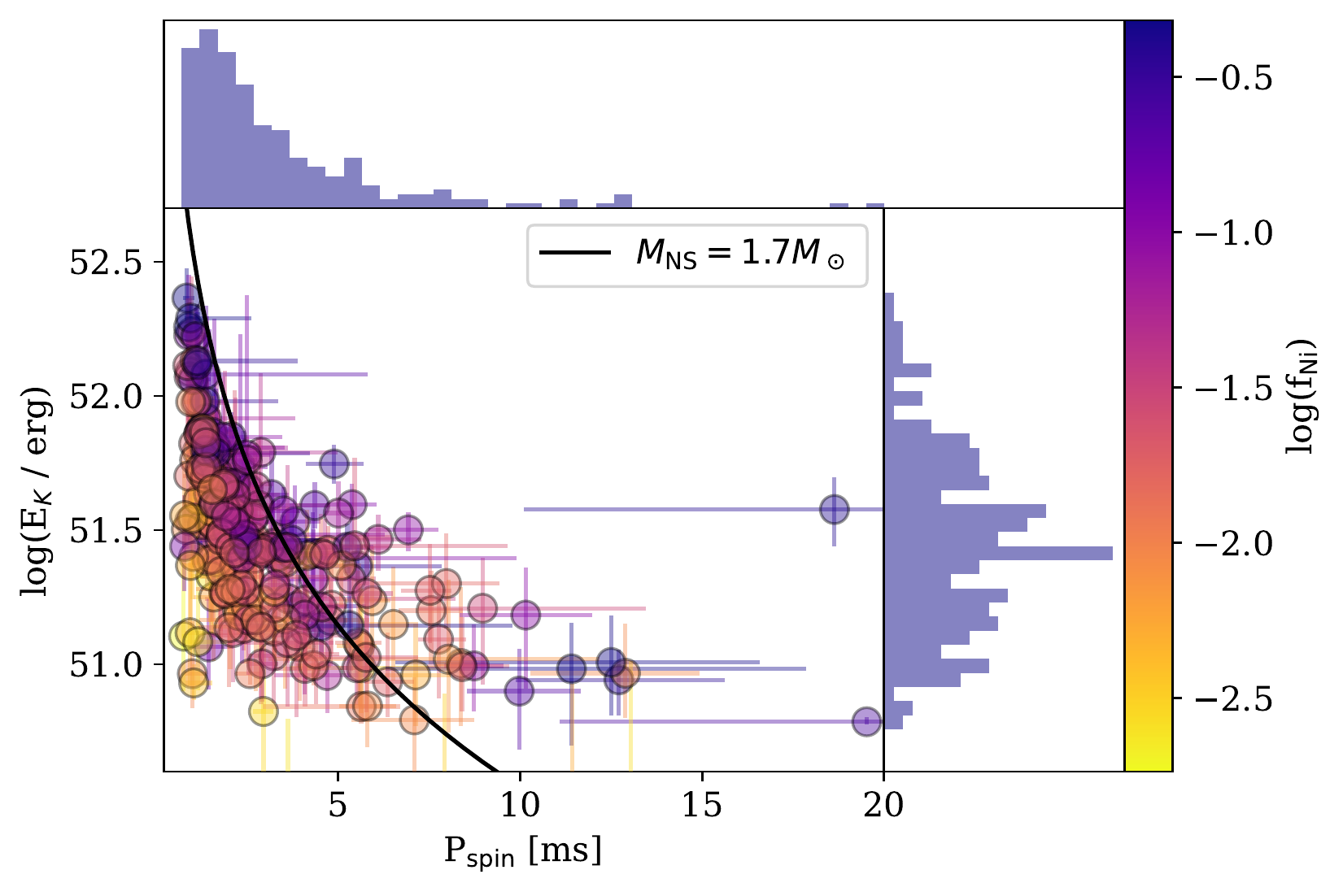}
		\caption{Kinetic energy versus spin period for the full sample of SLSNe, color coded by $^{56}$Ni fraction. The black line represents the magnetar energy for a 1.7 $M_\odot$ neutron star, which tracks the observed trend well. \label{fig:Pspin_log_KE}}
	\end{center}
\end{figure}

In Figure~\ref{fig:Pspin_log_KE} we show the total kinetic energy of the SLSN population as a function of the magnetar spin period $P_{\text{spin}}$. It is not surprising that this correlation tracks the evolution of the total energy of a magnetar with an assumed mass of 1.7 $M_\odot$, since most of the kinetic energy of the SLSNe is dominated by the magnetar contribution.

We calculate $f_{\rm mag}$, or the fraction of the total radiated energy that comes from the magnetar component during the first 200 days. For the SLSNe with a measurable radioactive decay contribution, we find that its fractional contribution tends to flatten after $\sim 100$ days. We aim to determine which SLSNe have a non-negligible contribution from radioactive decay. On the left panel of Figure~\ref{fig:bol_lum_frac} we show the $f_{\rm mag}$ value distribution for all SLSNe as a function of $P_{\text{spin}}$. We find that $\sim 6$\% of SLSNe have a contribution from radioactive decay in which the lower bound of the best fit $f_{\rm mag}$ value is still above $10$\%, and only $\sim 2.5$\% above $50$\%. The only six SLSNe with a mean magnetar contribution $f_{\rm mag} < 0.5$ are SN\,2020abjc, SN\,2020rmv, SN\,2020vpg, SN\,2021lwz, iPTF13bdl, and iPTF13bjz, all relatively dim and slowly-declining SLSNe. On the right panel of Figure~\ref{fig:bol_lum_frac} we show how SLSNe with high $\tau_{\rm rise}$ values are the most likely to be powered by at least some contribution from radioactive decay. This is particularly true for SLSNe with peak r-band magnitudes $\gtrsim -21$ mag.

\begin{figure}
	\begin{center}
        \includegraphics[width=\columnwidth]{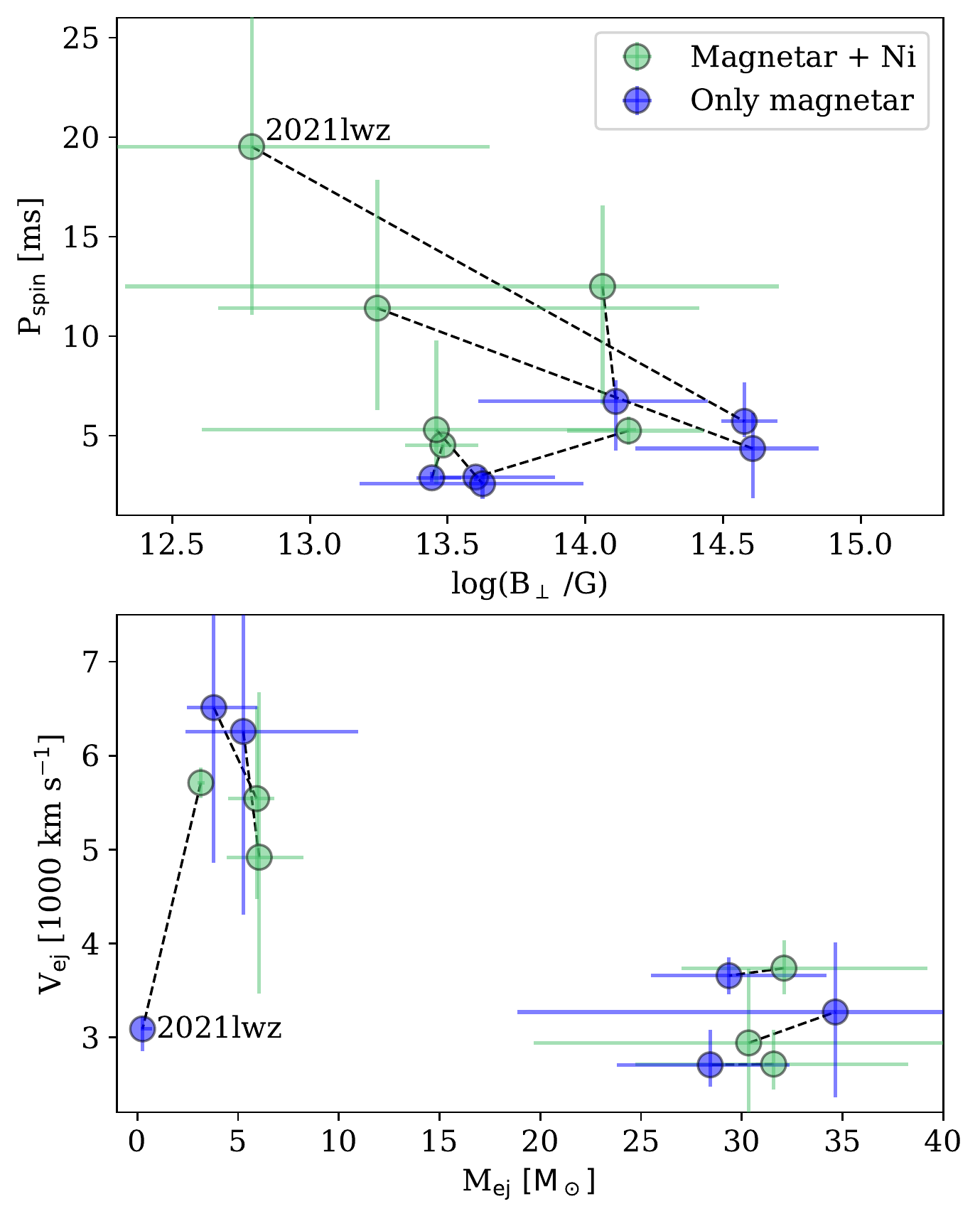}
        \caption{Key physical parameters for the six SLSNe with the smallest $f_{\rm mag}$ values (SN\,2020abjc, SN\,2020rmv, SN\,2020vpg, SN\,2021lwz, iPTF13bdl, and iPTF13bjz). In green we show the parameters from our fiducial magnetar plus radioactive decay model, and in blue the corresponding best-fit values after removing the radioactive decay component from the model. The parameters for the pure magnetar model deviate by more than $1 \sigma$ from the magnetar plus radioactive decay model only for SN\,2021lwz. This implies that a large radioactive decay contribution is not required for most SLSNe. \label{fig:no_radioactive}}
	\end{center}
\end{figure}

We test the effects of removing the radioactive decay component from the light curve models of the seven SLSNe with the lowest $f_{\rm mag}$ values, SN\,2020abjc, SN\,2020rmv, SN\,2020vpg, SN\,2021lwz, iPTF13bdl, and iPTF13bjz. In Figure~\ref{fig:no_radioactive} we show the best-fit parameters for these four SNe modeled with a pure magnetar model to explore how this model compares to the standard magnetar plus radioactive decay model. The magnetar-only models of all SLSNe, except for SN\,2021lwz, have values of $M_{\rm ej}$, $V_{\rm ej}$, $B_{\perp}$, and $P_{\rm spin}$ that remain consistent within $1 \sigma$ when compared to the magnetar plus radioactive decay models, although with systematically lower values of $P_{\rm spin}$ and higher values of $B_{\perp}$.

SN\,2021lwz is a relatively dim rapidly-evolving SLSN and the only one in our sample for which we do see a significant change in the best-fit parameters after removing the radioactive decay component. We find significantly lower values of $M_{\rm ej}$, $V_{\rm ej}$, $P_{\rm spin}$, and a significantly higher value of $B_{\perp}$. While both models provide good fits to the light curves of SN\,2021lwz, the standard magnetar plus radioactive decay model has a Bayesian Information Criterion (BIC; \citealt{Schwarz78}) value of 68, while the pure magnetar model has a BIC value of 63. This implies that the lower number of parameters in the pure magnetar model model is preferred over the marginal gains in likelihood from the standard magnetar plus radioactive decay model. Effectively this means that all SLSNe in our sample can be fit with a pure magnetar central engine model.

\begin{figure}
	\begin{center}
		\includegraphics[width=\columnwidth]{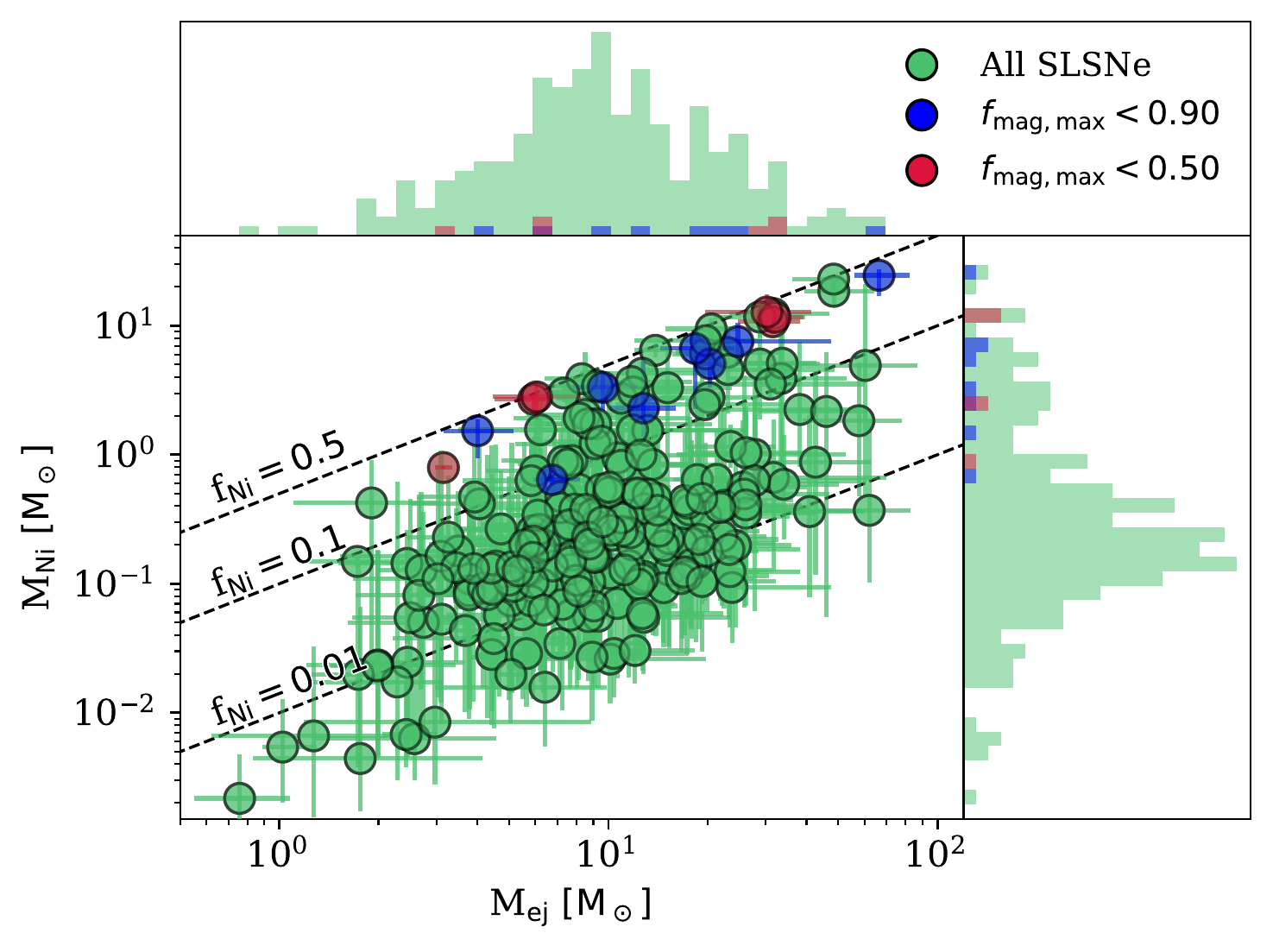}
		\caption{Nickel mass as a function of ejecta mass for the full sample. The dashed lines mark nickel fractions of 0.01, 0.1, and 0.5. The blue and red points show the SLSNe with a contribution from radioactive decay $>0.1$ and $>0.5$, respectively. A large nickel fraction or nickel mass does not necessarily imply the SN will be dominated by the contribution from radioactive decay. In some cases, large amounts of nickel can be hidden without having an impact on the magnetar dominated light curves. \label{fig:mejecta_nickel}}
	\end{center}
\end{figure}

In Figure~\ref{fig:mejecta_nickel} we show how the best-fit nickel mass varies as a function of ejecta mass, and mark in blue and red the SLSNe with the highest contributions from radioactive decay. A large value for the nickel fraction or nickel mass does not necessarily imply the light curve will be dominated by radioactive decay. Some SLSNe with high $f_{\rm Ni}$ values are still dominated by powerful magnetars with fast spin periods, whose contribution is able to overpower any contribution from radioactive decay. For these SLSNe, the posterior of the nickel mass and fraction are largely unconstrained, given that it is easy to hide large amounts of radioactive material under the large luminosity of the magnetar component.

\subsection{Progenitor Pre-explosion Mass Distribution}\label{sec:progenitor}

\begin{figure}
	\begin{center}
		\includegraphics[width=\columnwidth]{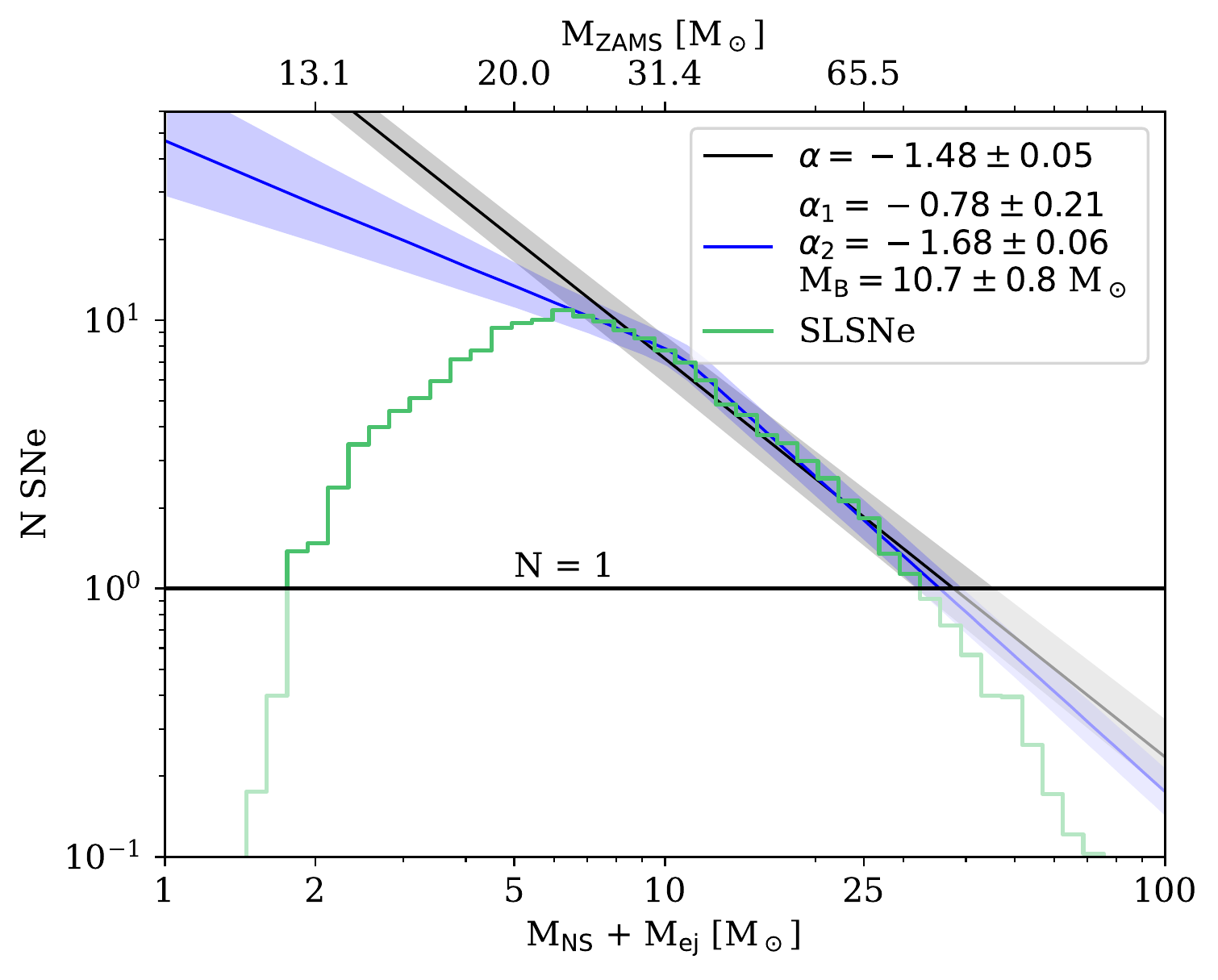}
		\caption{Progenitor pre-explosion mass distribution for the full sample of SLSNe. The green line shows a histogram of the joint posterior. The black line is the best-fit single power-law model, and the blue line is the best-fit broken power law to the tail of the distribution, with a break at $\approx 17$ M$_\odot$. Samples below the horizontal line at $N = 1$ are not statistically significant. The top axis is derived from the correlation between the CO core mass of low metallicity stars and their ZAMS from \cite{Sukhbold18}. \label{fig:distribution_mass}}
	\end{center}
\end{figure}

We measure the progenitor mass distribution for all SLSNe by summing the posterior of their ejecta mass and their corresponding posterior for the best-fit neutron star mass. The mass of the neutron star has little effect on the output light curves. For most SLSNe the neutron star mass posterior is largely unconstrained and dominated by the prior of $1.7 \pm 0.2$ M$_\odot$. We show the resulting progenitor mass distribution in Figure~\ref{fig:distribution_mass}, which has a peak of $\approx 6.5$ M$_\odot$, extending as low as $\sim 2$ M$_\odot$ and up to $\sim 40$ M$_\odot$. While there is a non-zero number of samples outside this range, these are not statistically significant. We fit the tail of the pre-explosion mass distribution, from 7 to $40$ M$_\odot$ with a single power law and find a best-fit slope of $\alpha = -1.48 \pm 0.05$. Alternatively, using a broken power law we find slopes of $\alpha_1 = -0.78 \pm 0.21$ and $\alpha_2 = -1.68 \pm 0.06$ with a break at $M_B = 10.7 \pm 0.8$ M$_\odot$. With a reduced $\chi^2$ value of $\chi^2 / d.o.f = 3.04$, the broken power law is a better fit than the single power law with $\chi^2 / d.o.f = 34.8$. The existence of a broken power law implies there might be multiple progenitor channels at play when producing SLSNe, leading to the distinct slopes in the high-end mass distribution. The slopes found here are steeper than the $\alpha_1 = -0.41 \pm 0.06$ and $\alpha_2 = -1.26 \pm 0.06$ found by \cite{Blanchard20}. We find that the decline at the low end of the distribution is shallower than the one seen in the sample of 62 SLSNe from \cite{Blanchard20}. We are able to recover the slopes found in \cite{Blanchard20} if we restrict our sample to the same 62 SLSNe used in that work, albeit with a break at $M_B = 10.6 \pm 0.8$ M$_\odot$, higher than the $M_B = 8.6$ M$_\odot$ found in \cite{Blanchard20}.

However, we note that the distribution of posteriors can vary with sample size, suggesting that there may be a sample size-dependent bias, due to the noisy mass measurements from our sample. Poorly constrained posteriors can ``smooth" the slope of this power law as the posteriors spread their support across wide mass ranges. To test for this bias, we measure the value of the best-fit slope as a function of number of SNe included in the sample. We run this test on 500 random sub-samples of 10 to \full\ SLSNe and find no correlation between sample size and best-fit slope. We determine that $N \sim 110$ SLSNe are necessary and sufficient to obtain a confident measurement of the slope and its corresponding uncertainty given that their values remain largely constant for samples larger than this. An exploration of possible observational biases was carried out in \cite{Blanchard20}, who found a slight bias against the lowest progenitors with $M \lesssim 6$ M$_\odot$, which effectively slightly flattens the distribution, but not significantly enough to change the results.

We compare our results with the CCSNe explosion models from \cite{Sukhbold18}, who evolve a set of stars with ZAMS from 12 to 60 M$_\odot$ at low metallicity, which should be the most representative of SLSNe progenitors. We equate our pre-explosion mass distribution measurements with the CO-mass measurements of the low-metallicity models from \cite{Sukhbold18} to derive an estimate of the ZAMS masses of the progenitors of SLSNe. We find a corresponding peak of the distribution of $M_{\rm ZAMS} \sim 23$ M$_\odot$, with a low end of $M_{\rm ZAMS} \sim 13$ M$_\odot$ and extending beyond the most massive 60 M$_\odot$ progenitors presented in \cite{Sukhbold18}.

In \cite{Gomez22_LSN} we found the pre-explosion mass distribution of LSNe peaks at $\sim 5$ M$_\odot$, and below $\sim 4$ M$_\odot$ for SNe Ic/Ic-BL. The fact that we see a turn-over below $\sim 6$ M$_\odot$ for SLSNe suggests there might be a fundamental limit or mechanism that governs the minimum mass of progenitors capable of producing SLSNe that becomes relevant for stars below $M_{\rm ZAMS} \sim 23$ M$_\odot$, or that stars with CO-cores below $\sim 6$ M$_\odot$ are less efficient at producing SLSNe. \cite{Blanchard20} studied the possibility that this could be due to an observational bias, and concluded that observational biases had little effect on the measured pre-explosion mass distribution. It is possible this is an effect of the low metallicity required to produce SLSNe. For example, \cite{Aguilera-Dena23} claim that Type Ic SN progenitors likely experience additional mass loss during their evolution, either from winds or a different mechanism, leading to their lower pre-explosion mass distribution.

\subsection{Lack of Redshift Dependence}\label{sec:redshift}

We find no strong correlation with redshift for any physical or observational parameter. The parameter with the strongest correlation with redshift is the spin period, which we find has a mean value of $P_{\rm spin} = 2.6^{+3.3}_{-1.3}$ ms for SLSNe closer than $z = 0.5$ and  $P_{\rm spin} = 2.1^{+0.9}_{-1.5}$ ms for SLSNe further than $z = 0.5$. This difference is not statistically significant enough to claim a redshift dependence. Moreover, the difference goes away completely when we consider only SLSNe brighter than $M_r = -21.5$ mag, suggesting this is simply an observational bias that makes SLSNe with slow $P_{\rm spin}$ values harder to observe at further distances. This is consistent with previous findings from \cite{Hsu21} and \cite{Cia18}.

\subsection{Velocity}\label{sec:velocity}

\begin{figure}
	\begin{center}
		\includegraphics[width=\columnwidth]{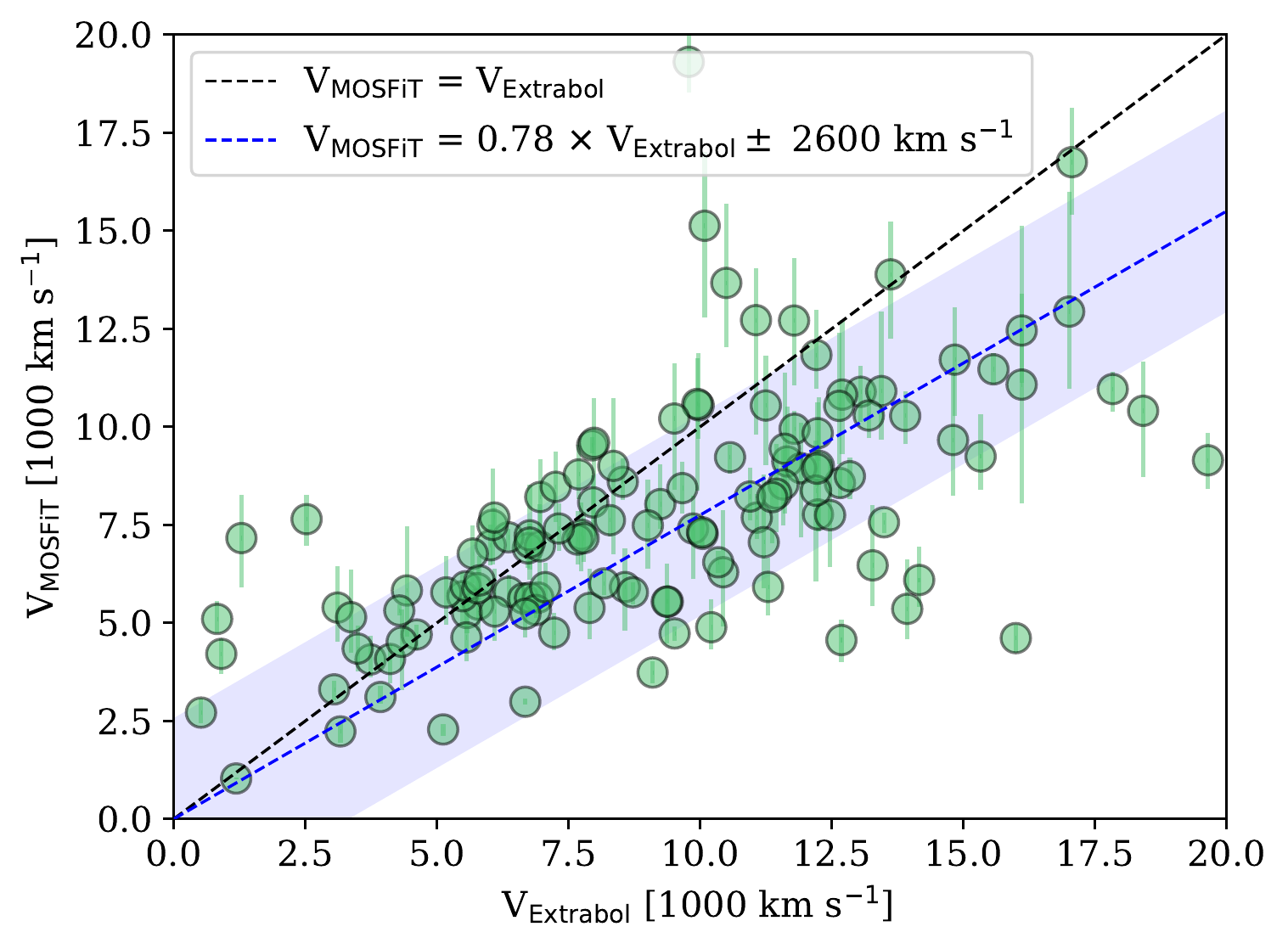}
		\caption{Velocity derived from the {\tt extrabol} models compared to the ejecta velocity derived from the \mosfit models a sample of 136 SLSNe. The black line shows the 1-to-1 correspondence between the two values, while the blue line is a best-fit line of the correlation with the y-intercept fixed at 0. We find the two measurements of velocity to be generally consistent. \label{fig:v_phot}}
	\end{center}
\end{figure}

In the bottom panel of Figure~\ref{fig:SEDs} we show how typical SLSNe begin with a blackbody velocity of $V \sim 10,000$~km\,s$^{-1}$ and remain constant before reaching peak, after which the photosphere begins to recede deeper into the ejecta. We compare the ejecta velocity from \mosfit to the blackbody velocity derived from the photospheric radius evolution measured with {\tt extrabol}, measured before peak. We are able to obtain this measurement with {\tt extrabol} for 136 SLSNe, other SLSNe do not have enough photometry to fit with {\tt extrabol}. In Figure~\ref{fig:v_phot} we show how these two measurements compare. We find good agreement between the two methods, but find that the ejecta velocity from \mosfit is around 20\% lower then the velocity derived from {\tt extrabol}, with an intrinsic scatter of $\sim 2,600$ km s$^{-1}$.

\subsection{Radiative Efficiency}\label{sec:efficiency}

We calculate the radiative efficiency $\epsilon$ by dividing the total radiated energy over the first 200 days of a SN by the total kinetic energy. SLSNe are known to have high radiative efficiencies, but if we use the ejecta velocity from \mosfit to calculate $\epsilon$ we find $\sim 15$ SLSNe with an $\epsilon > 1$. If we instead assume the higher velocity derived from the {\tt extrabol} models, the number of SLSNe with $\epsilon > 1$ goes down to 0, after accounting for the uncertainties in the value of $\epsilon$. Both the radiated and kinetic energy are dominated by the magnetar spin-down. Our models suggest that most of the energy goes into accelerating the ejecta, but for the SLSNe with rapid spin periods, a higher fraction of the magnetar energy goes into radiation, particularly if they have low ejecta velocity, making these SLSNe very efficient emitters.

\subsection{Comparison to Other Works}\label{sec:models}

\begin{figure}
	\begin{center}
		\includegraphics[width=\columnwidth]{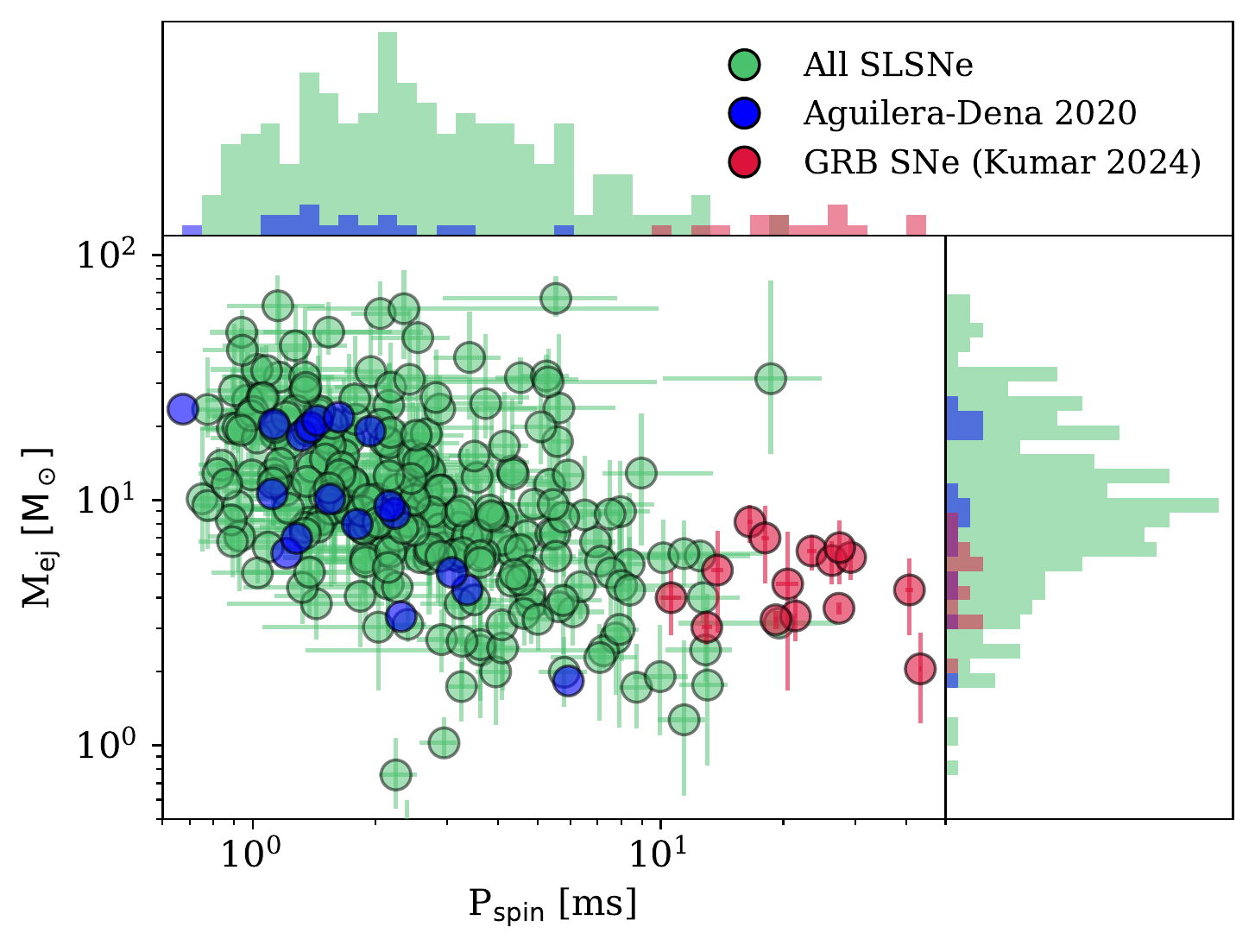}
		\caption{Ejecta mass versus spin period for the full sample shown in green. We find the trend seen previously in \citep{Blanchard20}, with a general absence of events with low ejecta mass and rapid spin. Also shown are models of SLSNe explosions from \cite{Aguilera20} (blue) and fits to GRB-SNe using a magnetar model from \cite{Kumar24} (red). \label{fig:Pspin_mejecta}}
	\end{center}
\end{figure}

The models from \cite{Aguilera20} evolve a set of low-metallicity, rapidly-rotating stars with pre-explosion masses ranging from $M_{\rm Init} = 4$ to 45 $M_\odot$. These models are thought to represent the progenitors of magnetar-powered transients such as SLSNe. In Figure~\ref{fig:Pspin_mejecta} we show how the spin period and ejecta mass measurements from \cite{Aguilera20} compare to the best-fit \mosfit parameter for our sample of SLSNe. We find that the general correlation between $P_{\rm Spin}$ and $M_{\rm ej}$ found in the \cite{Aguilera20} models, originally presented in \cite{Blanchard20}, is still true, with some additional scatter likely due to the inclusion of a radioactive decay component in our models.

In Figure~\ref{fig:Pspin_mejecta} we also show a comparison to the $P_{\rm Spin}$ and $M_{\rm ej}$ values measured for a set of GRB SNe from \cite{Kumar24}, who fit the light curves of these SNe with a magnetar central engine model. If GRB SNe are powered by SNe, we see that the general trend continues, as SNe with lower ejecta masses seem to have higher spin periods, or less powerful magnetars. The mean of the population of GRB SNe has a spin period $P_{\rm Spin} = 21.42$ ms, which is slower than the slowest spin period of $P_{\rm Spin} = 19.83$ ms found in the SLSN population. GRB SNe show a mean ejecta mass value of $M_{\rm ej} = 4.8$ M$_\odot$, which corresponds to the $\sim 13$th percentile of the SLSNe population.

For comparisons of SLSNe to the general population of SESNe, we direct the reader to \cite{Gomez22_LSN}. In that study we presented a connection between the ``normal'' Type Ic/Ic-BL SNe and SLSNe. We concluded that LSNe, which by definition span a range of peak magnitudes from $M_r = -19$ to $M_r = -20$, straddle the line between SNe Ic/Ic-BL and SLSNe in terms of not only their luminosities, but also their power sources, duration, and physical parameters.

\section{Conclusions}\label{sec:conclusions}

We have presented the largest sample of Type I superluminous supernovae (SLSNe) to date, encompassing all publicly known SLSNe and their photometry up until Dec 31, 2022, totalling \total\ events. Of those, we determined \full\ have enough photometry and spectroscopy to be robustly classified as ``gold'' or ``silver'' SLSNe. We use this sample of SLSNe to conduct our analysis, from which we derive our conclusions.

\begin{itemize}
    \item The SLSN population has a mean rise time of $\tau_{\rm rise} = 27^{+25}_{-13}$ days, and a mean e-folding decline time of $\tau_e = 44^{+38}_{-18}$ days.
    \item The mean peak rest-frame $r$-band absolute magnitude is $M_r = -21.3^{+0.9}_{-0.6}$, and the mean peak bolometric luminosity is $\log{L_{\rm max}} = 44.3^{+0.3}_{-0.5} {\rm\ erg \ s^{-1}}$.
    \item SLSNe have mean ejecta masses of $M_{\rm ej} = 9.3^{+12.9}_{-4.8}$ M$_\odot$ and mean ejecta velocities of $V_{\rm ej} = 6800^{+3400}_{-2000}$ km s$^{-1}$.
    \item We use the blackbody radius evolution from {\tt extrabol} to estimate the velocity of the photosphere and find these values to be $\sim 20$\% lower, but still consistent, with the ejecta velocities derived from \mosfit.
    \item The inferred mean blackbody radius reaches a peak of $\approx 5\times 10^{15}$ cm at $\approx 50$ days post peak, and then declines to $\approx 2 \times 10^{15}$ cm at $\gtrsim 200$ days after peak.
    \item The mean magnetar central engine parameters are $P_{\rm spin} = 2.4^{+3.0}_{-1.2}$ ms and $\log(B_{\perp} / G) = 14.2 \pm 0.4$.
    \item A magnetar central engine model is able to fit the light curves of effectively all SLSNe without additional energy sources.
    \item Radioactive heating from $^{56}$Ni decay does not have a noticeable contribution in most SLSNe, with the exception of a few of some slowly declining, relatively dim SLSNe.
    \item We do not find a correlation between ejecta mass and magnetic field strength, as suggested by \citet{Hinkle23}.
    \item We find no significant redshift-dependence for any physical or observational parameter.
    \item We find no strong evidence for sub-types of SLSNe when considering the full range of observed and physical properties. For example, SLSNe do not split between slow and rapid sub-populations, but instead form a smooth continuum.
    \item We find the late-time decline rates for most SLSNe to be faster than the expectation from radioactive $^{56}$Co decay. SLSNe with decline rates consistent with that of $^{56}$Co decay can be explained by simply representing the slow end of the distribution of decline rates.
    \item The progenitor pre-explosion mass distribution peaks at $\approx 6.5$ M$_\odot$, which roughly corresponds to a ZAMS mass of $M_{\rm ZAMS}\sim 23$ M$_\odot$.
    \item The progenitor pre-explosion mass distribution extends as low as $\approx 2$ M$_\odot$ and up to $\approx 40$ M$_\odot$, with a broken-power law distribution with a break at $M_B = 10.7 \pm 0.8$ M$_\odot$.
\end{itemize}

Starting in 2025, the Vera C.~Rubin Observatory Legacy Survey of Space and Time (Rubin; \citealt{LSST19}) is planned to commence and expected to discover $\sim 10^4$ SLSNe in the Wide-Fast-Deep survey \citep{Villar18}. Similarly, in 2027 the Nancy Grace Roman Space Telescope (\textit{Roman}; \citealt{Roman15}) is scheduled to launch and find SLSNe to $z\sim 5$ \citep{Moriya22_Roman, Gomez23_Roman}. Finding these SNe will allow us to constrain their event rate as a function of redshift, and test how star formation varies with redshift for the most massive progenitor stars (e.g., \citealt{Madau14, Frohmaier2020}). The catalog presented here can be used to simulate observations of SLSNe that Rubin and \textit{Roman} will observe, to help us better prepare for these surveys. Once these surveys begin, their discoveries will allow us to push the boundaries of our understanding of the SLSN population.

\onecolumngrid
\section{Catalog Description}\label{sec:catalog}

We provide the entire catalog, including all the photometry, models, and derived parameters as an open-source tool available on GitHub \href{https://github.com/gmzsebastian/SLSNe}{\faGithub}. The package can be easily installed via PyPI with a {\tt pip install slsne} command. Additionally, we provide Python examples on how to access different components of the catalog and use them to either reproduce plots on this paper or for comparison to other studies. The catalog is open-source and flexible enough for anyone to contribute their data via a pull request. While we list some examples of how to use the code here, we encourage users to reference the documentation\footnote{\url{https://slsne.readthedocs.io}} for the must up-to-date syntax and code examples.

In Listing~\ref{lst:curves} we provide examples on how to obtain the light curves of either individual SLSNe, or light curves for the full sample. These tools can be used to create comparisons such as the one shown in Figure~\ref{fig:SLSNe_LCs}. In Listing~\ref{lst:kcorr} and \ref{lst:bolcorr} we show examples on how to obtain an accurate K-correction or bolometric scaling as a function of wavelength and phase for any SLSN. In Listing~\ref{lst:params} we show an example on how to obtain a set of parameters from the SLSN population. For more examples, including scripts used to reproduce the plots in this paper, we encourage the reader to visit the GitHub repository of the catalog.

We encourage users to cite the original sources of data for all SNe used. Therefore, we provide a function named {\tt get\_references}, which can take in a list of SN names, and return all the bibcode entries used for those SNe. If no list is provided, the function will print all the bibcode entries used for all photometry used in this work.

\begin{lstlisting}[language=Python, label={lst:curves}, caption={Example of how to obtain the rest frame model light curves of all SLSNe, as well as a single SLSN.}]
from slsne.lcurve import get_all_lcs

# Get light curves of all SLSNe in r-band
(dim, mean, high), (time_samples, lightcurves) = get_all_lcs("r")

# Get a single light curve
time_samples, r_2018lfe = get_all_lcs("r", names = "2018lfe")

\end{lstlisting}

\begin{lstlisting}[language=Python, label={lst:kcorr}, caption=Example of how to calculate the K-correction for the observed photometry of a SLSN.]
from slsne.lcurve import get_kcorr, fit_map
from slsne.utils import get_lc

# Import a SLSN light curve, and define a redshift and peak date
phot = get_lc("2013dg")
redshift = 0.265
peak = 56447.62

# Fit the best scaling for the SLSN map
stretch, amplitude, offset = fit_map(phot, redshift, peak=peak)

# Get corresponding K-correction for the photometry
K_corr = get_kcorr(phot, redshift, peak = peak, stretch = stretch, offset = offset)

# Apply the K-correction to the photometry
corr_mag = phot["Mag"] - K_corr
\end{lstlisting}

\begin{lstlisting}[language=Python, label={lst:bolcorr}, caption=Example of how to calculate the bolometric scaling for a SLSN.]
from slsne.lcurve import get_bolcorr
from slsne.utils import calc_flux_lum, get_lc

# Import a SLSN light curve
phot = get_lc("2018lfe")
redshift = 0.35
peak = 58468.55

# Calculate the phase with respect to peak
phot["Phase"] = (phot["MJD"] - peak) / (1 + redshift)

# Measure the bolometric scaling
bol_scaling = get_bolcorr(phot, redshift, peak)

# The scaling can then be applied to the luminosity as
F_lambda, L_lambda = calc_flux_lum(phot, redshift)
L_bol = L_lambda.value / bol_scaling

\end{lstlisting}

\begin{lstlisting}[language=Python, label={lst:params}, caption={Example of how to obtain a sample of parameters. If no {\tt param\_names} is specified, all parameters will be returned.}]
from slsne.utils import get_params

# Get the parameters
params = get_params(param_names=["Pspin","mejecta"])

# The format of the output parameters is an Astropy Table
Pspin_med Pspin_up Pspin_lo mejecta_med mejecta_up mejecta_lo
--------- -------- -------- ----------- ---------- ----------
   3.8508   5.3963   2.4987       2.877     4.5832     1.2612
\end{lstlisting}

\vspace{0.5in}
\twocolumngrid

\acknowledgements

We thank Yuri Beletsky for observing most of the spectra and photometry from the Magellan Telescopes. S.G. is supported by an STScI Postdoctoral Fellowship. M.N. is supported by the European Research Council (ERC) under the European Union’s Horizon 2020 research and innovation programme (grant agreement No.~948381) and by UK Space Agency Grant No.~ST/Y000692/1. This project was supported in part by the Transients Science @ Space Telescope group. T. E. is supported by NASA through the NASA Hubble Fellowship grant HST-HF2-51504.001-A awarded by the Space Telescope Science Institute, which is operated by the Association of Universities for Research in Astronomy, Inc., for NASA, under contract NAS5-26555. This work is supported by the National Science Foundation under Cooperative Agreement PHY-2019786 (The NSF AI Institute for Artificial Intelligence and Fundamental Interactions, http://iaifi.org/). A.F. acknowledges the support by the State of Hesse within the Research Cluster ELEMENTS (Project ID 500/10.006). V.A.V. acknowledges support from the NSF through grant AST-2108676. R.K.T. is supported by the NKFIH/OTKA FK-134432 and the NKFIH/OTKA K-142534 grant of the National Research, Development and Innovation (NRDI) Office of Hungary. N.F. acknowledges support from the National Science Foundation Graduate Research Fellowship Program under Grant No. DGE-2137419. The LCO group is supported by NSF grants AST-1911151 and 1911225.

W. M. Keck Observatory and MMT Observatory access was supported by Northwestern University and the Center for Interdisciplinary Exploration and Research in Astrophysics (CIERA). This paper includes data gathered with the 6.5 meter Magellan Telescopes located at Las Campanas Observatory, Chile. Observations reported here were obtained at the MMT Observatory, a joint facility of the University of Arizona and the Smithsonian Institution. IRAF is written and supported by the National Optical Astronomy Observatories, operated by the Association of Universities for Research in Astronomy, Inc. under cooperative agreement with the National Science Foundation. This work makes use of observations from Las Cumbres Observatory global telescope network. Operation of the Pan-STARRS1 telescope is supported by the National Aeronautics and Space Administration under grant No. NNX12AR65G and grant No. NNX14AM74G issued through the NEO Observation Program. This work has made use of data from the European Space Agency (ESA) mission {\it Gaia} (\url{https://www.cosmos.esa.int/gaia}), processed by the {\it Gaia} Data Processing and Analysis Consortium (DPAC, \url{https://www.cosmos.esa.int/web/gaia/dpac/consortium}). Funding for the DPAC has been provided by national institutions, in particular the institutions participating in the {\it Gaia} Multilateral Agreement. This research has made use of the SIMBAD database, operated at CDS, Strasbourg, France. Based on observations obtained with MegaPrime/MegaCam, a joint project of CFHT and CEA/IRFU, at the Canada-France-Hawaii Telescope (CFHT) which is operated by the National Research Council (NRC) of Canada, the Institut National des Science de l'Univers of the Centre National de la Recherche Scientifique (CNRS) of France, and the University of Hawaii. This work is based in part on data products produced at Terapix available at the Canadian Astronomy Data Centre as part of the Canada-France-Hawaii Telescope Legacy Survey, a collaborative project of NRC and CNRS. This research has made use of the NASA/IPAC Extragalactic Database, which is funded by the National Aeronautics and Space Administration and operated by the California Institute of Technology. This research has made use of NASA’s Astrophysics Data System. This research has made use of the NASA/IPAC Infrared Science Archive, which is funded by the National Aeronautics and Space Administration and operated by the California Institute of Technology. This work made use of Swift/UVOT data reduced by P. J. Brown and released in the Swift Optical/Ultraviolet Supernova Archive (SOUSA). SOUSA is supported by NASA's Astrophysics Data Analysis Program through grant NNX13AF35G. This work makes use of the Weizmann Interactive Supernova Data Repository (WISeREP, \url{https://www.wiserep.org}).

\facilities{ADS, ASAS, Gaia, IRSA, OGLE, OSC, Las Cumbres, PS1, Sloan, SO, Swift, TNS, WISE}
\software{Astropy \citep{astropy}, extinction \citep{Barbary16}, Matplotlib \citep{matplotlib}, emcee \citep{Foreman13}, extrabol \citep{extrabol}, NumPy \citep{numpy}, scikit-learn \citep{Scikit-learn}, MOSFiT \citep{guillochon18}, PyRAF \citep{science12}, SAOImage DS9 \citep{Smithsonian00}, slsne \citep{slsne}, corner \citep{foreman16}, HOTPANTS \citep{Becker15}, FLEET \citep{Gomez20_FLEET}, George \citep{Foreman-Mackey15}}

\bibliography{references}

\clearpage
\newpage
\appendix
\setcounter{figure}{0}
\setcounter{table}{0}

The title of each sub-section shows the shorthand name adopted for each SN; other given names of each SN are listed in the each section. We specify the individual sources of photometry used for each source.


\section{``Gold" Superluminous Supernovae}\label{sec:golden_slsn}

\subsection{2005ap}

\subsection{2007bi}
SN\,2007bi (=SNF20070406-008) was classified as a luminous SN Ic by \cite{Nugent07} and suggested to be a pair-instability SN by \cite{Gal-Yam09}, but later reclassified as a SLSN-I by \cite{Nicholl13}. We include photometry from \cite{Gal-Yam09} and \cite{Young10}. The \cite{Gal-Yam09} photometry listed as R-band is a combination of photometry from the P60, P200, and P48 telescopes, synthetic Keck photometry taken from spectra, and CSS photometry calibrated to R-band. \cite{Dessart12} argue that SN\,2007bi is well reproduced by a model of delayed injection of energy by a magnetar. \cite{Yoshida14} claim the light curve is well reproduced by the aspherical explosion of a $> 100$ M$_\odot$ star. \cite{Moriya19} argue that 3 M$_\odot$ of CSM are likely to exist around the environment of SN\,2007bi. We include one spectrum from \cite{Gal-Yam09}, obtained from WISeREP.

\subsection{2009jh}
SN\,2009jh (=CSS090802:144910+292510 =PTF09cwl) was discovered by PTF and originally classified as a SN Ic, before the SLSN-I designation existed \citep{Drake09_09jh, Quimby09_09jh}. The SN was later classified as a SLSN-I by \cite{Quimby11}. \cite{Perley16} found an associated host galaxy at a redshift of $z = 0.3499$, which we adopt as the redshift of the SN. We include photometry from \cite{Cia18}, which the authors correct for extinction using $E(B-V) = 0.013$. We also include photometry from \cite{Quimby11}, which we correct for extinction. We include one spectrum from \cite{Quimby18}, obtained from WISeREP.

\subsection{2010gx}
SN\,2010gx (=CSS100313:112547-084941 =PS1-1000037 =PTF10cwr) was originally discovered by CRTS \cite{Mahabal10} with an erratum in \cite{Mahabal10b} and classified as a SLSN-I by \cite{Quimby10_10gx}. A detailed study of the SN was presented in \cite{Pastorello10}, who classify this as a Type-Ic SNe. The detailed study from \cite{Quimby18} later reclassified the source as a SLSN-I. \cite{Perley16} found an associated host galaxy at a redshift of $z = 0.2297$, which we adopt as the redshift of the SN. We include photometry from \cite{Cia18}, which the authors correct for extinction using $E(B-V) = 0.032$. We include photometry from \cite{Quimby11} and \cite{Pastorello10}, which we correct for extinction. We note that the early time $r$-band photometry from \cite{Quimby11} appears $\sim 0.2$ mag dimmer than the photometry from \cite{Pastorello10} and \cite{Cia18}. We include an additional upper limit from \cite{Quimby10_10gx}. There is UVOT photometry from the SOUSA archive, but not in a refereed publication and given that it does not match the $u$-band photometry from \cite{Pastorello10}, we do not include the UVOT photometry. We include one spectrum from \cite{Pastorello10}, obtained from WISeREP. \cite{Chen13} found the host galaxy of SN\,2010gx to be a dwarf galaxy with a high specific star formation rate. 

\subsection{2010hy}
SN\,2010hy (PTF10vwg) was discovered by PTF and originally classified as a luminous SN by \cite{Kodros10} and later as a luminous Type Ic SN by \cite{Vinko10_10hy}. The SN was ultimately classified as a SLSN-I in a detailed study by \cite{Quimby18}. \cite{Perley16} found an associated host galaxy at a redshift of $z = 0.1901$, which we adopt as the redshift of the SN. We include photometry from \cite{Cia18}, which the authors correct for extinction using $E(B-V) = 0.455$. We include one spectrum from \cite{Shivvers19}, obtained from WISeREP.

\subsection{2010kd}
SN\,2010kd was classified as an early Type-II supernova by \cite{Vinko10}, but later reclassified as a SLSN-I by \cite{Roy12}. We include photometry from Chapter 6 in \cite{Roy12} and from \cite{Kumar20}. The authors find a high intrinsic extinction value of $E(B-V) = 0.15$, in addition to a foreground $E(B-V) = 0.02$ extinction. We include one spectrum from \cite{Kumar20}, obtained from WISeREP.

\subsection{2010md}
SN\,2010md (=PTF10hgi =PSO J249.4461+06.2081) was discovered by PTF and originally classified as a SLSN-I by \cite{Quimby13_10md} with a detailed study presented in \cite{Inserra13}. \cite{Perley16} found an associated host galaxy at a redshift of $z = 0.0987$, which we adopt as the redshift of the SN. We include photometry from \cite{Cia18}, which the authors correct for extinction using $E(B-V) = 0.071$. We include photometry from \cite{Inserra13}, which we correct for extinction. \cite{Inserra13} transform the UVOT U-band to SDSS $u$-band. There is UVOT photometry from \cite{Quimby10}, but this is off by several magnitudes, possibly since these do not account for the host contribution; we therefore do not include it. This is one of the few SLSN-I that shows helium in its photospheric phase \citep{Yan20}. We include one spectrum from \cite{Quimby18}, obtained from WISeREP.

\subsection{2011ke}
SN\,2011ke (=CSS110406:135058+261642 =PS1-11xk =PTF11dij) was originally discovered by PS1 and classified as a SLSN-I \citep{Drake11_11ke, Quimby11_11ke, Smartt11}. A detailed study of the SN was presented in \cite{Inserra13}. \cite{Perley16} found an associated host galaxy at a redshift of $z = 0.1428$, which we adopt as the redshift of the SN. We include photometry from \cite{Cia18}, which the authors correct for extinction using $E(B-V) = 0.01$. We include photometry from \cite{Inserra13}, which we correct for extinction. \cite{Inserra13} transform the UVOT U-band to SDSS $u$-band. We include one spectrum from \cite{Inserra13}, obtained from WISeREP.

\subsection{2011kg}
SN\,2011kg (=PTF11rks) was discovered by PTF and originally classified as a SLSN-I by \cite{Quimby13} with a detailed study presented in \citep{Inserra13}. \cite{Perley16} found an associated host galaxy at a redshift of $z = 0.1924$, which we adopt as the redshift of the SN. We include photometry from \cite{Cia18}, which the authors correct for extinction using $E(B-V) = 0.035$; and photometry from \cite{Inserra13}, which we correct for extinction. \cite{Inserra13} transform the UVOT U-band to SDSS $u$-band. We do not include the late-time non-detections from \cite{Inserra13} given the multiple nearby detections from \cite{Cia18}. \cite{Inserra13} include two data points from \cite{Quimby11_11rks}, for which we assume an uncertainty of 0.1 mag. We include one spectrum from \cite{Quimby18}, obtained from WISeREP.

\subsection{2012il}
SN\,2012il (=CSS120121:094613+195028 =PS1-12fo) was originally discovered by \cite{Drake12} and classified as a SLSN-I by \cite{Smartt12}. A detailed study of the SN was presented in \cite{Inserra13}, from which we obtain the photometry. The photometry does not have the host flux subtracted, but we conclude its contribution should be negligible given the SN is in the outskirts of a dim galaxy with $r = 21.46$ mag. The host galaxy of SN\,2012il was modelled by \cite{Chen17_host}. We include one spectrum from \cite{Inserra13}, obtained from WISeREP.

\subsection{2013dg}
SN\,2013dg (=CSS130530:131841-070443 =MLS130517:131841-070443) was discovered by the CRTS and classified as a SLSN-I \cite{Drake13, Smartt13}. A detailed study of the SN was presented in \cite{Nicholl14}. We include photometry from \cite{Nicholl14}. We include one spectrum from \cite{Shivvers19}, obtained from WISeREP.

\subsection{2015bn}
SN\,2015bn (=PS15ae =CSS141223:113342+004332 =MLS150211:113342+004333) was discovered by ASAS-SN and originally classified as a SLSN-I by \citep{Guillou15, Guillou16}. A detailed study of the SN was presented in  \cite{Nicholl16_15bn}. We include photometry from \cite{Nicholl16_15bnnebular}, \cite{Nicholl16_15bn}, and \cite{Nicholl18_1000days}. \cite{Leloudas17} and \cite{Inserra18_polarimetry} presented polarimetry of SN\,2015bn and find an increased level of polarization after $\sim $ 20 days, which the authors interpret as a phase transition where the original outer layer of C and O becomes a more aspherical inner cored dominated by nucleosynthesized material. \cite{Bhirombhakdi18} present X-ray observations of SN\,2015bn and conclude that leakage of energy at late times is required to reproduce the X-ray upper limits. We do not include the late-time WISE photometry from \cite{Sun22}. We include one spectrum from \cite{Nicholl16_15bn}, obtained from WISeREP.

\subsection{2016ard}
SN\,2016ard (=PS16aqv =CSS160216:141045-100935) was classified as a SLSN-I by \cite{Blanchard18}. We include photometry from \cite{Blanchard18}. We include one spectrum from \cite{Blanchard18}, obtained from WISeREP.

\subsection{2016eay}
SN\,2016eay (=Gaia16apd) was discovered by WISE and originally classified as a SLSN-I by \cite{Elias-rosa16} with a detailed study presented in \cite{Nicholl17_16apd}. The authors transform the \textit{Gaia}-$G$ band to SDSS-$i$ band, and convert $ubv$ bands to $UBV$, respectively. We include our own PSF photometry of Las Cumbres GSP images after doing difference imaging to subtract the host contribution. We do not include the late-time WISE photometry from \cite{Sun22}. We include one spectrum from \cite{Yan17_16eay}, obtained from WISeREP.

\subsection{2016inl}
SN\,2016inl (=PS16fgt) was discovered by PS1 and classified as a SLSN-I by \cite{Blanchard21_16inl, Blanchard21_inl}. We include photometry from \cite{Blanchard21_16inl}. We include one spectrum from \cite{Blanchard21_16inl}.

\subsection{2017dwh}
SN\,2017dwh (=PS17dbf =ATLAS17fau =CSS170425:143443+312917) was discovered by the CRTS and classified as a SLSN-I by \cite{Blanchard19, Blanchard18_dwh}. We include photometry from \cite{Blanchard19}. We include one spectrum from \cite{Blanchard19}.

\subsection{2017egm}
SN\,2017egm (=Gaia17biu =PS18cn =iPTF17egm) was discovered by Gaia and originally classified as a Type II SN by \cite{Xiang17} but later re-classified as a SLSN-I by \cite{Dong17}. A detailed study of the SN was presented in \cite{Nicholl17_17egm}. \cite{Izzo18} studied the host galaxy of SN\,2017egm and report a redshift of $z = 0.03072$, which we adopt as the redshift of the SN. We include photometry from \cite{Nicholl17_17egm}, but exclude data before MJD = 57910 from the \mosfit model (The equivalent of Model 3 from \citealt{Nicholl17_17egm}). We also exclude the GSA, ATLAS, and PS1 photometry after MJD = 58105 from the \mosfit fit, since these show a peculiar late-time re-brightening. \cite{Saito20} find strong polarization at late times for SN\,2017egm. We do not include the late-time WISE photometry from \cite{Sun22}. We include one spectrum from \cite{Bose18}, obtained from WISeREP.

\subsection{2017ens}
SN\,2017ens (=ATLAS17gqa =CSS170614:120409-015552) was discovered by ATLAS and classified as a SLSN-I by \cite{Chen18}. We include photometry from \cite{Chen18}. The authors show that this SN developed hydrogen features at late times. We do not include the late-time WISE photometry from \cite{Sun22}. We include one spectrum taken by C. R. Angus obtained from WIseREP.

\subsection{2017gci}
SN\,2017gci (=Gaia17cbp) was discovered by Gaia and classified as a SLSN-I by \cite{Lyman17}, with a following detailed study in \cite{Fiore20}. We include photometry from \cite{Fiore20}, which the authors correct for extinction using $A_V = 0.36$. The authors also transform the GSA magnitudes to $g$-band. \cite{Stevance21} find that SN\,2017gci is well explained by a 30 M$_\odot$ binary system progenitor, as opposed to a single star progenitor. We do not include the late-time WISE photometry from \cite{Sun22}. We include one spectrum form \cite{Lyman17_17gci}, obtained from the TNS.

\subsection{2018avk}
SN\,2018avk (=ZTF18aaisyyp =Gaia18ayq =ATLAS18pcj) was discovered by ZTF and originally classified as a SLSN-I by \cite{Nicholl18_avk} with a detailed study presented in \cite{Lunnan20_four}. We include photometry from \cite{Lunnan20_four}, which the authors correct for extinction using $E(B-V) = 0.012$. We include our own PSF photometry of FLWO and Las Cumbres GSP images after doing difference imaging to subtract the host flux. We include our own Blue Channel spectrum from \cite{Nicholl18_avk}.

\subsection{2018bgv}
SN\,2018bgv (=ZTF18aavrmcg =Gaia18beg =PS18su =ATLAS18pko =MASTER OT J110230.30+553555.5) was discovered by ZTF and originally classified as a SLSN-I by \cite{Dong18, Dong18-TNS}. A detailed study of the SN was presented in \cite{Lunnan20_four}. We include photometry from \cite{Lunnan20_four}, which the authors correct for extinction using $E(B-V) = 0.008$ mag. We include data from ATLAS, \textit{Gaia}, and the CPCS \citep{Zielinski19}. We include our own PSF photometry of FLWO and Las Cumbres GSP images after doing difference imaging to subtract the host flux. We exclude CPCS photometry after MJD = 58285 due to a prominent bump not observed in the higher quality ZTF data. We do not include the late-time WISE photometry from \cite{Sun22}. We include a spectrum from \cite{Dong18, Dong18-TNS}.

\subsection{2018bsz}
SN\,2018bsz (=ASASSN-18km =ATLAS18pny) was discovered by ATLAS and originally classified as a SN II by \cite{Hiramatsu18} and \cite{Clark_18bsz}, but later re-classified as a SLSN-I by \citep{Anderson18}. We include photometry from ASAS-SN, ATLAS, and \citep{Anderson18}. The authors do not subtract the host flux from the UVOT photometry, but claim its contribution should be negligible. We exclude the long rising plateau before MJD = 58250 from the \mosfit fit. \cite{Maund21} provide polarimetry observations of SN\,2018bsz and find limits of $\lesssim 1-2$\%, except for one detection of $2\pm 0.5$\% at 11.4 days post maximum. \cite{Pursiainen22} studied the spectra of SN\,2018bsz in detail and found it to be consistent with having aspherical circumstellar material. We do not include the late-time WISE photometry from \cite{Sun22}. We include one spectrum form \cite{Clark_18bsz}, obtained from the TNS. 

\subsection{2018bym}
SN\,2018bym (=ZTF18aapgrxo =PS18aye =ATLAS18ohj =MLS180520:184313+451228) was discovered by ZTF and originally classified as a SLSN-I in \cite{Fremling18_bym} with a detailed study presented in \cite{Lunnan20_four}. A spectrum from \cite{Blanchard18_bym} shows a redshift of $z = 0.274$ based on host emission lines. We include the photometry from \cite{Lunnan20_four}, which the authors correct for extinction using $E(B-V) = 0.052$, plus LT photometry from \cite{Chen22a}, PS1 and ATLAS photometry. We include our own PSF photometry of FLWO and Las Cumbres GSP images after doing difference imaging to subtract the host flux. We exclude the data after MJD = 58372 from the \mosfit fit, since the SN shows a prominent late-time bump, although with a large scatter. We include one spectrum from \cite{Fremling18_bym}, obtained from the TNS.

\subsection{2018cxa}
We classified SN\,2018cxa (=ZTF18abfylqx =ATLAS18rsl =MLS180611:222835+113706) as a SLSN-I as part of FLEET \cite{Gomez21_TNS_18cxa} and determine a redshift of $z = 0.19$ based on the SN features. We include photometry from ATLAS and our own PSF photometry of FLWO and ZTF images after doing difference imaging to subtract the host flux. We include our own spectrum from LDSS3C.

\subsection{2018ffj}
SN\,2018ffj (=ZTF18abslpjy =ATLAS18tec =GRB180810.28 =MASTER OT J023059.78-172027.1) was discovered by ZTF and classified as a SLSN-I by \cite{Kostrzewa18}. We include photometry from \cite{Garcia18} and ATLAS. We include our own PSF photometry of Las Cumbres GSP and ZTF images, but do not subtract the host galaxy flux contribution, since this should be negligible. We include one spectrum from \cite{Kostrzewa18}, obtained from the TNS.

\subsection{2018ffs}
SN\,2018ffs (=ZTF18ablwafp =ATLAS18txu) was discovered by ZTF and classified as a SLSN-I by \cite{Gromadzki18}, \cite{Fremling18_18ffs}, and by ourselves as part of FLEET. We find a redshift of $z = 0.141$ from host emission lines. We include photometry from ATLAS and ZTF. We include our own PSF photometry of FLWO images after doing difference imaging to subtract the host flux. We include one spectrum from \cite{Fremling18_18ffs}, obtained from the TNS.

\subsection{2018gft}
SN\,2018gft (=ZTF18abshezu =ATLAS18uym) was discovered by ATLAS and classified as a SLSN-I by \cite{Fremling18_2018gft}. We include photometry from \cite{Chen22a} and ATLAS. Additionally, we include our own PSF photometry of Las Cumbres GSP, FLWO, and LDSS3C images. At the redshift of $z = 0.232$ determined by \cite{Chen22a} the peak magnitude of the SN is $M_r \sim -22.3$, comfortably in the SLSN-I regime. We include one spectrum from \cite{Fremling18_2018gft}, obtained from the TNS.

\subsection{2018hpq}
SN\,2018hpq (=ZTF18acapyww =Gaia18det =ATLAS18bcmq) was discovered by ZTF and classified as a SN Ic by \cite{Fremling18_hpq}, and then reclassified as a SLSN-I by \cite{Dahiwale20_beh} and presented in the \cite{Chen22a} sample. We include photometry from ATLAS, GSA, and ZTF, in addition to early-time upper limits from \cite{Chen22a}. We include our own PSF photometry from ZTF images after doing difference imaging to subtract the host flux. We include one spectrum from \cite{Chen22a}.

\subsection{2018hti}
SN\,2018hti (=Gaia19amt =PS19q =ATLAS18yff =MLS181110:034054+114637) was discovered by ATLAS and classified as a SLSN-I by \cite{Arcavi18_18hti}, a redshift of $z = 0.0612$ was determined by \cite{Lin20_hti} from host emission lines. We include photometry from ATLAS, \cite{Lin20_hti}, \cite{Fiore22}, and \cite{Chen22a}, in addition to our own PSF photometry of FLWO images after doing difference imaging to subtract the host flux. We include the UVOT data from \cite{Fiore22} as opposed to the one from \cite{Lin20_hti}, and include our own processed ATLAS data as opposed to the one from \cite{Fiore22}. We do not include the late-time WISE photometry from \cite{Sun22}. We include one spectrum from \cite{Fiore22}.

\subsection{2018ibb}
SN\,2018ibb (=ZTF18acenqto =Gaia19cvo =PS19crg =ATLAS18unu) was discovered by Gaia and classified as a SN Ia by \cite{Fremling18_2018ibb}, but later reclassified as a SLSN-I by \cite{Pursiainen18}. More recently, \cite{Schulze24} and \cite{Kozyreva24} presented an analysis of SN\,2018ibb as a pair-instability SN. The spectrum has a hint of H-alpha, which might be from the host galaxy. We include photometry from \textit{Gaia}, PS1, and ATLAS and our own PSF photometry of ZTF, LDSS3C, and Las Cumbres GSP images. We co-add the late time ZTF images after MJD = 58650 in bins of 5 days to increase the S/N of those epochs. We include one spectrum from \cite{Pursiainen18}, obtained from the TNS. 

\subsection{2018kyt}
SN\,2018kyt (=ZTF18acyxnyw =Gaia19afu) was discovered by ZTF and originally classified as a SLSN-I by \cite{Fremling19_2018kyt}. A detailed study of the SN was presented in \cite{Yan20}, who classify this as a SLSN-Ib/IIb. The SN shows a small amount of potential hydrogen at late times. We include photometry from ZTF, ATLAS, \cite{Chen22a}, and GSA. We do not subtract the host contribution from the GSA data, as this appears to have a negligible effect on the light curve. We include one spectrum from \cite{Fremling19_2018kyt}, obtained from the TNS.

\subsection{2018lfd}
SN\,2018lfd (=ZTF18acxgqxq =Gaia19afg =ATLAS18bcjv) was discovered by ATLAS and classified as a SLSN-I by \cite{Fremling19_2018lfd}. At the reported redshift of $z = 0.2686$ in \cite{Chen22a} the peak magnitude of the SN is $M_r \sim -22.3$, comfortably in the SLSN-I regime. We include photometry from ZTF, GSA, ATLAS, \cite{Chen22a}, and our own PSF photometry of FLWO images after doing difference imaging to subtract the host flux. We include one spectrum from \cite{Chen22a}.

\subsection{2018lfe}
SN\,2018lfe (=ZTF18acqyvag =PS18cpp) was discovered by ZTF and classified as a SLSN-I by \cite{Gomez19_lfe} with a detailed study presented in \cite{Yin22} and \cite{Chen22a}. We include photometry from \cite{Yin22} and ATLAS. We include one spectrum from \cite{Yin22}.

\subsection{2019aamp}
SN\,2019aamp (=ZTF19aantokv) was discovered by ZTF and classified as a SLSN-I by \cite{Chen22a, Yan_19aamp}. We include photometry from \cite{Chen22a} and ATLAS. We include one spectrum from \cite{Chen22a}.

\subsection{2019aamq}
SN\,2019aamq (=ZTF19aayclnm) was discovered by ZTF and classified as a SLSN-I by \cite{Chen22a}. We include photometry from \cite{Chen22a} and ATLAS. We include one spectrum from \cite{Chen22a}, which shows a peculiar reddening feature, maybe due to extinction.

\subsection{2019aamt}
SN\,2019aamt (=ZTF19abzoyeg) was discovered by ZTF and classified as a SLSN-I by \cite{Chen22a}. We include photometry from ATLAS and \cite{Chen22a}, plus our own PSF photometry of ZTF images after subtracting the host contribution. We include one spectrum from \cite{Chen22a}.

\subsection{2019aamv}
SN\,2019aamv (=ZTF20aagikvv) was discovered by ZTF and classified as a SLSN-I by \cite{Chen22a}. We include photometry from ATLAS and \cite{Chen22a}, plus our own PSF photometry of ZTF images. We include one spectrum from \cite{Chen22a}.

\subsection{2019bgu}
SN\,2019bgu (=ZTF19aaknqmp =PS19cma =ATLAS19dor) was discovered by ATLAS and classified as a SLSN-I by \cite{Fremling19_2019bgu}. At the reported redshift of $z = 0.148$ the peak magnitude of the SN is $M_r \sim -20.7$, within the range of SLSNe. We include photometry from \cite{Chen22a}, PS1, and ATLAS. We include one spectrum from \cite{Chen22a}.

\subsection{2019cca}
SN\,2019cca (=ZTF19aajwogx) was discovered by ZTF and classified as a SLSN-I by \cite{Perley19}. We include photometry from \cite{Chen22a} and \cite{Chen19_19cca}. We include one spectrum form \cite{Gal-yam_19cca}, obtained from the TNS.

\subsection{2019cdt}
SN\,2019cdt (=ZTF19aanesgt =Gaia19bll =ATLAS19ekt) was discovered by ZTF and classified as a SLSN-I by \cite{Fremling19_2019cdt}. We include photometry from GSA, \cite{Chen22a}, and ATLAS. The spectrum from \cite{Fremling19_2019cdt} shows significant absorption from Fe group elements. We include one spectrum from \cite{Fremling19_2019cdt}, obtained from the TNS.

\subsection{2019cwu}
SN\,2019cwu (=ZTF19aapaeye) was discovered by ZTF and classified as a SLSN-I by \cite{Perley19} and \cite{Yan19_slsne}. We include photometry from ATLAS, LT photometry from \cite{Chen22a}, and our own PSF photometry of ZTF images. The late time spectra is consistent with either a SLSN-I or a SN Ic, but given the peak absolute magnitude of $M_g = -22$, we adopt a SLSN-I classification. We include one spectrum from \cite{Smith19}, obtained from the TNS.

\subsection{2019dgr}
SN\,2019dgr (=ZTF19aamhast =PS19cwg =ATLAS19geq) was discovered by ZTF and classified as a SLSN-I by \cite{Chen22a}. We include photometry from \cite{Chen22a}, ZTF, PS1, and ATLAS, and our own PSF photometry of one late-time ZTF image. We include one spectrum from \cite{Chen22a}.

\subsection{2019dlr}
SN\,2019dlr (=ZTF19aaohuwc) was discovered by ZTF and classified as a SLSN-I by \cite{Yan19_slsne} and \cite{Perley19_fiy}. We include photometry from \cite{Chen22a}, ATLAS, and our own PSF photometry of ZTF images. The late time spectrum is consistent with a SLSN-I or SNe Ic, but given the peak absolute magnitude of $M_g = -21.7$, we adopt a SLSN-I classification. We include one spectrum from \cite{Perley19_fiy}, obtained from the TNS.

\subsection{2019enz}
SN\,2019enz (=Gaia19bty =PS19cys =ATLAS19ine) 2019enz was discovered by ATLAS and classified as a SLSN-I by \cite{Short19_enz}. The authors find a redshift of $z = 0.22$, but we find a redshift of $z = 0.255$ to be a better fit to the spectral features of the SN. We include photometry from \textit{Gaia}, ZTF, and ATLAS. We include one spectrum from \cite{Short19_enz}, obtained from the TNS.

\subsection{2019eot}
SN\,2019eot (=ZTF19aarphwc =ATLAS19kes) 2019eot was discovered by ZTF and classified as a SLSN-I by \cite{Fremling19_2019eot}. We include photometry from \cite{Chen22a}, ZTF and ATLAS. We include one spectrum from \cite{Fremling19_2019eot}, obtained from the TNS.

\subsection{2019gfm}
SN\,2019gfm (=ZTF19aavouyw =PS19ave =ATLAS19may) 2019gfm was discovered by ATLAS and classified as a SLSN-I by \cite{Perley19}. We adopt the redshift of the SDSS host galaxy of $z = 0.18167$ \citep{Stoughton02}. We include photometry from \cite{Chen22a}, PS1, ATLAS, and ZTF. We include one spectrum from \cite{Chen19}, obtained from the TNS.

\subsection{2019gqi}
SN\,2019gqi (=ZTF19aasdvfr =ATLAS19mas) 2019gqi was discovered by ZTF and classified as a SLSN-I by \cite{Yan19_slsne, Perley19_fiy}. We include photometry from ATLAS, \cite{Chen22a}, and two late time upper limits from our own PSF photometry of FLWO images after doing difference imaging to subtract the host flux. The spectra is consistent with a SLSN-I or a SN Ic, but given the peak absolute magnitude of $M_r = -21.9$, we adopt a SLSN-I classification. We include one spectrum from \cite{Perley19_fiy}, obtained from the TNS.

\subsection{2019hno}
SN\,2019hno (=ZTF19aawsqsc =ATLAS19ndu) 2019hno was discovered by ZTF and classified as a SLSN-I by \cite{Yan19_slsne} and \cite{Perley19_fiy}. We include photometry from \cite{Chen22a}, ATLAS and ZTF. We include one spectrum from \cite{Perley19_fiy}, obtained from the TNS.

\subsection{2019itq}
We classified SN\,2019itq (=ZTF19abctjtj =PS19bsr) as a SLSN-I as part of FLEET \cite{Gomez21_TNS}, originally discovered by ZTF. We determine a redshift of $z = 0.481$ based on the host galaxy emission lines. We include photometry from ATLAS and our own PSF photometry of FLWO and ZTF images, without subtracting the negligible host galaxy contribution. We co-add ZTF images in bins of 5 days to increase the S/N during the decline of the SN. For ZTF images after MJD = $59000$ we co-add images in bins of $30$ days, but do not detect the SN and only report the upper limits.

\subsection{2019kcy}
SN\,2019kcy (=ZTF19abaeyqw =PS19dmu =ATLAS19oho) 2019kcy was discovered by ZTF and classified as a SLSN-I by \cite{Yan19_slsne} and \cite{Perley19_fiy}. We include photometry from \cite{Chen22a}, PS1, ATLAS, and ZTF. We include one spectrum from \cite{Perley19_fiy}, obtained from the TNS.

\subsection{2019kwq}
SN\,2019kwq (=ZTF19aalbrph) 2019kwq was discovered by ZTF and classified as a SLSN-I by \cite{Yan19_slsne} and \cite{Perley19_fiy}. We include photometry from \cite{Chen22a}, ATLAS and our own PSF photometry of ZTF images. We co-add individual ZTF images taken on the same calendar day to increase the S/N, for images taken after MJD = 58690, we co-add images within 5 days of each other. The late time spectrum of SN\,2019kwq is consistent with a SLSN-I or SNe Ic, but given the peak absolute magnitude of $M_g = -23.1$, we adopt a SLSN-I classification. We include one spectrum from \cite{Perley19_fiy}, obtained from the TNS.

\subsection{2019kws}
SN\,2019kws (=ZTF19aamhhiz =ATLAS19gkz) 2019kwq was discovered by ZTF and classified as a SLSN-I by \cite{Yan19_slsne, Perley19_fiy} with a detailed study presented in \cite{Yan20}. The SN is one of the few with a helium rich spectra. We include photometry from \cite{Chen22a}. We include one spectrum from \cite{Chen22a}.

\subsection{2019kwt}
SN\,2019kwt (=ZTF19aaqrime) 2019kwq was discovered by ZTF and classified as a SLSN-I by \cite{Yan19_slsne} and \cite{Perley19_fiy}. We include photometry from \cite{Chen22a}, ATLAS and our own PSF photometry of FLWO and ZTF images. We co-add the ZTF photometry before MJD = 58620 in bins of 1 day to increase the S/N of the detections during the rise. We include one spectrum from \cite{Perley19_fiy}, obtained from the TNS.

\subsection{2019kwu}
SN\,2019kwu (=ZTF19aaruixj) 2019kwq was discovered by ZTF and classified as a SLSN-I by \cite{Yan19_slsne} and \cite{Perley19_fiy}. We include photometry from \cite{Chen22a}, ATLAS plus our own PSF photometry of ZTF images. The late time spectrum is consistent with a SLSN-I or SN Ic, but given the peak absolute magnitude of $M_g = -23.0$, we adopt a SLSN-I classification. We include one spectrum from \cite{Perley19_fiy}, obtained from the TNS.

\subsection{2019lsq}
SN\,2019lsq (=ZTF19abfvnns =Gaia19dop =ATLAS19prf) 2019kwq was discovered by ZTF and classified as a SLSN-I by \cite{Fremling19_2019lsq}. We include photometry from \cite{Chen22a}, ATLAS and our own PSF photometry of Las Cumbres GSP images after doing difference imaging to subtract the host flux. We include one spectrum from \cite{Chen22a}.

\subsection{2019neq}
SN\,2019neq (=ZTF19abpbopt =ATLAS19sph =PS19eov) 2019kwq was discovered by ZTF and classified as a SLSN-I by \cite{Perley19_19neq} and \cite{Fremling19_19neq}, with a detailed study presented in \cite{Fiore24}. We adopt the redshift of $z = 0.1059$ determined by \cite{Konyves20}. We include photometry from \cite{Chen22a}, ATLAS and our own PSF photometry of Las Cumbres GSP images after doing difference imaging to subtract the host flux. We include one spectrum from \cite{Konyves20}.

\subsection{2019nhs}
SN\,2019nhs (=ZTF19abnacvf =ATLAS19typ =PS19eok) 2019kwq was discovered by ZTF and classified as a SLSN-I by \cite{Perley19_2019nhs} and \cite{Fremling19_19neq}. We include photometry from \cite{Chen22a}, PS1 and ATLAS, in addition to a healthy amount of our own PSF photometry of Las Cumbres GSP images. We include one spectrum from \cite{Perley19_2019nhs}, obtained from the TNS.

\subsection{2019otl}
We classified SN\,2019otl (=ZTF19abkfshj =ATLAS19tup =PS19fsp) as a SLSN-I as part of FLEET \cite{Gomez21_TNS}, originally discovered by ZTF. We determine a redshift of $z = 0.514$ based on the SN features. We include photometry from \cite{Chen22a}, ATLAS, and PS1. We include our own PSF photometry of IMACS and LDSS3C images, without subtracting the negligible host galaxy contribution. We have no early time spectra of the source, but late time spectra is consistent with either a SNe Ic or SLSN-I. We include our own LDSS3C spectrum.

\subsection{2019pud}
SN\,2019pud (=ZTF19abxgmzr =Gaia19eri =ATLAS19bfto) 2019pud was discovered by ATLAS and classified as a SLSN-I by \cite{Fremling19_pud}. We include photometry from ZTF, \textit{Gaia}, and ATLAS. The spectra shows some absorption from Fe group elements and is consistent with either a SN Ic or a SLSN-I. But given the peak magnitude of $M_r = -20.7$, we adopt a SLSN-I classification. We include one spectrum from \cite{Fremling19_pud}.

\subsection{2019sgg}
SN\,2019sgg (=ZTF19abuyuwa) 2019sgg was discovered by ZTF and classified as a SLSN-I by \cite{Yan19_2019sgg}. At the reported redshift of $z = 0.5726$, the peak magnitude of the SN is $M_r = -22.7$, comfortably in the SLSN-I regime. We include photometry from ATLAS and \cite{Chen22a}, plus our own PSF photometry of FLWO and LDSS3C images. We include one spectrum from \cite{Chen22a}.

\subsection{2019sgh}
SN\,2019sgh (=ZTF19abzqmau) 2019sgg was discovered by ZTF and classified as a SLSN-I as part of FLEET \cite{Gomez21_TNS}. We determine a redshift of $z = 0.344$ based on host galaxy emission lines. We include photometry from ATLAS, PS1 and \cite{Chen22a}. We include our own PSF photometry of FLWO and MMTCam images after doing difference imaging to subtract the host flux. We include our own Binospec spectrum \citep{Gomez21_TNS}.

\subsection{2019szu}
SN\,2019szu (=ZTF19acfwynw =Gaia19fcb =ATLAS19ynd) 2019sgg was discovered by ZTF and classified as a SLSN-I by \cite{Nicholl19_szu} and \cite{Dahiwale19}. We include photometry from ATLAS, GSA, and \cite{Chen22a}. We do not subtract the host flux from the \textit{Gaia} photometry since this is from a galaxy of magnitude $m_r = 21.9$ and its contribution should not negligible. We include one spectrum from \cite{Nicholl19_szu}, obtained from the TNS.

\subsection{2019ujb}
We classified SN\,2019ujb (=ZTF19ackjrru =Gaia19fne =ATLAS19bach) as a SLSN-I as part of FLEET \cite{Gomez21_TNS}, originally discovered by ZTF. We include photometry from GSA, \cite{Chen22a}, and ATLAS. We include our own PSF photometry of FLWO images, without subtracting the negligible host galaxy contribution. We include our own Binospec spectrum \citep{Gomez21_TNS}.

\subsection{2019xaq}
We classified SN\,2019xaq (=ZTF19acyjzbe =Gaia19fue =ATLAS19bfnq) as a SLSN-I as part of FLEET \cite{Gomez21_TNS}, originally discovered by \textit{Gaia}. We determine a redshift of $z = 0.2$ based on the presence of host emission lines. We include photometry from ATLAS, \textit{Gaia}, and ZTF. We include our own PSF photometry of FLWO and MMTCam images, without subtracting the negligible host galaxy contribution. We include our own Binospec spectrum \citep{Gomez21_TNS}.

\subsection{2019zbv}
We classified SN\,2019zbv (=ZTF19adaivcf =ATLAS19bfmr) as a SLSN-I as part of FLEET \cite{Gomez21_TNS}, originally discovered by ZTF. We adopt the redshift based on host galaxy emission lines of $z = 0.3785$ from \cite{Chen22a}. We include photometry from ATLAS and \cite{Chen22a}. We include our own PSF photometry of FLWO images, without subtracting the negligible host galaxy contribution. We include our own Binospec spectrum \citep{Gomez21_TNS}.

\subsection{2019zeu}
We classified SN\,2019zeu (=ZTF19adajybt =ATLAS19bfmg =PS20dr) as a SLSN-I as part of FLEET \cite{Gomez21_TNS}, originally discovered by ZTF. We determine a redshift of $z = 0.39$ based on the SN features. We include photometry from ATLAS and PS1. We include our own PSF photometry of FLWO and ZTF images, without subtracting the negligible host galaxy contribution. We include our own Blue Channel spectrum \citep{Gomez21_TNS}.

\subsection{2020abjc}
We classified SN\,2020abjc (=ZTF20acpyldh =PS20mny) as a SLSN-I as part of FLEET \citep{Blanchard20_2020abjc}, originally discovered by ZTF. We determine a redshift of $z = 0.219$ based on host emission lines. We include photometry from ATLAS, PS1, and ZTF photometry, in addition to our own PSF photometry of FLWO and Las Cumbres GSP images. We exclude photometry between MJD = 59723 and 59750 from the \mosfit fit due to an apparent re-brightening of the light curve. We include our own Binospec spectrum \citep{Blanchard20_2020abjc}.

\subsection{2020adkm}
We classified SN\,2020adkm as a SLSN-I \citep{Blanchard21_TNS_20adkm}, originally discovered by ZTF. We include photometry from ATLAS and ZTF, in addition to our PSF photometry of FLWO, Las Cumbres GSP, and Binospec images. We include our own LDSS3C spectrum, which shows a possible P-Cygni profile at the location of H$\alpha$ \cite{Blanchard21_TNS_20adkm}.

\subsection{2020afag}
SN\,2020afag (=ZTF20abisijg) was discovered by ZTF and classified as a SLSN-I by \cite{Chen22a}. We include photometry from \cite{Chen22a}. We include one spectrum from \cite{Chen22a}.

\subsection{2020afah}
SN\,2020afah (=ZTF20aawkgxa) was discovered by ZTF and classified as a SLSN-I by \cite{Chen22a}. We include photometry from \cite{Chen22a} and ATLAS. We include one spectrum from \cite{Chen22a}.

\subsection{2020ank}
SN\,2020ank (=ZTF20aahbfmf =ATLAS20dzr =PS20eyd) was discovered by ZTF and classified as a SLSN-I by \cite{Poidevin20} and \cite{Dahiwale20_ank}, and later presented in \cite{Kumar21}. Polarimetry observations from \cite{Lee20} find very low polarization, consistent with a very spherical explosion. We include photometry from \cite{Chen22a}, ATLAS, PS1, and \cite{Kumar21}. We include one spectrum from \cite{Poidevin_20b}.

\subsection{2020aup}
SN\,2020aup (=ZTF20aahrxgw =ATLAS20cvq) was discovered by ZTF and classified as a SLSN-I by \cite{Chen22a}. We include photometry from \cite{Chen22a} and ATLAS. We include one spectrum from \cite{Chen22a}.

\subsection{2020auv}
SN\,2020auv (=ZTF20aaifybu =ATLAS20eaj) was discovered by ZTF and classified as a SLSN-I by \cite{Yan20_auv}. We include photometry from \cite{Chen22a} and ATLAS, as well as our own PSF photometry of FLWO images. We include one spectrum from \cite{Chen22a}.

\subsection{2020dlb}
SN\,2020dlb (=ZTF20aaoqwpo =PS20air =ATLAS20ism) was discovered by ZTF and classified as a SLSN-I by \cite{Chen22a}. We include photometry from \cite{Chen22a}, PS1, and ATLAS. We include one spectrum from \cite{Chen22a}.

\subsection{2020exj}
SN\,2020exj (=ZTF20aattyuz =PS20bxx =ATLAS20irs) was discovered by ATLAS and classified as a SLSN-I by \cite{Dahiwale20}. We include photometry from \cite{Chen22a}, ATLAS, and PS1, in addition to our own PSF photometry of FLWO images. We have no early time spectra of the source, but late time spectra are consistent with either an SN Ic or SLSN-I. Given the peak absolute magnitude of $M_r = -20.49$ mag, we adopt a SLSN-I classification. We include one spectrum from \cite{Dahiwale20}, obtained from the TNS.

\subsection{2020fvm}
SN\,2020fvm (=ZTF20aadzbcf =PS20eum) was discovered by ZTF and classified as a SLSN-I by \cite{Chen22a}. We include photometry from \cite{Chen22a}, PS1, and ATLAS. We include one spectrum from \cite{Chen22a}.

\subsection{2020htd}
SN\,2020htd (=ZTF20aauoudz =PS20ccj =ATLAS20mzx) was discovered by ZTF and classified as a SLSN-I by \cite{Chen22a}. We include photometry from \cite{Chen22a}, PS1, and ATLAS. We exclude photometry after MJD = 59070 from the \mosfit fit due to a prominent secondary peak in the light curve. We include one spectrum from \cite{Chen22a}.

\subsection{2020iyj}
SN\,2020iyj (=ZTF20aavfbqz =ATLAS20mbx) was discovered by ZTF and classified as a SLSN-I by \cite{Chen22a}. We include photometry from \cite{Chen22a} and ATLAS. We include one spectrum from \cite{Chen22a}.

\subsection{2020jii}
We classified SN\,2020jii (=ZTF20aawfxlt =PS20cvb =ATLAS20mmz) as a SLSN-I as part of FLEET \cite{Gomez20_TNS_20jii}, originally discovered by ATLAS. We include photometry from \cite{Chen22a}, PS1, and ATLAS, plus our own PSF photometry of FLWO images. We determine a redshift of $z = 0.396$ based on the SN features. We include our own Binospec spectrum \cite{Gomez20_TNS_20jii}.

\subsection{2020qef}
SN\,2020qef (=ZTF20ablkuio =PS20hhb =ATLAS20ulh) was discovered by ATLAS and classified as a SLSN-I by \cite{Terreran20} with a TNS classification report \citep{Terreran20_TNS}. The authors find a spectrum similar to the SLSNe-Ib presented in \cite{Yan20}. We include photometry from PS1, ATLAS, and \cite{Chen22a}. Instead of the ZTF photometry from \cite{Chen22a}, we include our own reduction of ZTF images with sets of images co-added in bins of 1-day. The light curve shows a peculiar late-time flattening after MJD = 59121. Therefore, we exclude detections after this date from the \mosfit fit. We include one spectrum from \cite{Terreran20_TNS}, obtained from the TNS.

\subsection{2020qlb}
SN\,2020qlb (=ZTF20abobpcb =Gaia20ekn =ATLAS20vmc) was discovered by ZTF and classified as a SLSN-I by \cite{Perez20}. We include photometry from GSA, ATLAS, ZTF, and \cite{West22}. We determine a redshift of $z = 0.1585$ from host emission lines. The light curve shows multiple light curve bumps studied in \cite{West22}, who argue these are the result of interaction of the ejecta with CSM of varying density. We include our own Binospec spectrum.

\subsection{2020rmv}
SN\,2020rmv (=ZTF20abpuwxl =PS20nxm =ATLAS20xqi) was discovered by ZTF and classified as a SLSN-I by \cite{Terreran20_rmv,Terreran20_rmv_TNS}. We include photometry from \cite{Chen22a} and ATLAS, and upper limits from our own FLWO PSF photometry. We include one spectrum from \cite{Terreran20_rmv_TNS}, obtained from the TNS.

\subsection{2020tcw}
SN\,2020tcw (=ZTF20abzumlr =Gaia20ewr =PS20jci =ATLAS20zst) was discovered by ZTF and classified as a SLSN-I by \cite{Perley20_tcw}. We determine a redshift of $z = 0.064$ from host emission lines of our own spectrum with higher resolution than the one presented in \cite{Perley20_tcw}. We include photometry from GSA, ATLAS, and the ASAS-SN Sky Patrol V2.0 \citep{Hart23}, in addition to our own PSF photometry of ZTF and FLWO images after subtracting the contribution of the host. We subtract a nominal magnitude of $g = 17.6$ from the ASAS-SN photometry to account for the zero-point contribution of the host.

\subsection{2020uew}
SN\,2020uew (=Gaia20eme =PS20ldz =ATLAS20bbum) was discovered by ATLAS and classified as a SLSN-I by \cite{Jaeger20}. We include photometry from GSA, PS1, and ATLAS, in addition to our PSF photometry of Las Cumbres GSP, IMACS and DECam images after subtracting the host contribution. We include one spectrum form \cite{Ihanec_20uew}, obtained from the TNS.

\subsection{2020vpg}
SN\,2020vpg (=ZTF20acjmsdu =PS20ikn) was discovered by ZTF and classified as a SLSN-I by \cite{Terreran20_vpg}. We include photometry from ATLAS, and PS1, and our own PSF photometry of ZTF images after subtracting the contribution of the host. We include one spectrum from \cite{Terreran20_vpg}, obtained from the TNS.

\subsection{2020wnt}
SN\,2020wnt (=ZTF20acjeflr =ATLAS20beko) was discovered by ZTF and identified by \cite{Tinyanont20} as a supernova with a spectrum similar to a super-Chandrasekhar SN Ia, but also similar to a SN Ic or SLSN-I. The authors disfavor the SLSN-I interpretation due to the low luminosity at the time of discovery. Nevertheless, the SN evolved to a peak absolute magnitude of $M_r \sim -20.5$, within the range for SLSNe. The SN was studied in more detail by \cite{Gutierrez22}, who argue SN\,2020wnt is consistent with being powered by radioactive decay and point out the peculiar spectra of SN\,2020wnt lacks the distinctive \ion{O}{2} lines seen in SLSNe. \cite{Tinyanont22} explain the pre-maximum peak as being due to interaction with the CSM and explain the rest of the light curve as magnetar-powered. We include photometry from ATLAS and photometry from seven different telescopes from \cite{Gutierrez22}, in addition to our own PSF photometry of Binospec and FLWO images.

\subsection{2020xga}
SN\,2020xga (=ZTF20acilzkh =PS20jxm =ATLAS20beys) was discovered by ZTF and classified as a SLSN-I by \cite{Gromadzki20}. We include photometry from PS1, ATLAS, and ZTF. We include our own PSF photometry from IMACS, Binospec, and DECam. We include one spectrum from \cite{Gromadzki20}, obtained from the TNS.

\subsection{2020xgd}
SN\,2020xgd (=ZTF20aceqspy =PS20jxq =ATLAS20beed) was classified as a SLSN-I by \cite{Weil20} and \cite{Gomez20_TNS_20xgd}. We include photometry from PS1, ATLAS and LT photometry from \cite{Chen22a}, in addition to own PSF photometry of ZTF images. We also co-add 18 ZTF $r$-band images between MJD = 59247 and 59253 to recover a late-time detection. We include one spectrum from \cite{Weil20}, obtained from the TNS.

\subsection{2020xkv}
SN\,2020xkv (=ZTF20abzaacf =ATLAS20bdpf) was discovered by ZTF and classified as a SLSN-I by \cite{Chen22a}. We include photometry from ZTF, ATLAS, and \cite{Chen22a}. We include one spectrum from \cite{Chen22a}.

\subsection{2020znr}
SN\,2020znr (=ZTF20acphdcg =Gaia20fkx =PS20lkc =ATLAS20bgae) was discovered by ZTF and classified as a SLSN-I by \cite{Ihanec20_znr}. We include photometry from PS1, ATLAS, ZTF, and GSA. We include our own PSF photometry from FLWO, LDSS3C, and IMACS images. The light curve has an early time bump, so we exclude detections before MJD = 59165 from the \mosfit fit. \cite{Poidevin22} provide optical imaging polarimetry of SN\,2020znr and find null-polarization detection. A more detailed study of SN\,2020znr will be presented in Chen, T. W., et al., in prep. We include our own LDSS3C spectrum.

\subsection{2020zzb}
SN\,2020zzb (=ZTF20acjaayt =PS21adu =ATLAS20bfng) was discovered by ZTF and classified as a SLSN-I by \cite{Yan20_2020abjx}. We obtained a spectrum of the source with LDSS and confirm the classification and redshift from host emission lines. We include photometry from ATLAS and our own PSF photometry of ZTF images after doing image subtraction. We combine late-time ZTF images after MJD = 59250 taken on the same day. Additionally, we combined 16 ZTF $r$-band images between MJD = 59260 and MJD = 59265 to recover a late-time detection. We include our own LDSS3C spectrum.

\subsection{2021bnw}
SN\,2021bnw (=ZTF21aagpymw =Gaia21caf =PS21ajy =ATLAS21dpf) was discovered by ZTF and classified as a SLSN-I by \cite{Magee21} and \cite{Terwel21}. \cite{Poidevin23} recently presented polarimetry observations of SN\,2021bnw and found no evidence for polarization. We include photometry from ZTF, ATLAS, PS1, and GSA. Additionally, we include our own PSF photometry of Las Cumbres GSP, FLWO, IMACS, and LDSS3C images. A more detailed study of SN\,2021bnw will be presented in Fiore, A., et al., in prep. We include one spectrum from \cite{Magee21}.

\subsection{2021een}
SN\,2021een (=ZTF21aakjkec =Gaia21btp =ATLAS21iyb) was discovered by ZTF and classified as a SLSN-I by \cite{Dahiwale21}. We include photometry from ATLAS, GSA, ZTF, as well as our own PSF photometry of Binospec and LDSS3C images. We exclude images from the CPCS due to the lack of information about their source and poor match to the rest of the photometry. We include one spectrum from \cite{Dahiwale21}.

\subsection{2021ejo}
We classified SN\,2021ejo (=ZTF21aaherjf) a SLSN-I \citep{Gomez21_21ejo}, originally discovered by ATLAS. We include photometry from ZTF and our own PSF photometry of FLWO images. We include photometry from ZTF and ATLAS. We include our own Binospec spectrum.

\subsection{2021ek}
SN\,2021ek (=ZTF21aaarmti =PS21fo =ATLAS21ajr) was discovered by ZTF and classified as a SLSN-I by \cite{Gillanders21} and \cite{Srivastav21}. We include photometry from ZTF, ATLAS, and PS1. Additionally, we include our own PSF photometry of Las Cumbres GSP, Binospec, and DECam images. We include one spectrum from \cite{Gillanders21}.

\subsection{2021fpl}
SN\,2021fpl (=ZTF21aaxwpyv =Gaia21ckf =PS21evf =ATLAS21iao) was discovered by ATLAS and classified by \cite{Deckers21b, Deckers21} as a SLSN-I at a redshift between 0.11 and 0.12. \cite{Poidevin23} recently presented polarimetry observations of SN\,2021fpl and found evidence for polarization. We adopt a redshift of 0.121 based on the SN features. We include photometry from Gaia, ATLAS, PS1, and ZTF. Additionally, we include our own PSF photometry from FLWO, LDSS3C, and Las Cumbres GSP. We include our own spectrum from FAST.

\subsection{2021gtr}
We classified SN\,2021gtr (=ZTF21aagdezv =ATLAS21jbn) as a SLSN-I \citep{Gomez21_21gtr}, originally discovered by ZTF. We include images from ATLAS, as well as our own PSF photometry from FLWO and ZTF images. We include our own Binospec spectrum \citep{Gomez21_21gtr}.

\subsection{2021hpc}
We classified SN\,2021hpc (=ZTF21aaqawpd =PS21cui =ATLAS21lxa) as a SLSN-I \citep{Gomez21_21hpc}, originally discovered by ATLAS. We include photometry from ZTF, ATLAS, as well as our own PSF photometry from FLWO images. We include our own Binospec spectrum \citep{Gomez21_21hpc}.

\subsection{2021hpx}
SN\,2021hpx (=ZTF21aappdnv =Gaia21bwa =ATLAS21jis) was discovered by ATLAS and classified as a SLSN-I by \cite{Gonzalez21}. We include photometry from Gaia, ZTF, and ATLAS, as well as our own PSF photometry from Las Cumbres GSP, FLWO, and LDSS3C images. We include our own spectrum from Las Cumbres Observatory taken as part of the GSP. We include one spectrum from \cite{Gonzalez21}.

\subsection{2021kty}
SN\,2021kty (=PS21eya =ZTF21aavdqgf =ATLAS21nrn) was discovered by ZTF and classified as a SLSN-I by \cite{Yao21} after retracting a classification as a TDE \citep{Yao21b}. We include photometry from ZTF, ATLAS, and PS1. Additionally, we include our own PSF photometry of FLWO and LDSS3C images. We include one spectrum from \cite{Yao21}.

\subsection{2021mkr}
SN\,2021mkr (=ZTF21abbqeea =Gaia21dbn =PS21fax =ATLAS21rem) was discovered by ZTF and classified as a SLSN-I by \cite{Chu21} and \cite{Poidevin21}. We exclude data between MJD = 59420 and MJD = 59500 from the \mosfit fit due to a very prominent second peak in the light curve. We include photometry from ATLAS, ZTF and our own PSF photometry from FLWO images. We include one spectrum from \cite{Chu21}.

\subsection{2021nxq}
SN\,2021nxq (=ZTF21abcpsjy =PS21etq =ATLAS21rdl) was discovered by PS2 and classified as a SLSN-I by \cite{Weil21b, Weil21}. We include photometry from ATLAS, ZTF and our own PSF photometry from FLWO images. We include one spectrum from \cite{Weil21b}.

\subsection{2021txk}
We classified SN\,2021txk (=ZTF21abjgzhn) as a SLSN-I and determine a redshift of $z = 0.46$ based on the SN features \cite{Gomez21_21txk}. The SN was originally discovered by ZTF. We include photometry from ATLAS, ZTF and our own PSF photometry from FLWO. We include our own Binospec spectrum.

\subsection{2021vuw}
We classified SN\,2021vuw (=ZTF21abrqria =Gaia21ead =ATLAS21bimf) as a SLSN-I and determine a redshift of $z = 0.2$ based on the SN features \cite{Gomez21_21vuw}. The SN was originally discovered by ZTF. We include photometry from ATLAS, ZTF and our own PSF photometry from FLWO images. We include our own Binospec spectrum.

\subsection{2021xfu}
We classified SN\,2021xfu (=ZTF21absyjff =ATLAS21bhce) as a SLSN-I and determine a redshift of $z = 0.32$ based on the SN features \cite{Gomez21_21xfu}. The SN was originally discovered by ZTF. We include photometry from ATLAS, ZTF and our own PSF photometry from FLWO images. We co-add images in bins of 5 days after MJD = 59869.0 to increase their S/N. We include our own Binospec spectrum.

\subsection{2021ynn}
We classified SN\,2021ynn (=ZTF21acaqcrw =PS21jzd =ATLAS21bjwc) as a SLSN-I. We determine a redshift of $z \approx 0.22$ by cross-matching a spectrum of SN\,2021ynn taken at a phase of 42 days with one of the SLSN-I 2016eay at 46 days. The SN was originally discovered by ZTF. At this redshift the peak absolute magnitude of the transient is $M_r \sim -20.8$ mag. Given this and its strong similarity to SN\,2016eay, we adopt a SLSN-I classification. We include photometry from ZTF, PS1, ATLAS, and our own PSF photometry of FLWO and DECam images. We include our own LDSS3C spectrum.

\subsection{2021yrp}
We classified SN\,2021yrp (=ZTF21abwzpme =ATLAS21bioa) as a SLSN-I and determine a redshift of $z = 0.3$ based on the SN features \cite{Gomez21_21yrp}. The SN was originally discovered by ZTF. We include photometry from ATLAS, ZTF and our own PSF photometry from FLWO images. We include our own Binospec spectrum.

\subsection{2021zcl}
SN\,2021zcl (=ZTF21accwovq =PS21kev =ATLAS21bjql) was discovered by ZTF and classified as a SLSN-I by \cite{Gromadzki21}. We include photometry from ZTF, PS1, and ATLAS. We include one spectrum from \cite{Gromadzki21}.

\subsection{2022ful}
SN\,2022ful (=ZTF22aadeuwu =Gaia22bon =PS22ger) was discovered by ATLAS and classified as a SN Ia by \cite{Chu_22ful} then reclassified as a SLSN-I by \cite{Sollerman_22ful}. We include photometry from ZTF, Gaia, ATLAS, and PS1. We exclude detections after MJD = from the \mosfit fit due to a strong late-time flattening. We include one spectrum from \cite{Sollerman_22ful}, obtained from the TNS.

\subsection{2022le}
We classified SN\,2022le (=ZTF21acrbbwi =PS22cd) as a SLSN-I and determine a redshift of $z = 0.2491$ based on the host emission lines \cite{Gomez22_22le}. The SN was originally discovered by ZTF. We include photometry from ATLAS, ZTF and our own PSF photometry from FLWO and ZTF images. We include our own Binospec spectrum.

\subsection{2022ljr}
SN\,2022ljr (=ZTF22aalzjdc =ATLAS22pxh =PS22ezd) was discovered by PS2 and classified as a SLSN-I by \cite{Davis_22ljr}. We include photometry from ZTF, ATLAS, PS1, and upper limits from our own PSF photometry of FLWO images. We include one spectrum from \cite{Davis_22ljr}, obtained from the TNS.

\subsection{2022lxd}
SN\,2022lxd (=ZTF22aaljlzq =ATLAS22rdp) was discovered by ZTF and classified as a SLSN-I by \cite{Angus_22lxd}. We include photometry from ZTF and ATLAS, and upper limits from our own PSF photometry of FLWO images. We include one spectrum from \cite{Angus_22lxd}, obtained from the TNS.

\subsection{2022npq}
SN\,2022npq (=PS22fqn =ATLAS22vfs =ZTF22aarqrxf) was discovered by PS2 and classified as a SLSN-I by \cite{Ayala_22npq,Ayala_22npq_b}. We include photometry from ZTF, ATLAS, PS1, and our own PSF photometry of FLWO images. We include one spectrum from \cite{Ayala_22npq_b}, obtained from the TNS.

\subsection{2022pjq}
SN\,2022pjq (=PS22ggy =ATLAS22xpx =ZTF22aausnwr) was discovered by ZTF and classified as a SLSN-I by \cite{Fulton22_pjq}. We include photometry from PS1, and ATLAS, and our own PSF photometry of ZTF images. We include our own LDSS3C spectrum.

\subsection{2019aamu}
SN\,2019aamu (=ZTF19acvxquk) was discovered by ZTF and classified as a SLSN-I by \cite{Chen22a}. We include photometry from \cite{Chen22a}, ATLAS, and our own PSF photometry of ZTF images. We include one spectrum from \cite{Chen22a}.

\subsection{2019vvc}
SN\,2019vvc (=ZTF19acucxij =Gaia19fnw =ATLAS19bcfc) was discovered by ZTF and classified as a SLSN-I by \cite{Chen22a}. We note the authors quote ZTF19acucxij as the name of the ZTF source, but this name is not available in any of the ZTF alert brokers. We include photometry from \cite{Chen22a}, and ATLAS. We include one spectrum from \cite{Chen22a}.

\subsection{2022abdu}
SN\,2022abdu (=ATLAS22bmme) was discovered by ATLAS and classified as a SLSN-I by \cite{Gromadzki_22abdu}. We include photometry from ATLAS. We include one spectrum from \cite{Gromadzki_22abdu}, obtained from the TNS.

\subsection{2022ued}
SN\,2022ued (=ZTF22abexkqi =ATLAS22bete =PS22knc) was discovered by ATLAS and classified as a SN Ib by \cite{Perley22}. Nevertheless, we also find good spectral matches to SNe Ic and SLSNe-I, and at the reported redshift of $z = 0.1087$, the peak absolute magnitude is $M_r \sim -20.4$, within the SLSNe-I regime. We include photometry from ZTF and ATLAS. We include one spectrum from \cite{Perley22}, obtained from the TNS.

\subsection{2018lzv}
SN\,2018lzv (=ZTF18aazgrfl) was discovered by ZTF and classified as a SLSN-I by \cite{Perley22_lzv}. We include photometry from \cite{Chen22a}. We include one spectrum from \cite{Chen22a}.

\subsection{2018gbw}
SN\,2018gbw (=ZTF18acslpji) was discovered by ZTF and classified as a SLSN-I by \cite{Chen22a}. We include photometry from \cite{Chen22a}. We include one spectrum from \cite{Chen22a}.

\subsection{DES14S2qri}
DES14S2qri was discovered by DES and classified as a ``gold" SLSN-I by \cite{Angus19}. We include photometry form \cite{Angus19}. We exclude the first three data points from the \mosfit fit before MJD = 56982 due to an early time flattening. We include one spectrum from \cite{Angus19}.

\subsection{DES14X2byo}
DES14X2byo was discovered by DES and classified as a ``gold" SLSN-I by \cite{Angus19}. We include photometry form \cite{Angus19}. We include one spectrum from \cite{Angus19}.

\subsection{DES14X3taz}
DES14X3taz was discovered by DES and classified as a ``gold" SLSN-I by \cite{Angus19}. We include photometry form \cite{Angus19}. We exclude the first data points before MJD = 57035 from the \mosfit fit since these are from a pre-cooling shock \citep{Smith16}. We include one spectrum from \cite{Angus19}.

\subsection{DES15E2mlf}
DES15E2mlf was discovered by DES and classified as a ``gold" SLSN-I by \cite{Angus19}. We include photometry form \cite{Angus19}. We exclude the first $g$-band data point before MJD = 57328 since this is from a precursor of the SN \cite{Pan17}. We include one spectrum from \cite{Angus19}.

\subsection{DES15X1noe}
DES15X1noe was discovered by DES and classified as a ``gold" SLSN-I by \cite{Angus19}. We include photometry form \cite{Angus19}. We exclude the first $z$-band data point before MJD = 57350 since this is too bright to be consistent with a smooth rise. We include one spectrum from \cite{Angus19}.

\subsection{DES15X3hm}
DES15X3hm was discovered by DES and classified as a ``gold" SLSN-I by \cite{Angus19}. We include photometry form \cite{Angus19}. We include one spectrum from \cite{Angus19}.

\subsection{DES16C2aix}
DES16C2aix was discovered by DES and classified as a ``gold" SLSN-I by \cite{Angus19}. We include photometry form \cite{Angus19}. We include one spectrum from \cite{Angus19}.

\subsection{DES16C3dmp}
DES16C3dmp was discovered by DES and classified as a ``gold" SLSN-I by \cite{Angus19}. We include photometry form \cite{Angus19}. \cite{Angus19} report a small bump during the rise in the bluer bands, which is why we exclude the data before MJD = 57705 from the \mosfit fit. We include one spectrum from \cite{Angus19}.

\subsection{DES17X1amf}
DES17X1amf was discovered by DES and classified as a ``gold" SLSN-I by \cite{Angus19}. We include photometry form \cite{Angus19}. We exclude the data before MJD = 58018 since these are from an early bump. We include one spectrum from \cite{Angus19}.

\subsection{DES17X1blv}
DES17X1blv was discovered by DES and classified as a ``gold" SLSN-I by \cite{Angus19}. We include photometry form \cite{Angus19}. We exclude the first $z$-band point before MJD = 58030 since this appears to be from a pre-cooling peak. We include one spectrum from \cite{Angus19}.

\subsection{iPTF13ajg}
iPTF13ajg was discovered by PTF and classified as a SLSN-I by \cite{Vreeswijk14}. We include photometry from \cite{Vreeswijk14}. The light curve shows what could be a flattening or bump in $Rs$-band after MJD = 56600. We include one spectrum from \cite{Vreeswijk14}.

\subsection{iPTF13ehe}
iPTF13ehe was discovered by PTF and classified as a SLSN-I by \cite{Yan15}, who find the supernova shows late time hydrogen emission. We include photometry from \cite{Yan15}, who do not subtract the host contribution from the photometry. \cite{Wang16} modelled the SN and argues for a triple power source (Radioactive decay + Magnetar + CSM Interaction). We include one spectrum from \cite{Yan15}.

\subsection{iPTF15eov}
iPTF15eov was discovered by PTF and classified as a SN Ic by \cite{Taddia19_broadlined}, but noted to be significantly luminous. \citep{gullin2019} present an analysis of the SN and conclude it is closer to a SLSN-I. The peak absolute magnitude of $M_r \sim -21.2$ is well within the range for SLSN-I. We include photometry from \cite{Taddia19_broadlined}. We include one spectrum from \cite{Taddia19_broadlined}.

\subsection{iPTF16eh}
iPTF16eh was classified as a SLSN-I by \cite{Lunnan18_iPTF16eh}. We include photometry from \cite{Lunnan18_iPTF16eh}. We include one spectrum from \cite{Lunnan18_iPTF16eh}.

\subsection{LSQ12dlf}
LSQ12dlf was discovered by PESSTO classified as a SLSN-I by \cite{Nicholl14}. We include photometry from \cite{Nicholl14}. We include one spectrum from \cite{Nicholl14}.

\subsection{LSQ14bdq}
LSQ14bdq was discovered by PESSTO classified as a SLSN-I by \cite{Nicholl15_LSQ14bdq}. We include photometry from \cite{Nicholl15_LSQ14bdq}. The supernova has an early time bump before MJD = 56740, which we exclude from the \mosfit fit. We include one spectrum from \cite{Nicholl15_LSQ14bdq}.

\subsection{LSQ14mo}
LSQ14mo was classified as a SLSN-I by \cite{Chen17}. We include photometry from \cite{Chen17} and UVOT photometry from the SOUSA. Additionally, we include our own PSF photometry of Las Cumbres GSP images after doing difference imaging to subtract the host flux. We include a spectrum from \cite{Chen17}, obtained from WISeREP.

\subsection{OGLE15qz}
OGLE15qz was discovered by OGLE and classified as a SLSN-I by \cite{Kostrzewa15}. We include photometry from OGLE, and from \cite{Kostrzewa15}, the latter already corrected for extinction. We include a spectrum obtained from WISeREP.

\subsection{PS110awh}
PS1-10awh (=PSc090022) was discovered by MDS and classified as a SLSN-I by \cite{Chomiuk11} and presented in \citep{Lunnan18}. The authors correct the photometry for extinction. We include photometry from \cite{Lunnan18}, plus $y$-band photometry from \cite{Chomiuk11}. The SN was reported as a spectroscopically classified SLSN-I as part of the PS1 Medium Deep Survey \citep{Hosseinzadeh20, Villar20}. We include one spectrum from \cite{Chomiuk11}.

\subsection{PS110bzj}
PS1-10bzj (=PSc110405) was discovered by MDS and classified as a SLSN-I by \cite{Lunnan13} and presented in \citep{Lunnan18}. The authors correct the photometry for extinction. The SN was reported as a spectroscopically classified SLSN-I as part of the PS1 Medium Deep Survey \citep{Hosseinzadeh20, Villar20}. We include one spectrum from \cite{Lunnan13}.

\subsection{PS110ky}
PS1-10ky (=PSc060270) was discovered by MDS and classified as a SLSN-I by \cite{Chomiuk11} and presented in \citep{Lunnan18}. We include the photometry from \citep{Lunnan18}, which the authors correct for extinction. The SN was reported as a spectroscopically classified SLSN-I as part of the PS1 Medium Deep Survey \citep{Hosseinzadeh20, Villar20}. We include one spectrum from \cite{Chomiuk11}.

\subsection{PS110pm}
PS1-10pm (=PSc030129) was discovered by MDS and classified as a SLSN-I by \cite{McCrum15} and presented in \citep{Lunnan18}. The authors correct the photometry for extinction. We include photometry from \citep{Lunnan18}, but exclude the late-time WHT and GN data from \cite{McCrum15}, since these do not have the host galaxy host subtracted. The SN was reported as a spectroscopically classified SLSN-I as part of the PS1 Medium Deep Survey \citep{Hosseinzadeh20, Villar20}. We include one spectrum from \cite{McCrum15}.

\subsection{PS111afv}
PS1-11afv (=PSc160103) was discovered by MDS and classified as a SLSN-I by \cite{Lunnan14} and presented in \citep{McCrum15} and \cite{Lunnan18}. We include photometry from \cite{Lunnan18}, which the authors correct for extinction. The SN was reported as a spectroscopically classified SLSN-I as part of the PS1 Medium Deep Survey \citep{Hosseinzadeh20, Villar20}. We include one spectrum from \cite{Lunnan18}.

\subsection{PS111aib}
PS1-11aib (=PSc180279) was discovered by MDS and classified as a SLSN-I by \cite{Lunnan14} and presented in \citep{McCrum15} and \cite{Lunnan18}. We include photometry from \cite{Lunnan18}, which the authors correct for extinction. The light curve shows a possible early time bump. The SN was reported as a spectroscopically classified SLSN-I as part of the PS1 Medium Deep Survey \citep{Hosseinzadeh20, Villar20}. We include one spectrum from \cite{Lunnan18}.

\subsection{PS111ap}
PS1-11ap (=PSc120031) was discovered by MDS and classified as a SLSN-I by \cite{Lunnan18}, reported as a spectroscopically classified SLSN-I as part of the PS1 Medium Deep Survey \citep{Hosseinzadeh20, Villar20}. We include photometry from \cite{McCrum14} and from \cite{Lunnan18}. Photometry from \cite{Lunnan18} is already corrected for extinction. We include one spectrum from \cite{McCrum14}.

\subsection{PS111bdn}
PS1-11bdn (=PSc340195) was discovered by MDS and classified as a SLSN-I by \cite{Lunnan14} and presented in \citep{Lunnan18}. We include photometry from UVOT and from \citep{Lunnan18}, which the authors correct for extinction. The SN was reported as a spectroscopically classified SLSN-I as part of the PS1 Medium Deep Survey \citep{Hosseinzadeh20, Villar20}. We include one spectrum from \cite{Lunnan18}.

\subsection{PS112bqf}
PS1-12bqf (=PSc440176) was discovered by MDS and classified as a SLSN-I by \cite{Lunnan14} and presented in \citep{Lunnan18}. We include photometry from \citep{Lunnan18}, which the authors correct for extinction. This is one of the lowest luminosity objects in the \cite{Lunnan18} sample. The SN was reported as a spectroscopically classified SLSN-I as part of the PS1 Medium Deep Survey \citep{Hosseinzadeh20, Villar20}. We include one spectrum from \cite{Lunnan18}.

\subsection{PS112cil}
PS1-12cil (=PSc460103) was discovered by MDS and classified as a SLSN-I by \cite{Lunnan18}. The authors correct the photometry for extinction. The light curve has a pronounced second peak after MJD = 56333, we therefore exclude these data from the \mosfit fit. We also exclude the first $y$-band data point from the \mosfit fit due to it being unusually bright. The SN was reported as a spectroscopically classified SLSN-I as part of the PS1 Medium Deep Survey \citep{Hosseinzadeh20, Villar20}. We include one spectrum from \cite{Lunnan18}.

\subsection{PS113or}
PS1-13or (=PSc480552) was discovered by MDS and classified as a SLSN-I by \cite{Lunnan18}. We include photometry from \citep{Lunnan18}, which the authors correct for extinction. The SN was reported as a spectroscopically classified SLSN-I as part of the PS1 Medium Deep Survey \citep{Hosseinzadeh20, Villar20}. We include one spectrum from \cite{Lunnan18}.

\subsection{PS114bj}
PS1-14bj (=PSc590123) was discovered by MDS and classified as a SLSN-I by \cite{Lunnan16}. We include photometry from \citep{Lunnan18}, which the authors correct for extinction. The redshift in \cite{Lunnan18} has a typo quoted as $z = 0.5125$, where it should be $z = 0.5215$. The SN was reported as a spectroscopically classified SLSN-I as part of the PS1 Medium Deep Survey \citep{Hosseinzadeh20, Villar20}.

\subsection{PS15cjz}
PS15cjz (=DES15S2nr) was discovered by DES and classified as a ``gold" SLSN-I by \cite{Angus19}. We include photometry from PS1 and \cite{Angus19}. We exclude the detections before MJD = 57257 from the \mosfit fit due to a early-time bump. We include one spectrum from \cite{Angus19}.

\subsection{PTF09atu}
PTF09atu was discovered by PTF and classified as a SLSN-I by \cite{Perley16}, the authors find an associated host galaxy at a redshift of $z = 0.5015$. We include photometry from \cite{Cia18}, as opposed to the original presentation of the data from \cite{Quimby11}. The photometry from \cite{Cia18} is already corrected for extinction using $E(B-V) = 0.042$. We include one spectrum from \cite{Quimby18}.

\subsection{PTF09cnd}
PTF09cnd was discovered by PTF and classified as a SLSN-I by \cite{Quimby11} and presented in \cite{Perley16}, who find an associated host galaxy at a redshift of $z = 0.2584$. We include photometry from \cite{Cia18}, which the authors correct for extinction using $E(B-V) = 0.019$. We include one spectrum from \cite{Quimby18}.

\subsection{PTF10uhf}
PTF10uhf was discovered by PTF and classified as a SLSN-I by \cite{Quimby18}. \cite{Perley16} found an associated host galaxy at a redshift of $z = 0.2882$. We include photometry from \cite{Cia18}, which the authors correct for extinction using $E(B-V) = 0.016$. We include one spectrum from \cite{Quimby18}.

\subsection{PTF10vqv}
PTF10vqv was discovered by PTF and classified as a SLSN-I by \cite{Quimby18}. \cite{Perley16} found an associated host galaxy at a redshift of $z = 0.4518$. We include photometry from \cite{Cia18}, which the authors correct for extinction using $E(B-V) = 0.055$. We include one spectrum from \cite{Quimby18}.

\subsection{PTF12dam}
PTF12dam was discovered by PTF and classified as a SLSN-I by \cite{Nicholl13}. \cite{Perley16} found an associated host galaxy at a redshift of $z = 0.1073$. We include photometry from \cite{Cia18}, which the authors correct for extinction using $E(B-V) = 0.03$. We also include photometry from \cite{Nicholl13} and \cite{Chen15}, which we correct for extinction. We include one spectrum from \cite{Quimby18}.

\subsection{PTF12mxx}
PTF12mxx was discovered by PTF and classified as a SLSN-I by \cite{Quimby18}. \cite{Perley16} found an associated host galaxy at a redshift of $z = 0.3296$. We include photometry from \cite{Cia18}, which the authors correct for extinction using $E(B-V) = 0.039$. We include one spectrum from \cite{Quimby18}.

\subsection{SCP06F6}
SCP06F6 (=J143227.42+333225.1) was discovered by \cite{Barbary09} and classified as a new class of transient, and subsequently as a SLSN-I by \cite{Quimby11}. \cite{Chatzopoulos09} model the light curve of SCP06F6 with either a pair-instability model or a CSM interaction model. All photometry is in the Vega magnitude system. We exclude the first $z$-band data point before MJD = 53750 due to an early time excess. We include one spectrum from \cite{Barbary09}.

\subsection{SNLS06D4eu}
SNLS 06D4eu was classified as a SLSN-I by \cite{Howell13}. All the photometry is presented in the Vega magnitude system. We exclude the first three detections before MJD = 53950 from the \mosfit model due to an early time excess. We include one spectrum from \cite{Howell13}.

\subsection{SNLS07D2bv}
SNLS 07D2bv was classified as a SLSN-I by \cite{Howell13}. All the photometry is presented in the Vega magnitude system. We include one spectrum from \cite{Howell13}.

\subsection{2018lzx}
SN\,2018lzx (=ZTF18abszecm) was discovered by ZTF and classified as a SLSN-I by \cite{Chen22a, Yan_18lzx}. We include photometry from \cite{Chen22a} and ATLAS. We include one spectrum from \cite{Chen22a}.

\subsection{2019aamr}
SN\,2019aamr (=ZTF19abdlzyq) was discovered by ZTF and classified as a SLSN-I by \cite{Chen22a}. We include photometry from \cite{Chen22a}, ATLAS, and our own PSF photometry of early-time ZTF images. We include one spectrum from \cite{Chen22a}.

\subsection{2019aams}
SN\,2019aams (=ZTF19abnqqdp) was discovered by ZTF and classified as a SLSN-I by \cite{Chen22a}. We include photometry from \cite{Chen22a} and ATLAS. We include one spectrum from \cite{Chen22a}.

\subsection{2019aamx}
SN\,2019aamx (=ZTF19abcvwrz) was discovered by ZTF and classified as a SLSN-I by \cite{Chen22a}. We include photometry from \cite{Chen22a} and ATLAS. We include one spectrum from \cite{Chen22a}.

\subsection{2019qgk}
SN\,2019qgk (=ZTF19abuolvj) was discovered by ZTF and classified as a SLSN-I by \cite{Chen22a}. We include photometry from from \cite{Chen22a} and ATLAS. We include one spectrum from \cite{Chen22a}.

\subsection{2019xdy}
SN\,2019xdy (=ZTF19acsajxn =PS19jja) was discovered by ZTF and classified as a SLSN-I by \cite{Chen22a}. We include photometry from ATLAS, PS1, and upper limits from \cite{Chen22a}, additionally we include our own PSF photometry of ZTF images. Given the high cadence of ZTF observations we co-add the images in bins of 1 hour, or 1 day. We include one spectrum from \cite{Chen22a}.

\subsection{2020kox}
SN\,2020kox (=ZTF20aavqrzc =ATLAS20nev) was discovered by ZTF and classified as a SLSN-I by \cite{Chen22a}. We include photometry from \cite{Chen22a}, ATLAS, and our own PSF photometry of late time ZTF $r$-band images. We include one spectrum from \cite{Chen22a}.

\subsection{2020zbf}
SN\,2020zbf (=ATLAS20bfee) was discovered by ATLAS and classified as a SLSN-I by \cite{Ihanec20} and \cite{Lunnan21}, with a detailed study in \cite{Gkini24}. We include photometry from ATLAS. We include one spectrum from \cite{Ihanec20}.


\section{``Silver" Superluminous Supernovae}\label{sec:silver_slsn}

\subsection{2019unb}
We presented SN\,2019unb (=ZTF19acgjpgh =Gaia19fbu =PS19isr =ATLAS19bari) as a SLSN-like LSN in \cite{Gomez22_LSN}. The SN was originally was discovered by ZTF and classified by \cite{Dahiwale20_unb} with a detailed study presented in \cite{Yan20}, and presented in \cite{Prentice19,Prentice21}. We include additional photometry from ATLAS and \cite{Chen22a}. We assign a silver label given that this is a LSN. We include one spectrum from \cite{Dahiwale20_unb}.

\subsection{2019hge}
We presented SN\,2019hge (=ZTF19aawfbtg =Gaia19est =ATLAS19och =PS19elv) as a SLSN-like LSN in \cite{Gomez22_LSN}. The SN was discovered by ZTF, classified by \cite{Yan20} and presented in \cite{Prentice21}. We include additional photometry from ATLAS and \cite{Chen22a}, but exclude detections after MJD = 58770 from the \mosfit fit due to a late-time re-brightening of the light curve. We assign a silver label given that this is a LSN. We include one spectrum from \cite{Dahiwale19_hge}.

\subsection{2019gam}
We presented SN\,2019gam (=ZTF19aauvzyh =ATLAS19lsz) as a SLSN-like LSN in \cite{Gomez22_LSN}. The SN was discovered by ZTF and classified by \cite{Yan20}. We include additional photometry from ATLAS, and late-time LT and SEDM photometry from \cite{Chen22a}. We assign a silver label given that this is a LSN. We include one spectrum from \cite{Chen22a}.

\subsection{PTF12gty}
We presented PTF12gty as a SLSN-like LSN in \cite{Gomez22_LSN}. The SN was originally discovered by PTF, classified by \cite{Quimby18,Barbarino20}, and presented in \cite{Cia18}. We assign a silver label given that this is a LSN. We include one spectrum from \cite{Quimby18}.

\subsection{2016aj}
SN\,2016aj (=PS16op) was classified as a SLSN-I by \cite{Young16}. We include PS1 photometry and one epoch of UVOT photometry from the SOUSA archive. We do not correct the UVOT photometry for the contribution of the host, since the host magnitude is below the PS1 detection limit and is therefore likely negligible in the UV. We assign a silver label since the photometry available is very sparse and there is no data during the rise. We include one spectrum from \cite{Young16}.

\subsection{OGLE15xx}
OGLE15xx was discovered by OGLE and classified as a SN Ib/c by \cite{Wyrzykowski15_15xx}, but the spectrum and peak magnitude match those of a SLSN-I. We include OGLE photometry and exclude the first detection before MJD = 57375 due to a pre-explosion bump. We assign a silver label given there is only one band of photometry available. We include a spectrum obtained from WISeREP.

\subsection{1991D}
We presented 1991D as a SN with spectra consistent with either a Type Ib SN or a SLSN in \cite{Gomez22_LSN}. The SN was originally classified by \cite{Benetti02} and presented in \cite{Matheson01}. We assign a silver label given that this is a LSN. We include one spectrum from \cite{Matheson01}.

\subsection{2006oz}
SN\,2006oz (=SDSS-II SN 15557) was classified by \cite{Leloudas12} as a SLSN-I. We include photometry from \cite{Leloudas12}. We exclude photometry before MJD = 54036 from the \mosfit fit since these appear to be from an initial bump. \cite{Leloudas12} pointed out this bump could be powered by a recombination wave in the circumstellar medium. Alternatively, \cite{Ouyed13} explain the bump as part of a dual-shock quark nova. We assign a silver label given the lack of photometry after peak. We include one spectrum from \cite{Leloudas12}, obtained from WISeREP.

\subsection{2009cb}
We presented SN\,2009cb (=CSS090319:125916+271641 =PTF09as) as a LSN in \cite{Gomez22_LSN}. The SN was originally discovered by PTF, classified by \cite{Quimby18}, and presented in \cite{Cia18}. We assign a silver label given that this is a LSN. We include one spectrum from \cite{Quimby18}.

\subsection{2011kl}
We presented SN\,2011kl (=GRB111209A) as a SLSN-like LSN in \cite{Gomez22_LSN}. The SN was originally classified by \cite{Greiner15,Mazzali16,Kann19}. We assign a silver label given that this is a LSN. We include one spectrum from \cite{Greiner15}.

\subsection{2012aa}
We presented SN\,2012aa (=PSN J14523348-0331540 =Howerton-A20) as a SN with spectra consistent with either a Type Ibc SN or a SLSN in \cite{Gomez22_LSN}. The SN was originally discovered by CRTS, classified by \cite{Roy16}, and studied in \cite{Yan17,Shivvers19}. We assign a silver label given that this is a LSN. We include one spectrum from \cite{Shivvers19}.

\subsection{2013hy}
We presented SN\,2013hy (=DES13S2cmm) as a SLSN-like LSN in \cite{Gomez22_LSN}. The SN was originally discovered by DES and classified by \cite{Papadopoulos15,Angus19}. We assign a silver label given that this is a LSN. We include one spectrum from \cite{Angus19}.

\subsection{2018beh}
We presented SN\,2018beh (=ZTF18aahpbwz =ASASSN-18ji =PS18ats =ATLAS18nvb) as a SLSN-like LSN in \cite{Gomez22_LSN}. Here we include additional photometry from ATLAS. The SN was originally classified by \cite{Mcbrien18,Dahiwale20_beh}. We assign a silver label given that this is a LSN. We include one spectrum from \cite{Mcbrien18}.

\subsection{2018don}
We presented SN\,2018don (=ZTF18aajqcue =PS18aqo =ATLAS18nxb) as a SLSN-like LSN in \cite{Gomez22_LSN}. The SN was originally discovered by ZTF, classified by \cite{Lunnan20_four}, and presented in \cite{Fremling19}. We include additional photometry from ATLAS, and late-time LT images from \cite{Chen22a}. We assign a silver label given that this is a LSN. We include one spectrum from \cite{Fremling19}.

\subsection{2019J}
We presented SN\,2019J (=ZTF19aacxrab =PS18crs =ATLAS19cay) as a SLSN-like LSN in \cite{Gomez22_LSN}. The SN was originally classified by \cite{Fremling19_19J}. We adopt a redshift of $z = 0.1346$ determined by \cite{Chen22a}. We include additional photometry from \cite{Chen22a}. We assign a silver label given that this is a LSN. We include one spectrum from \cite{Chen22a}.

\subsection{2019dwa}
We presented SN\,2019dwa (=ZTF19aarfyvc =Gaia19bxj) as a SN with spectra consistent with either a Type Ic SN or a SLSN in \cite{Gomez22_LSN}. The SN was originally discovered by ZTF, classified by \cite{Fremling_19dwa}, and presented in \cite{Prentice21}. We include additional photometry from ATLAS. We assign a silver label given that this is a LSN. We include one spectrum from \cite{Fremling_19dwa}.

\subsection{2019ieh}
SN\,2019ieh (=ZTF19abauylg =Gaia19cxd =PS19bil =ATLAS19nsv) was discovered by ZTF and classified as a SN Ic-BL by \cite{Zheng_19ieh} and as a SN Ic by \cite{Dahiwale_19ieh}. We adopt the redshift from \cite{Zheng_19ieh}, derived from host emission lines. At this redshift the SN peaks at a magnitude of $M_r \sim -19.3$, which makes this a LSN. We include photometry from ZTF, ATLAS, GSA, and PS1. We include one spectrum from \cite{Zheng_19ieh}, obtained from the TNS. We assign a silver label given that this is a LSN.

\subsection{2019obk}
We presented SN\,2019obk (=ZTF19abrbsvm =PS19eqz =ATLAS19tvm) as a SLSN-like LSN in \cite{Gomez22_LSN}. The SN was originally discovered by ZTF and classified by \cite{Yan20}. We include additional photometry from ATLAS and \cite{Chen22a}. We assign a silver label given that this is a LSN. We include one spectrum from \cite{Chen22a}.

\subsection{2019pvs}
We presented We classified SN\,2019pvs (=ZTF19abuogff =PS19fbe) as a SLSN-like LSN in \cite{Gomez22_LSN}. The SN was originally discovered by ZTF and classified in \cite{Gomez21_TNS}. We include additional photometry from ATLAS. We assign a silver label given that this is a LSN. We include one spectrum from \cite{Gomez22_LSN}.

\subsection{2021lwz}
We presented SN\,2021lwz as a SLSN-like LSN in \cite{Gomez22_LSN}. The SN was originally discovered by ATLAS and classified by \cite{Perley21}. We include additional photometry from ATLAS. We assign a silver label given that this is a LSN. We include one spectrum from \cite{Perley21}.

\subsection{DES14C1rhg}
We presented DES14C1rhg as a SLSN-like LSN in \cite{Gomez22_LSN}. The SN was originally discovered by DES and classified by \cite{Angus19}. We assign a silver label given that this is a LSN. We include photometry and one spectrum from \cite{Angus19}.

\subsection{DES15C3hav}
We presented DES15C3hav as a SLSN-like LSN in \cite{Gomez22_LSN}. The SN was originally discovered by DES and classified in \cite{Angus19}. We assign a silver label given that this is a LSN. We include photometry and one spectrum from \cite{Angus19}.

\subsection{OGLE15xl}
We presented OGLE15xl as a SLSN-like LSN in \cite{Gomez22_LSN}. The SN was originally discovered by OGLE and classified by \cite{Breton15}. We assign a silver label given that this is a LSN. We include one spectrum from \cite{Breton15}.

\subsection{PTF10iam}
We presented PTF10iam as a SLSN-like LSN in \cite{Gomez22_LSN}. The SN was originally classified by \cite{Arcavi16}. We assign a silver label given that this is a LSN. We include one spectrum from \cite{Arcavi16}.

\subsection{PTF12hni}
We presented PTF12hni as a SN with spectra consistent with either a Type Ic SN or a SLSN in \cite{Gomez22_LSN}. The SN was originally disovered by PTF and classified by \cite{Quimby18} and presented in \cite{Cia18}. We assign a silver label given that this is a LSN. We include one spectrum from \cite{Quimby18}.

\subsection{PTF10bjp}
PTF10bjp was discovered by PTF and classified as a SLSN-I by \cite{Quimby18}. \cite{Perley16} found an associated host galaxy at a redshift of $z = 0.3584$. We include photometry from \cite{Cia18}, which the authors correct for extinction using $E(B-V) = 0.055$. We assign a silver label given there is only one band of photometry available. We include one spectrum from \cite{Quimby18}.

\subsection{2017jan}
SN\,2017jan (=OGLE17jan) was discovered by OGLE and classified as a SLSN-I by \cite{Angus17}. We include photometry from OGLE \citep{Wyrzykowski14} between MJD = 58005 and MJD = 58200, excluding noisy late-time photometry and an apparent early-time bump. We assign a silver label given there is only one band of photometry available. We include one spectrum from \cite{Angus17}, obtained from the TNS.

\subsection{OGLE16dmu}
OGLE16dmu was discovered by OGLE and classified as a SLSN-I by \cite{Prentice16} and presented in \cite{Cikota18}. We assign a silver label given there is only one band of photometry available. We include photometry from OGLE. We include a spectrum from \cite{Cikota18}.

\subsection{OGLE15sd}
OGLE15sd was discovered by OGLE and classified as a SLSN-I by \cite{Wyrzykowski15}. We include photometry from OGLE. We assign a silver label given that there is only one band of photometry available. We include a spectrum obtained from WISeREP.

\subsection{SSS120810}
SSS120810:231802-560926 was originally discovered by \cite{Wright12} and classified as a SLSN-I by \cite{Inserra12}. A detailed study of the SN was presented in \cite{Nicholl14} and a study of its host galaxy in \cite{Leloudas15}. We include photometry from \cite{Nicholl14}. We assign a silver label given the lack of photometry before peak.

\subsection{2020myh}
We classified SN\,2020myh (=ZTF20abgbxby =ATLAS20pxs), originally discovered by ATLAS, as a SLSN-I as part of FLEET \cite{Gomez20_TNS_20jii}. We determine a redshift of $z = 0.283$ based on host emission lines. We include photometry from ATLAS and ZTF. We have no early time spectra of the source, but one low S/N late time spectrum consistent with either a SNe Ic or a SLSN-I, but given the peak absolute magnitude of $M_r \sim -21.3$, we adopt a SLSN-I classification. We include photometry from ZTF and ATLAS. We assign a silver label given the lack of photometry before peak. We include our own Binospec spectrum from \cite{Gomez20_TNS_20jii}.

\subsection{CSS160710}
CSS160710:160420+392813 (=MLS160616:160420+392813) was discovered by the CRTS and classified as a SLSN-I by \cite{Drake16}. We include photometry from the CRTS and MLS, and our own PSF photometry of Las Cumbres GSP images. We assign a silver label given there is only one band of photometry available.

\subsection{LSQ14an}
LSQ14an was classified as a SLSN-I by \cite{Inserra17}. We include photometry from \cite{Inserra17}. Additionally, we include our own PSF photometry of Las Cumbres GSP images after doing difference imaging to subtract the host flux. We assign a silver label given the lack of photometry before peak. We include one spectrum from \cite{Inserra17}.

\subsection{PTF10bfz}
PTF10bfz was discovered by PTF and classified as a SLSN-I by \cite{Quimby18}. \cite{Perley16} found an associated host galaxy at a redshift of $z = 0.1701$. We include photometry from \cite{Cia18}, which the authors correct for extinction using $E(B-V) = 0.016$. The spectra from \cite{Quimby18} show a systematic blueshift of $\sim 12,000$ km s$^{-1}$ that might be the result of an asymmetrical explosion or a viewing angle effect. We assign a silver label given the lack of photometry before peak. We include one spectrum from \cite{Quimby18}.

\subsection{2016els}
SN\,2016els (=PS16dnq) was classified as a SLSN-I by \cite{Fraser16} and \cite{Mattila16}. We include PS1 photometry and UVOT photometry from the SOUSA archive. There is no photometry available before peak. We assign a silver label given the lack of photometry before peak. We include one spectrum from \cite{Fraser16}, obtained from WISeREP.

\subsection{2016wi}
SN\,2016wi (=PS16yj =iPTF15esb) was classified as a SLSN-I by \cite{Yan17}. We include photometry from PS1 and \cite{Yan17}. The light curve shows three distinct bumps at early times, which we include in the \mosfit fit. We assign a silver label given the lack of photometry before peak. We include one spectrum from \cite{Fraser16}, obtained from WISeREP.

\subsection{2018gkz}
SN\,2018gkz (=ZTF18abvgjyl =PS18ced) was discovered by ZTF and classified as a SLSN-I by \cite{Yan20_gkz}. We include photometry from PS1 and \cite{Chen22a}, in addition to our own PSF photometry of ZTF images after doing difference imaging to subtract the host flux. We assign a silver label given the lack of photometry before peak. We include one spectrum from \cite{Fremling18}, obtained from the TNS.

\subsection{iPTF16bad}
iPTF16bad was classified as a SLSN-I by \cite{Yan17}. \cite{Gal-Yam19_spec} show that iPTF16bad lacks hydrogen at early times, but that tentative hydrogen features emerge at late times. We include photometry from \cite{Yan17}. We assign a silver label given the lack of photometry before peak. We include one spectrum from \cite{Fraser16}.

\subsection{2018fd}
We classified SN\,2018fd (=ZTF18accdszm =PS18df =MLS171011:091036+354318), originally discovered by CRTS, as a SLSN-I as part of FLEET \cite{Gomez21_TNS_18cxa}. We find a redshift of $z = 0.263$ based on host emission lines. We include CRTS $C$-band photometry, but do not subtract any host galaxy contribution from the CRTS photometry since the host galaxy contribution should be minimal given its magnitude of $m_r = 22.13$ mag. We include our own PSF photometry of FLWO images after doing difference imaging to subtract the host flux. There is no photometry before peak. We assign a silver label given the lack of photometry before peak. We include our own spectrum from Binospec \citep{Gomez21_TNS_18cxa}.

\subsection{PS113gt}
PS1-13gt (=PSc480107) was discovered by MDS and classified as a SLSN-I by \cite{Lunnan14} and presented in \cite{Lunnan18}. We include photometry from \cite{Lunnan18}, who correct the photometry for extinction. This SN shows signs of reddening, since it has a red continuum and \ion{O}{2} features that require high temperatures \citep{Lunnan18}. There is no photometry at or before peak. The SN was reported as a spectroscopically classified SLSN-I as part of the PS1 Medium Deep Survey \citep{Hosseinzadeh20, Villar20}. We assign a silver label given the lack of photometry before peak. We include one spectrum from \cite{Lunnan18}.

\subsection{DES16C2nm}
DES16C2nm was discovered by DES and classified as a SLSN-I by \cite{Smith18} and presented as a ``gold" SLSN-I at a very high redshift of $z = 1.998$ by \cite{Angus19}. We include photometry form \cite{Angus19}. We assign a silver label given the lack of photometry before peak. We include one spectrum from \cite{Angus19}.

\subsection{2011kf}
SN\,2011kf (=CSS111230:143658+163057) was originally classified as a SLSN-I by \cite{Prieto12}. A detailed study of the SN was presented in \cite{Inserra13}. We include photometry from \cite{Inserra13}, which does not have the host contribution subtracted, but we conclude this should be negligible. There is no photometry available before peak. We assign a silver label given the lack of photometry before peak. We include one spectrum from \cite{Inserra13}, obtained from WISeREP.

\subsection{PS112bmy}
PS1-12bmy (=PSc440420) was discovered by MDS and classified as a SLSN-I by \cite{Lunnan14} and presented in \cite{Lunnan18}. We include photometry from \cite{Lunnan18}, who correct the photometry for extinction. There is no photometry available before peak. The SN was reported as a spectroscopically classified SLSN-I as part of the PS1 Medium Deep Survey \citep{Hosseinzadeh20, Villar20}. We assign a silver label given that there are no data before peak. We include one spectrum from \cite{Lunnan18}.

\subsection{PS111bam}
PS1-11bam (=PSc330114) was discovered by MDS and classified as a SLSN-I by \cite{Berger12} and presented in \citep{Lunnan18}. We include photometry from \citep{Lunnan18}, which the authors correct for extinction. The SN was reported as a spectroscopically classified SLSN-I as part of the PS1 Medium Deep Survey \citep{Hosseinzadeh20, Villar20}. We assign a silver label given the lack of photometry before peak. We include one spectrum from \cite{Berger12}.

\subsection{1999as}
SN\,1999as was classified as a SN Ia by \cite{Knop99}, but the authors noted that the source was blue, with broad absorption features, and about 2 magnitudes brighter than a typical SN Ia. The source was later presented in \cite{Deng01} as the most luminous supernova discovered at the time. \cite{Hatano01} classified SN\,1999as as a Type Ic hypernova. \cite{Moriya19} finally present a comparison between SN\,1999as and SN\,2007bi, and classify SN\,1999as as a SLSN-I. We assign a silver label given the lack of data before peak. We include photometry and one spectrum from \cite{Kasen04}.

\subsection{2002gh}
SN\,2002gh was discovered by \cite{Cartier22} two decades after the actual SN explosion. The authors classify the object as one of the most luminous SLSN-I ever discovered with $M_V = -22.4$ mag. Although we note that given the high redshift of $z = 0.3653$, applying an additional cosmological K-correction of +($2.5 \times \log(1 + z)$) brings the peak magnitude down to a less extreme value of $M_V = -22.2$ mag. We include photometry from \cite{Cartier22}. We assign a silver label given the lack of data before peak. We include one spectrum from \cite{Cartier22}.

\subsection{iPTF13bjz}
iPTF13bjz was discovered by PTF and classified as a SLSN-I by \cite{Cia18}. We include photometry from \cite{Cia18}, which the authors correct for extinction using $E(B-V) = 0.019$. We assign a silver label given that the light curve only has $r$-band data and no data during the decline. We include one spectrum from \cite{Schulze20}.

\subsection{PS110ahf}
PS1-10ahf (=PSc080079) was discovered by MDS and classified as a SLSN-I by \cite{McCrum15} and presented in \cite{Lunnan18}. We include photometry from \cite{Lunnan18}, who correct the photometry for extinction. Even though \cite{McCrum15} present late time data, these are not included since they are not corrected for the flux contribution of the host. There are two late-time detections that appear to suggest a second peak in the light curve. The spectrum shows some significant absorption bluewards of $\sim 2800$\AA. The SN was reported as a spectroscopically classified SLSN-I as part of the PS1 Medium Deep Survey \citep{Hosseinzadeh20, Villar20}. We assign a silver label given the lack of photometry after peak. We include one spectrum from \cite{McCrum15}.

\subsection{2017beq}
SN\,2017beq (=PS17bek =iPTF17beq) was classified as a SLSN-I by \cite{Gal-Yam17_17beq}, the authors provide two points of photometry. We include an additional data point from \cite{Cikota18}. We assign a silver label since there are only a total of three photometry points available. We include one spectrum, obtained from WISeREP.

\subsection{2018lzw}
SN\,2018lzw (=ZTF18abrzcbp) was discovered by ZTF and classified as a SLSN-I by \cite{Chen22a, Yan22_18lzw}. We include photometry from \cite{Chen22a}, ATLAS, and our own upper-limits from pre-explosion ZTF images. We assign a silver spectrum since there is no data during the rise. We include one spectrum from \cite{Chen22a}.

\subsection{PTF10aagc}
PTF10aagc was discovered by PTF and classified as a SLSN-I by \cite{Quimby18}. \cite{Perley16} found an associated host galaxy at a redshift of $z = 0.206$. We include photometry from \cite{Cia18}, which the authors correct for extinction using $E(B-V) = 0.023$. The spectra from \cite{Quimby18} shows hydrogen and helium, but the spectrum does not look like a SLSN-II. \cite{Yan15} suggest PTF10aagc is a SLSN-I with ejecta that interacts with a H-rich CSM at late times. We assign a silver label given the spectral uncertainties. We include one spectrum from \cite{Quimby18}.

\subsection{2020abjx}
SN\,2020abjx (=ZTF20aceicyy =ATLAS20bfww) was discovered by ZTF and classified as a SLSN-I by \cite{Yan20_2020abjx}. At the reported redshift of $z = 0.39$, the peak magnitude of the SN is $M_r = -22.3$, comfortably in the SLSN-I regime. Nevertheless, we assign a silver label given the single noisy available spectrum. We include photometry from ZTF and ATLAS, as well as our own PSF photometry of LDSS3C images. We include our own spectrum from Binospec.

\subsection{2021rwz}
SN\,2021rwz (=ZTF21abezyhr =PS21hfp =ATLAS21tkh) was discovered by ZTF and classified as a SLSN-I by \cite{Weil21_rwz, Weil21b_rwz}. We include photometry from ZTF, ATLAS, and PS1. Additionally, we include our own PSF photometry of FLWO images. We assign a silver label given the highly noisy single spectrum available from \cite{Weil21b_rwz}.

\subsection{2017hbx}
SN\,2017hbx (=Gaia17cna) was discovered by Gaia and classified as a SN Ic by \cite{Neill17}. We include photometry from GSA, and upper limits from ATLAS. The redshift of $z = 0.1652$ reported by \cite{Neill17} is accurate to $\sim 0.02$. The spectrum is consistent with either a SLSN-I or a SN Ic, but we assign a SLSN-I classification given the peak absolute magnitude of $M_G \sim -21.3$. We assign a silver label given the uncertain redshift. We include one spectrum from \cite{Neill17}.

\subsection{iPTF13dcc}
iPTF13dcc (=CSS130912:025702-001844) was classified as a SLSN-I by \cite{Vreeswijk17}. We include photometry from \cite{Vreeswijk17}. The light curve shows a flat profile at early times, which is hard to fit with our current model. The authors explain the complex light curve structure by invoking CSM interaction, as did \cite{Liu18}. We assign a silver label given the lack of photometry before peak. We include one spectrum from \cite{Schulze20}, obtained from WISeREP.

\subsection{2022aawb}
SN\,2022aawb (=ZTF22abvcnnl) was discovered by ZTF and classified as a SLSN-I by \cite{Poidevin22_a, Poidevin22_b}. We include data from ATLAS and ZTF. We include one spectrum from \cite{Poidevin22_a}, obtained from the TNS. We assign a silver label given the very low S/N spectrum.

\subsection{2022gyv}
SN\,2022gyv (=ZTF22aadqgoa =PS22diw) was classified as a SLSN-I by \cite{Poidevin_22gyv}. We include photometry from ATLAS, ZTF and PS1. We exclude the first $i$-band detection from the \mosfit fits since this is well before the nominal SN explosion. We assign a silver label given the slightly uncertain redshift measurement of $z = 0.38 - 0.40$. We include one spectrum from \cite{Poidevin_22gyv}.

\subsection{DES14C1fi}
DES14C1fi was discovered by DES and classified as a ``silver" SLSN-I by \cite{Angus19}. We include photometry and one spectrum from \cite{Angus19}.

\subsection{DES14E2slp}
DES14E2slp was discovered by DES and classified as a ``silver" SLSN-I by \cite{Angus19}. We include photometry and one spectrum from \cite{Angus19}.

\subsection{2019aamw}
SN\,2019aamw (=ZTF19acujvsi) was discovered by ZTF and classified as a SLSN-I by \cite{Chen22a}. We include photometry from \cite{Chen22a} and ATLAS. We assign a silver label given the very noisy spectra available. We include one spectrum from \cite{Chen22a}.

\subsection{DES15S1nog}
DES15S1nog was discovered by DES and classified as a ``silver" SLSN-I by \cite{Angus19} and as a possible SLSN-I by \cite{Casas16}. We include photometry and one spectrum from \cite{Angus19}.

\subsection{DES17C3gyp}
DES17C3gyp was discovered by DES and classified as a ``silver" SLSN-I by \cite{Angus19}. We include photometry from \cite{Angus19}. We exclude the two detections before MJD = 58100 from the \mosfit fit since these are much earlier than the nominal explosion date. We include one spectrum from \cite{Angus19}.

\subsection{iPTF13bdl}
iPTF13bdl was discovered by PTF and classified as a SLSN-I by \cite{Cia18}. We include photometry from \cite{Cia18}, which the authors correct for extinction using $E(B-V) = 0.042$. The light curve shows a large scatter, making it hard to constrain its physical parameters. We assign a silver label given the uncertain light curve. We include one spectrum from \cite{Cia18}.

\subsection{2020fyq}
SN\,2020fyq (=ZTF20aapaecd =PS20arv =ATLAS20kwv) was discovered by ZTF and classified as a SLSN-I by \cite{Chen22a}. We include photometry from \cite{Chen22a}, PS1, and ATLAS. We assign a silver label given the similarty to LSNe given the peak magnitude of $M_r \sim -19.9$ and spectrum consistent with a SN Ic. We include one spectrum from \cite{Chen22a}.

\subsection{PS111tt}
PS1-11tt (=PSc150381) was discovered by MDS and classified as a SLSN-I by \cite{Lunnan14} and presented in \cite{McCrum15}. The latter authors provide photometry corrected for extinction. The first data point might be indicative of a precursor according to \citep{Lunnan18}. We assign a silver label since there is only one spectrum available of the source with low S/N with significant absorption bluewards of $\sim 2800$\AA, nevertheless this is consistent with a SLSN-I. The SN was reported as a spectroscopically classified SLSN-I as part of the PS1 Medium Deep Survey \citep{Hosseinzadeh20, Villar20}. We include one spectrum from \cite{Lunnan18}.

\subsection{SNLS07D3bs}
SNLS-07D3bs was classified as an uncertain SLSN-I by \cite{Prajs17}, given the low S/N spectrum used for classification. At the redshift of $z = 0.757$ provided by \cite{Fremling18_2018gft}, the peak magnitude of the SN is $M_r \sim -21.1$, within the SLSN-I regime. The spectra of the source are not available for download and only late time spectra exist, hence the silver label.

\subsection{2020onb}
We classified SN\,2020onb (=ZTF20abjwjrx =Gaia20dub =ATLAS20bkdh), originally discovered by ZTF, as a SLSN-I as part of FLEET \cite{Gomez20_TNS_20jii}. We determined a redshift of $z = 0.153$ based on host galaxy emission lines. We include photometry from ATLAS, GSA, and \cite{Chen22a}. The spectra at early times are significantly redder than normal SLSN-I. It is possible the spectra are heavily reddened by extinction in the host galaxy, hence why we assign a silver label. We include our own Binospec spectrum from \cite{Gomez20_TNS_20jii}.

\subsection{DES16C3ggu}
DES16C3ggu was discovered by DES and classified as a ``gold" SLSN-I by \cite{Angus19}. We include DES photometry from \cite{Angus19}. We exclude the first $g$-band point before MJD = 57740 from the \mosfit fit since this is well before the expected SN explosion. We assign a silver label given the lack of photometry after peak. We include one spectrum from \cite{Angus19}.

\subsection{PTF10nmn}
PTF10nmn was discovered by PTF and classified as a SLSN-I by \cite{Quimby18}. \cite{Perley16} found an associated host galaxy at a redshift of $z = 0.1237$. We include photometry from \cite{Cia18}, which the authors correct for extinction using $E(B-V) = 0.14$. We assign a silver label given the very sparse photometry. We include one spectrum from \cite{Quimby18}.

\subsection{iPTF13cjq}
iPTF13cjq was discovered by PTF and classified as a SLSN-I by \cite{Cia18}. At the reported redshift of $z = 0.3962$, the peak magnitude of the SN is $M_r = -21.6$, in the SLSN-I regime. We include photometry from \cite{Cia18}, which the authors correct for extinction using $E(B-V) = 0.042$. The light curve has peculiar flat early-time photometry with no data before the peak. Therefore, we assign a silver label. We include one spectrum from \cite{Schulze20}, obtained from WISeREP.

\subsection{iPTF16asu}
We presented iPTF16asu as a SLSN-like LSN in \cite{Gomez22_LSN}. The SN was originally classified by \cite{Whitesides17} and presented in \cite{Taddia19_broadlined}. We assign a silver label given that this is a LSN. We include one spectrum from \cite{Taddia19_broadlined}.


\section{``Bronze" Superluminous Supernovae}\label{sec:bronze_slsn}

\subsection{1999bz}
SN\,1999bz (=AAVSO 1359+69) was classified as a SN Ic by \cite{Berlind99}. We adopt a redshift of $z = 0.0846$, established from a spectrum of the host galaxy by \cite{Kirshner83}. At this redshift the peak magnitude of the SN is $M_C \sim -20.4$, within the range of SLSN-I. We include a single unfiltered photometry point from \cite{Pearce99} and assume an uncertainty of 0.1 mag. We assign a bronze label given the near total absence of photometry and single noisy spectrum available.

\subsection{2011ep}
SN\,2011ep (=CSS110414:170342+324553) was discovered by the CRTS and classified as a SN Ic-BL by \cite{Graham11_11ep} and as a SN Ic by \cite{Graham11}. Given the peak absolute magnitude of $M_C \sim -21.9$, the transient instead might be a SLSN-I. An analysis of the host galaxy of SN\,2011ep was presented in \cite{Schulze18}, who also considered SN\,2011ep to be a likely SLSN-I. Nevertheless, there are no public spectra of the source and we are unable to confirm this classification, hence the bronze label. We include photometry from CRTS, without subtracting any host flux, since we find this to be negligible.

\subsection{2014bl}
SN\,2014bl (=PSN J13253881+2557339) was discovered by the CRTS and discovered by \cite{Li14, Li14_atel} and classified as a Type Ic SN at a $z = 0.0377$. The SN is somewhat luminous with a peak magnitude of $M \sim -19$. There is only one spectrum with low S/N of uncertain classification. The photometry is only from CRTS, which we correct for extinction using $E(B-V) = 0.0152$. We subtract a nominal host magnitude of $m_C = 19.36$ from all photometry. We assign a bronze label given the sparse photometry and uncertain spectral classification.

\subsection{2018jfo}
SN\,2018jfo (=ZTF18achdidy =MLS181220:112339+255952) was discovered by ZTF and classified as a SLSN-I by \cite{Fremling19_2018lfd}. The authors report a redshift of $z = 0.163$, which makes the peak magnitude of the SN $M_r \sim -20.8$, within the range of SLSN-I. Nevertheless, we assign a bronze label since there are no public spectra of the source and we are unable to confirm this classification. We include photometry from CRTS, ATLAS, and our own PSF photometry from ZTF images.

\subsection{2018fcg}
SN\,2018fcg (=ZTF18abmasep =Gaia18cms =ATLAS18ucc) was discovered by ZTF and classified as a SLSN-I by \citep{Fremling18_fcg} and \cite{Lunnan18_fcg}. We presented SN\,2018fcg as a Ic-like LSN in \cite{Gomez22_LSN}. We include additional photometry from \cite{Chen22a}. We assign a bronze label given that SN\,2018fcg most closely resembles a Type Ic SN.

\subsection{2018hsf}
SN\,2018hsf (=ZTF18acbvpzj =ATLAS18yer) was discovered by ZTF and classified as a SN Ic-BL by \cite{Fremling18_2018ibb}. We include photometry from ZTF and ATLAS, plus our own PSF photometry of ZTF images after subtracting the host. At the quoted redshift of $z = 0.119$ the peak absolute magnitude is $M_r \sim -19.8$, within the range of SLSNe. Nevertheless, the source has no public spectra and we therefore assign a bronze label.

\subsection{2019une}
We classified SN\,2019une (=ZTF19acmbjmp =ATLAS19balq =PS20bus), originally discovered by ZTF, as a SLSN-I as part of FLEET. We include photometry from ATLAS, ZTF, PS1, and our own PSF photometry of FLWO images after doing difference imaging to subtract the host flux. Even though the spectrum resembles that of a SLSN-I, the redshift is too uncertain to provide a confident classification, and we therefore assign a bronze label.

\subsection{2020aewh}
SN\,2020aewh (=ZTF20acwmyzx) was discovered by ZTF and originally classified as a SLSN-I by \cite{Yan23} with a detailed study presented in \cite{Yan21}. We include photometry from ATLAS and ZTF, as well as one late-time detection from PSF photometry of IMACS images. Nevertheless, we asign a bronze label given that there are no public spectra of the source.

\subsection{2021uvy}
SN\,2021uvy was discovered by ZTF and classified as a SLSN-I by \cite{Poidevin21}, as a SN Ib/c by \cite{Ridley21}, and as a peculiar SN Ib by \cite{Chu21_uvy}. We presented SN\,2021uvy as a Ic-like LSN in \cite{Gomez22_LSN}. We assign a bronze label given the spectra most closely resembles a Type Ic SN. We include additional photometry from ATLAS. We exclude detections after MJD = 59476 due to a very prominent secondary peak.

\subsection{2020wfh}
SN\,2020wfh (=ZTF20acitbmf =PS20kqu =ATLAS20bdjz) was discovered by ZTF and classified as a SLSN-I by \cite{Yan20_2020abjx}. At the reported redshift of $z = 0.33$, the peak magnitude of the SN is $M_r = -22.1$, in the SLSN-I regime. Nevertheless, we assign a bronze label since there are no public spectra of the source and we are unable to confirm this classification. We include photometry from PS1 and ATLAS, as well as our own PSF photometry of FLWO and ZTF images after subtracting the host contribution.

\subsection{2021ybf}
SN\,2021ybf was discovered by ZTF and classified as a SLSN-I by \cite{Bruch21}. We presented SN\,2021ybf as a Ic-like LSN in \cite{Gomez22_LSN}. We assign a bronze label given the spectra most closely resembles a Type Ic SN. We include additional photometry from ATLAS.

\subsection{2022ojm}
SN\,2022ojm (=ZTF22aapjqpn) was discovered by ZTF and classified as a SLSN-I by \cite{Perez_22ojm}. We include photometry from ZTF and ATLAS. We assign a bronze label given the very uncertain redshift estimate of either $z \sim 0.28$ to $z \sim 0.48$.

\subsection{PSNJ000123}
PSN J000123+000504 was discovered by \cite{Kostrzewa13} and classified as a possible SLSN-I based on the light curve alone. We assign a bronze label given that there is no spectra from the supernova. We include photometry from \cite{Kostrzewa13}, which the authors correct for extinction.

\subsection{CSS140925}
CSS140925:005854+181322 was discovered by the CRTS and classified as a Type-I by \cite{Campbell14}, and the host galaxy study presented in \cite{Schulze18} as a SLSN-I. Based on the reported redshift of $z = 0.46$ the peak absolute magnitude of the SN is $M_C = -23.12$. Nevertheless, there are no public spectra of the source and we are unable to confirm this classification. We include photometry from the CRTS and Las Cumbres GSP. Given the lack of spectra and sparse photometry we assign a bronze label.

\subsection{ASASSN15no}
ASASSN-15no was discovered by ASAS-SN and presented in \cite{Benetti18} as a SN with features of SLSN-I, normal SNe Ib/c, but also signs of interaction, such as hydrogen emission lines. The nature of this source appears very distinct from normal SLSNe, and we therefore assign a bronze label.

\subsection{DES16C3cv}
DES16C3cv was discovered by DES and classified as a ``silver" SLSN-I by \cite{Angus19}. We presented DES16C3cv as a Ic-like LSN in \cite{Gomez22_LSN} at a redshift of $z = 0.727$. We assign a bronze label given the spectra and light curve most closely resemble a Type Ic SN.

\subsection{LSQ14fxj}
LSQ14fxj was classified as a SLSN-I by \cite{Smith14}, with an erratum from \cite{Galbany14}. \cite{Margutti18} present Swift-XRT observations of the source, which are all upper limits. The light curve has very sparse optical data with only one detection from \cite{Galbany14}, which is why we assign a bronze label. We include one spectrum from WISeREP.

\subsection{MLS121104}
MLS121104:021643+204009 (=LSQ12fzb) was discovered by the CRTS and classified as a SN SLSN-Ic by \cite{Fatkhullin12} and later classified as a SLSN-I by \cite{Lunnan14}. Nevertheless, there are no public spectra of the source, so we are unable to confirm this classification. There is only one epoch of photometry from the CRTS. Given the lack of spectra and photometry, we assign a bronze label.

\subsection{PS112zn}
PS1-12zn (=PSc380044) was included in the SLSN-I host galaxy sample of \cite{Lunnan14}. However, \cite{Lunnan18} later excluded it from their sample since the SN spectrum lacks the O II and broad UV features typical in SLSNe. Since the spectrum does not cover H$\alpha$, the authors can not rule out a hydrogen-rich event. We therefore assign a bronze label. At the given redshift, the peak magnitude of the is $M_r \sim -21.0$, consistent with a SLSN. We include photometry from \cite{Hosseinzadeh20, Villar20}, excluding data before MJD = 56010, before the nominal SN explosion.

\subsection{SDSS17789}
SDSS-II SN 17789 was classified as a SLSN by \cite{Sako18} as part of the Sloan Digital Sky Survey-II Supernova Survey. The authors do not specify if the source is hydrogen rich or poor, and there is no public spectra of the source. We include photometry from \cite{Sako18}. Given the lack of public information on the source, we assign a bronze label.

\subsection{UID30901}
UID30901 was presented in \cite{Hueichapan22} as a SLSN-I discovered in the UltraVISTA survey. We include photometry from \cite{Hueichapan22}. Nevertheless, we assign a bronze label since there are no spectra of the SN or the host galaxy.

\subsection{SN1000}
SN1000+0216 was classified as a SLSN-I by \cite{Cooke12}. Given that the SN was discovered only after stacking long baselines of photometry, only a late-time spectrum exists, which makes it hard to verify the SLSN nature of the source and we assign a bronze label. The peak absolute magnitude of $M_r \sim -21.6$ is within the SLSN range. We include photometry from \cite{Cooke12}.

\subsection{SN2213}
SN2213-1745 was classified as a SLSN-I by \cite{Cooke12}. Given that the SN was discovered only after stacking long baselines of photometry, only a late-time spectrum exists, which makes it hard to verify the SLSN nature of the source and we assign a bronze label. The peak absolute magnitude of $M_r \sim -21.1$ is within the SLSN range. We include photometry from \cite{Cooke12}.


\section{Not Superluminous Supernovae}\label{sec:not_slsn}

\subsection{2009bh}
SN\,2009bh (=PTF09q) was classified as a possible SLSN-I by \cite{Quimby18} and as a SN Ic by \cite{Kasliwal09}. The peak absolute magnitude of $M_R \sim -18.6$ is not consistent with being a SLSN-I. Moreover, there is only one photometry data point publicly available.

\subsection{2019fiy}
SN\,2019fiy (=ZTF19aauiref =PS19agg) was classified as a SLSN-I by \cite{Yan19_slsne} and \cite{Perley19_fiy}, as well as included in the SLSN-I sample from \cite{Chen22a}. \cite{Yan19_slsne} claim a tentative redshift of $z = 0.67$ due to the low S/N spectrum, but also find a match to lower redshift SNe. A redshift of $z = 0.67$ would make SN\,2019fiy the most luminous SLSN-I ever discovered by a large margin, we find a spectral match to a SN Ia at $z \sim 0.13-0.14$, a much more likely redshift and classification.

\subsection{2019hcc}
SN\,2019hcc (=Gaia19cdu =ATLAS19mgw) was classified as a SLSN-I by \cite{Frohmaier19}. However, the peak absolute magnitude of the SN is $\sim -17.7$, much too faint to be consistent with a SLSN. The source has since been reclassified as a Type II SN (Inserra et al., in prep).

\subsection{2021ahpl}
SN\,2021ahpl (=ZTF21aalkhot =Gaia22asc =PS22bca =ATLAS22fga) was classified as a SLSN-I by \cite{Smith22_ahpl}. We include photometry from ZTF, GSA, PS1, and ATLAS. And our own PSF photometry from FLWO and Las Cumbres GSP images. We exclude three possibly spurious detections from ZTF and GSA $sim 400$ days before explosion. Nevertheless, the peak absolute magnitude of this SN is $M_r \sim -19.7$ and the spectrum does not resemble a normal SLSN, and we therefore argue this is likely a transient of a different nature.

\subsection{2020jhm}
SN\,2020jhm (=ZTF20aayprqz =PS20dfm =ATLAS20luz) was classified as a SLSN-I at a redshift of $z = 0.06$ by \cite{Perley20}, which would make the SN peak magnitude $M_r = -20.3$. Instead, we find a good spectral match to a SN Ia at $z = 0.05$, with a corresponding peak magnitude of $M_r = -19.9$, within the typical magnitudes of SNe Ia. The light curve also appears very similar to typical SNe Ia.

\subsection{2022aig}
SN\,2022aig (=ZTF22aaagqvw =ATLAS22cpm) was classified as a relatively luminous SN Ic at a redshift of $z = 0.4$ by \cite{Aamer22}. Nevertheless, we determine a redshift of $z = 0.31$ from host emission lines in a late-time LDSS3C spectrum. At this redshift the peak magnitude is $M_r \sim -18.9$ and therefore no longer superluminous.

\subsection{2022csn}
SN\,2022csn	(=ZTF22aabimec =Gaia22ayp =PS22bju =ATLAS22ggz) was classified as a SLSN-I by \cite{Srivastav22, Srivastav22b}, but later reclassified as a TDE by \cite{Arcavi22}. We update the redshift of $z = 0.15$ quoted by \cite{Srivastav22} to $z = 0.147$, determined from narrow host emission lines from our own late-time Binospec spectrum.

\subsection{2022czy}
SN\,2022czy (=PS22bvf =ATLAS22gtw) was classified as a SLSN-I by \cite{Fulton22b, Fulton22a} and \cite{Hinkle22}, but later reclassified as a TDE by \cite{Blanchard22b_czy, Blanchard22_czy} on the basis of a broad H-alpha emission component.

\subsection{2022vxc}
SN\,2022vxc (=ZTF22abcvfgs =Gaia22ebz =PS22kdy) was classified as a SLSN-I by \cite{Harvey22, Harvey22b}. We include photometry from ZTF, ATLAS, Gaia, and PS1, and our own PSF photometry of FLWO and Las Cumbres GSP images. The light curve of SN\,2022vxc shows a peculiar triangular-shape, which we are unable to reproduce using standard models. While this object has a spectrum consistent with a SLSN interpretation, its light curve nature is too distinct to be considered within the standard SLSN population. We include one spectrum from \cite{Harvey22}, obtained from the TNS.

\subsection{PTF11mnb}
PTF11mnb was classified as a possible SLSN-I by \cite{Quimby18} since the spectra are consistent with both a SLSN-I or a SN Ic. The source was presented in a detailed study by \cite{Taddia18}, who conclude this is a SN Ic, a 2005bf analog with a double peaked light curve. The peak absolute magnitude of $M_r \sim -18.5$ is most consistent with the SN Ic interpretation.

\end{document}